\documentclass[11pt, onehalfspacing]{article} 

\usepackage[T1]{fontenc}

\usepackage{threeparttable}
\usepackage{multirow}

\usepackage{libertine}

\usepackage{geometry} 
\usepackage{xcolor, graphicx} 

\usepackage{tikz}
\usetikzlibrary{arrows,chains,matrix,positioning,scopes,automata}
\tikzset{
  invisible/.style={opacity=0},
  visible on/.style={alt={#1{}{invisible}}},
  alt/.code args={<#1>#2#3}{%
    \alt<#1>{\pgfkeysalso{#2}}{\pgfkeysalso{#3}} 
  },
}

\usepackage{multicol}

\usepackage{caption}

\usepackage{subfigure}  

\usepackage{pgfplots}
\pgfplotsset{compat=1.16}

\usepackage{float}

\usepackage{booktabs} 
\usepackage{array} 
\usepackage{paralist} 
\usepackage{verbatim} 
\usepackage{amsfonts, amsmath, amssymb, amsthm, euscript}
\usepackage{mathrsfs}
\usepackage{mhsetup, mathtools}
\usepackage{thmtools}
\usepackage{halloweenmath}

\theoremstyle{definition}

\renewcommand\thmcontinues[1]{Continued}

\definecolor{purple}{RGB}{85, 6,139}
\definecolor{teal}{RGB}{2,108,128}
\definecolor{lavender}{RGB}{129, 102, 122}
\definecolor{carolina blue}{RGB}{68, 157, 209}
\definecolor{phthalo blue}{RGB}{2, 8, 135}
\definecolor{purple2}{RGB}{149, 96, 219}
\definecolor{green1}{RGB}{96, 219, 117}
\definecolor{darkblue}{RGB}{0,0,102}

\usepackage{natbib}
\usepackage{hyperref}
\usepackage{nameref}
\hypersetup{
 breaklinks=true,
  colorlinks   = true, 
  urlcolor     = purple, 
  linkcolor    = darkblue, 
  citecolor   = darkblue, 
}

\usepackage{fancyhdr} 
\usepackage{sectsty}
\usepackage{setspace}
\allsectionsfont{\upshape \raggedright \fontsize{13}{15} \selectfont} 

\usepackage[nottoc,notlof,notlot]{tocbibind} 
\usepackage[titles,subfigure]{tocloft} 

\begin{document}

\title{\textsc{Can an LLM Learn Preferences from Choice Data?}\footnote{
The authors thank the editor and two anonymous referees for many helpful suggestions. We also thank Syngjoo Choi, Federico Echenique, Keaton Ellis, Daeyoung Jeong, Dukgyoo Kim, Seo-young Silvia Kim, R. Vijay Krishna, Jinhyuck Lee, Po-Hsuan Lin, Tracy Xiao Liu, Ryan Oprea, Erkut Ozbay, Leeat Yariv, and seminar participants at Florida State University, Hanyang University, KAIST, Korea Development Institute, Korea University, Sungkyunkwan University, UC Santa Barbara, Yonsei University, and the North American ESA meeting for helpful comments and suggestions. The author names are in alphabetical order and all have equally contributed to this paper. A previous verison of this manuscript was circulated under the title "Learning to be Homo Economicus: Can an LLM Learn Preferences from Choice Data?"}}

\author{
\begin{tabular}[t]{c@{\hskip 3em}c@{\hskip 3em}c}
 Jeongbin Kim\footnote{Department of Economics, Florida State University, \href{mailto:jkim33@fsu.edu}{jkim33@fsu.edu}} &
 Matthew Kovach\footnote{Department of Economics, Purdue University, \href{mailto:mlkovach@purdue.edu}{mlkovach@purdue.edu}} &
 Kyu-Min Lee\footnote{Department of Management Information Systems, Chungbuk National University, \href{mailto:kyuminlee@kaist.ac.kr}{kyumin.lee@cbnu.ac.kr}} \\[0.5em]
 \multicolumn{3}{c}{
   Euncheol Shin\footnote{College of Business, Korea Advanced Institute of Science and Technology, \href{mailto:eshin.econ@kaist.ac.kr}{eshin.econ@kaist.ac.kr}} 
   \hskip 3em
   Hector Tzavellas\footnote{Department of Economics, Virginia Tech, \href{mailto:hectortz@vt.edu}{hectortz@vt.edu}}
 }
\end{tabular}
}
 
\date{\today}



\maketitle

\vspace{5 mm}

\noindent{\textbf{Abstract:} Can large language models (LLMs) learn a decision maker’s preferences from observed choices and generate preference-consistent recommendations in new situations? We propose a portable Simulate-Recommend-Evaluate framework that tests preference learning from revealed-choice data by comparing LLM recommendations with optimal choices implied by known preference primitives. We apply the framework to choice under uncertainty using the disappointment aversion model. Recommendation accuracy improves as models observe more choices, but learning is heterogeneous across preference types and LLMs: GPT learns risk aversion better than disappointment aversion, Gemini performs best in high disappointment-aversion regions, and Claude shows the broadest effective learning across parameter regions.

\vspace{5 mm}


\vspace{3 mm}


\vspace{10 mm}

\onehalfspacing

\pagebreak


\section{Introduction}\label{section:introduction}

Large language models (LLMs) are increasingly used as decision aids in domains such as investment advice, insurance choice, and consumption planning. In these settings, optimal decisions depend critically on individual preferences, which are often heterogeneous and not directly observed. From an economic perspective, this implies that a key criterion for evaluating the recommendations of such systems is whether they \emph{align} with the preferences of the user. In the standard framework, choices reveal these preferences, so welfare-improving recommendations must be consistent with the structure reflected in observed behavior. If an LLM cannot infer this structure from limited revealed-choice data, even recommendations that appear reasonable may lead to systematically suboptimal decisions for particular users, especially when preferences differ in economically meaningful ways.

This motivates the following question: \emph{can an LLM learn a decision maker’s preferences from observed choices and generate preference-consistent recommendations in novel decision problems?} Our aim is not to assess whether LLMs behave “rationally” or “human-like,” but whether they can recover and act on an individual user’s preference structure when deployed as a recommendation system.

A growing literature studies LLMs as economic agents, examining whether their behavior resembles that of humans or satisfies standard notions of rationality. For example, \citet{horton2023} shows that LLMs can reproduce behavioral patterns from classic experiments, while \citet{Chenetal:2023:PNAS} document LLMs consistency with utility maximization. While this body of work has advanced our understanding along two important dimensions—replicating human behavior and exhibiting internally consistent choice—it has largely left aside whether LLMs can align with the preferences of a particular decision maker. Yet this dimension is crucial for LLM deployment: performance that appears rational or human-like is not sufficient if recommendations do not reflect the user’s preferences. By focusing on this dimension, our paper complements the existing literature and, together with it, provides a more complete picture of LLMs as economic agents.

To address our question, we propose the \emph{Simulate-Recommend-Evaluate} (SRE) framework, a portable methodology for evaluating preference learning from revealed-choice data. We start by specifying a decision environment together with an explicit preference representation $\theta$, and proceed to generate a revealed-choice dataset $D(\theta)$ implied by that preference representation. The LLM is then placed into the same environment through a fixed interface and provided only with revealed choices—not the preference parameters or underlying utility representation themselves. The LLM is asked to generate recommendations in new, counterfactual decision problems. These recommendations are evaluated using transparent, welfare-relevant metrics, distinguishing between non-parametric measures of recommendation quality and parametric measures that diagnose which aspects of preferences the model has learned. \autoref{figure:procedure} summarizes this framework.

\begin{figure}[ht]
\centering
\includegraphics[width=0.75\linewidth]{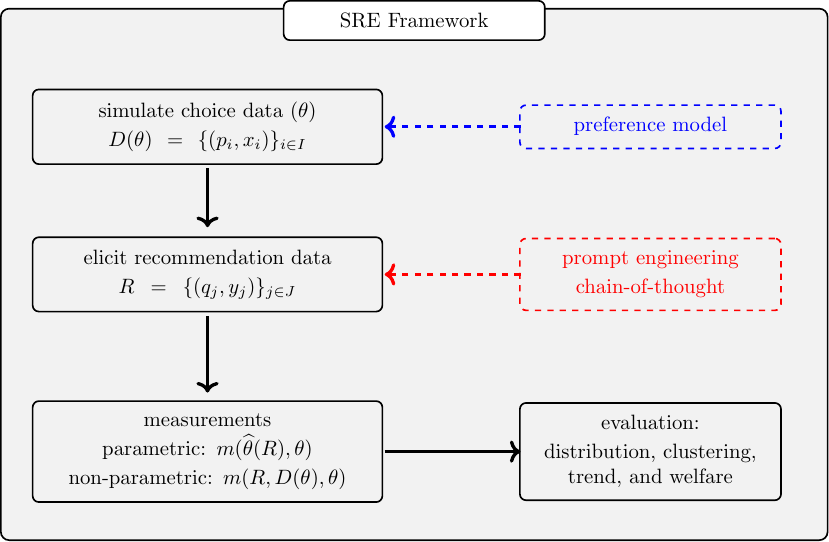}
\caption{Summary of the SRE procedure.}
\label{figure:procedure}
\end{figure}

Our approach establishes a standard for identifying preference learning in our context. Improvement in recommendation quality alone is not sufficient: learning must be demonstrated out of sample and evaluated relative to the true optimal choices implied by the decision maker’s preferences. By separating the generation of choice data, the production of recommendations, and their economic evaluation, the framework provides a disciplined way to study whether LLMs internalize preference primitives rather than merely interpolating past behavior.

We illustrate the framework using choice under uncertainty, a canonical environment in economics with well-understood preference representations. Specifically, we study the disappointment aversion model of \citet{Gul:1991:ECMA}. The model is characterized by two parameters: risk aversion, which governs curvature over outcomes, and disappointment aversion, which governs sensitivity to downside realizations relative to expectations. This environment is well suited to our framework because the preference parameters are behaviorally and structurally interpretable and can be cleanly identified from choice data. This allows us to evaluate not only recommendation quality but also which underlying preference components the LLM successfully recovers.

We construct a two-dimensional parameter grid spanning both disappointment-averse and elation-seeking regions, with the risk aversion range consistent with the experimental findings of \citet{Choietal:2007:AER}. For each combination of risk aversion and disappointment aversion parameters, we generate revealed-choice datasets of varying size and provide these data to the LLM. We then evaluate the model’s recommendations on new budget problems not seen during training. Our evaluation combines non-parametric measures of recommendation quality—such as distance from the true optimal allocation, utility loss relative to the optimum, and
deviation from risk-neutral allocations—with parametric recovery exercises (nonlinear least squares) that ask which preference parameters would best rationalize the LLM’s recommendations.

Our analysis reveals two primary findings regarding LLM learning capacity:

\begin{itemize}
\item \textbf{Heterogeneous learning performance:} Although recommendation quality generally improves as the size of the revealed-choice dataset increases, the rate and magnitude of this improvement vary substantially across the preference space. We identify specific regions where the LLM reliably learns user preferences and others where it persistently fails to align with the decision maker. 

\item \textbf{Asymmetric parameter recovery:} Using parametric decomposition, we find that the GPT learns risk aversion  significantly better than it learns disappointment aversion. This distinction is economically meaningful because disappointment aversion governs sensitivity to downside outcomes, which is central in contexts such as insurance choice and portfolio allocation under tail risk. Gemini's learning is concentrated in the high disappointment aversion region rather than the elation-seeking one, while Claude exhibits broad effective learning.
\end{itemize}

The framework, therefore, reveals which preference features are currently learnable from observed choices and where learning breaks down across models. More broadly, our primary contribution is methodological. While risky choice serves as a theoretically clean and economically relevant setting, the approach is not specific to uncertainty. It can be applied to any decision environment in which preferences can be specified and recommendations can be evaluated using transparent economic criteria, including intertemporal choice, social preferences, and other domains where personalized decision support is increasingly automated. By providing a systematic way to evaluate preference learning from revealed-choice data, this paper offers a toolkit for assessing the economic reliability of LLM-based recommendation systems.

\section{Related Literature}

\noindent \textbf{Revealed preference and disappointment aversion.} We build on the revealed preference literature that provides methods to identify preferences from choice data. \citet{Choietal:2007:AER} propose an experimental design for portfolio choice under risk that allows researchers to test consistency with utility maximization and estimate preference parameters. We adopt this design to generate simulated choice data from known preference parameters and evaluate whether LLMs can recover these parameters from recommendations. We focus on the disappointment aversion model of \citet{Gul:1991:ECMA}, which captures sensitivity to downside risk as well as conventional risk aversion. Disappointment aversion has been shown to be quantitatively important in financial markets, with direct implications for portfolio allocation, insurance demand, and robo-advising \citep[e.g.,][]{Angetal:2005:JFE, Jouinietal:2014:MS, Augustinetal:2021:MS, Routledge2010}. Our key contribution is to provide a systematic procedure to measure LLMs' ability to learn disappointment aversion from choice data, finding that while LLMs can learn risk aversion reasonably well, they systematically fail to learn disappointment aversion.

\vskip+1em

\noindent \textbf{LLMs as economic agents.} Recent research examines whether LLMs can replicate human behavior across a range of domains \citep[e.g.,][]{LeMens:2023:PNAS, Webbetal:2023:NHB}, including psychology \citep{Meietal:2024:PNAS}, portfolio choice \citep{romanko2023chatgptbased, koandlee2023}, financial literacy \citep{NiszczotaandAbbas2023}, and strategic games \citep{Guo:2023:WP, BrookinsandDeBacker:2023:WP, Akata:2023:WP, Meietal:2024:PNAS, Xieetal:2025:PNAS}. \citet{goli2024frontiers} show that LLMs do not fully capture human time preferences; we complement this evidence by documenting an analogous asymmetry in the learning of risk aversion and disappointment aversion from revealed-choice data. \citet{horton2023} proposes using LLMs as simulated economic agents and shows that GPT-3 reproduces several behavioral patterns observed in classic experiments. The paper most closely related to ours is \citet{Chenetal:2023:PNAS}, which uses the experimental design of \citet{Choietal:2007:AER} to study whether GPT’s choices are consistent with utility maximization.

This body of work provides important insights into the behavioral properties of LLMs when they are treated as autonomous decision makers. Our focus is different. Rather than evaluating the LLM’s own choices, we study whether an LLM can serve as a decision aid by learning another agent’s preferences from observed choice data. In our setting, the relevant criterion is not whether the LLM’s behavior is internally consistent or rational, but whether its recommendations align with the decision maker’s true preferences under welfare-relevant economic metrics. This shift in perspective gives rise to a different methodological problem: how to evaluate preference learning out of sample against known preference primitives, and how to diagnose which dimensions of preference are and are not being learned.

\vskip+1em

\noindent \textbf{Human-AI interaction.} Our paper is broadly related to the recent literature on human-AI interaction \citep[e.g.,][]{DAcuntoandRossi2023, FedykEtAl2025, Luetal:2023:Advisory, NoyandZhang:2023:Science}. This literature has primarily focused on processing natural language text to provide valuable insights and feedback to humans. For example, \citet{FedykEtAl2025} study AI-generated investment advice and show that AI systems can reproduce aspects of human financial reasoning while exhibiting perception biases depending on prompting context. Our paper contributes to this literature by evaluating the learning capability of LLMs through a controlled economic experiment. One promising application is financial robo-advising \citep[e.g.,][]{DAcuntoandRossi2023, Luetal:2023:Future}. In the current paper, we provide a systematic procedure to evaluate whether LLMs can learn preferences from observed choice data, a natural alternative to the simple questionnaires typically used by robo-advisors. Our main findings show that LLMs can learn risk aversion but systematically fail to learn disappointment aversion. Since disappointment aversion captures sensitivity to downside risk, this suggests the need for caution in deploying LLMs as personalized financial advisors.


\section{Procedure and Estimation Method}
\label{sec:methodology}

We propose the systematic SRE framework illustrated in \autoref{figure:procedure}. This procedure consists of three stages: (i) generating simulated choice data for preference learning from a known utility model, (ii) prompting an LLM to function as a recommendation system using this choice data, and (iii) evaluating the generated recommendations against the ground-truth preferences using both parametric and non-parametric metrics.


\subsection{Decision Environment: Portfolio Choice Experiment}
\label{subsec:choice_experiment}

We adopt the experimental design of \citet{Choietal:2007:AER} concerning portfolio choice under risk. In each decision round, the decision maker is presented with two equally likely states of the world, $s \in \{A, B\}$, and a budget to allocate between two corresponding Arrow securities. Let $x_A$ and $x_B$ denote the demand for the securities paying 1 unit of numeraire in state A and state B, respectively. The budget constraint is given by $p_A x_A + p_B x_B = 100$, where $p_A$ and $p_B$ are the prices for the two securities, respectively. Prices are randomly drawn independently from a uniform distribution $U[1,10]$, subject to the constraint that $p_{\max} \geq 5$ to ensure sufficient variation in budget sets. The budget size is set to $100$ to avoid very small values of the choice vector $(x_A, x_B)$.\footnote{Our setting differs from \citet{Chenetal:2023:PNAS} In their design, the original task in \citet{Choietal:2007:AER} is converted into a choice problem over returns: in each round, GPT allocates 100 points between two assets with different payoffs, and only one asset's return is realized with equal probability. This transformation was motivated by earlier GPT models' failure to satisfy the budget constraint. In contrast, we find that GPT-5 meets the budget constraint without such modifications, so we use the original description of the choice environment in \citet{Choietal:2007:AER}.}


\subsection{Data Generation and Behavioral Model}

To construct the data for preference learning, we simulate optimal choices using the disappointment aversion (DA) model of \citet{Gul:1991:ECMA}. This model allows us to capture sensitivity to downside outcomes, a critical dimension for financial decision-making \citep[e.g.,][]{Angetal:2005:JFE, Routledge2010, Jouinietal:2014:MS, Augustinetal:2021:MS}. The utility function is defined as:
\begin{align*}
U(x) = \frac{1}{2+\beta} u(x_{\max}) + \left( 1 - \frac{1}{2+\beta} \right) u(x_{\min}),
\end{align*}
where $x_{\max} = \max\{x_A, x_B\}$, $x_{\min} = \min\{x_A, x_B\}$, and the disappointment aversion parameter $\beta > -1$ that governs the weight assigned to the worse (disappointing) outcome. $\beta > 0$ indicates disappointment aversion, while $\beta < 0$ indicates elation seeking. $u: \mathbb{R}_+ \to \mathbb{R}$ is the Bernoulli utility function that is assumed to be the constant relative risk aversion (CRRA) utility function, $u(z) = \frac{z^{1-\rho} - 1}{1-\rho}$, admitting a risk aversion parameter $\rho > 0$ that governs the curvature of the indifference curves. Thus, the DA model is fully summarized by the preference parameter vector $\theta = ( \beta, \rho)$.

Using the DA model, we generate simulated datasets for a grid of parameter $\theta = (\beta, \rho)$. Specifically, as illustrated in \autoref{figure:parameter_space}, we set preference $\theta = (\beta, \rho) \in  [-0.9, 0.9] \times [0.2, 0.95]$, which includes the median preference vector of $(\beta_{\text{median}}, \rho_{\text{median}}) = (0.121, 0.451)$, estimated from the data of \citet{Choietal:2007:AER}.\footnote{This calculation is based on the dataset generated by the symmetric treatment (i.e., equally likely states) in \citet{Choietal:2007:AER}. They report a median $\rho$ of $0.438$ and a median $\beta$ of $0.179$, imposing the restriction that $\beta \geq 0$ using the same NLLS method. As we do not require $\beta$ to be positive, our median estimates are lower than theirs.} The red box in \autoref{figure:parameter_space} indicates the region of the preference parameters in which typical individuals' preferences lie, as most people exhibit disappointment aversion. For each parameter $\theta$, we generate a \textit{history} of optimal choices $D(\theta) = \{(p_t, x_t)\}_{t=1}^{h}$, where $h \in H = \{ 0, 1, 5, 25, 125 \}$ denotes the size of the training history provided to an LLM.

\begin{figure}[ht]
    \centering
    \includegraphics[width=0.95\linewidth]{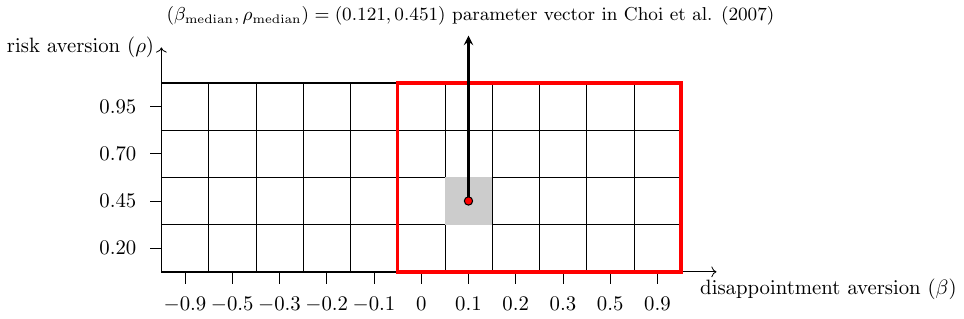}
\caption{Parameter space and grids.}
\label{figure:parameter_space}
\end{figure}

\subsection{LLM Recommendation Task}

We employ GPT-5 (OpenAI), Gemini (Google), and Claude (Anthropic), three of the most widely adopted commercial LLMs, as the AI recommendation systems.\footnote{We use the most recent version of the three models at the time of data collection.} All three APIs share a common prompting structure that organizes prompts into system and user roles.\footnote{Detailed prompts and screenshots are provided in \autoref{sec:prompts}. We also conducted the recommendation tasks without choice data using the prompt in \autoref{sec:prompts}. Without data, the estimated parameters are both zero for GPT, which implies that, without data, GPT recommends choices that maximize expected return.} In particular, the system role instructs the model to act as a ``recommendation system for our valuable customers''. As such, we present results for GPT-5 in the main text and gather detailed results from other LLMs in \autoref{sec:additional_tables_and_figures}. The prompt structure is as follows:
\begin{itemize}

\item \textbf{Context:} A description of the decision environment and the objective to provide recommendations the customer ``likes most''.

\item \textbf{Learning data:} A simulated choice data table summarizing $h$ rounds of the customer's past choices (prices and allocations).

\item \textbf{Recommendation ask:} A new set of 25 budget problems (prices) for which the LLM must output recommended allocations.

\end{itemize}

The first part (context) is written in the system role, while the second and third parts (learning data and recommendation ask) are written in the user role.\footnote{Note that our single-turn prompt design does not require an explicit assistant role, as these models perform chain-of-thought reasoning before generating a response \citep{openai2025gpt5}.} Notably, the learning data prompt reveals neither the DA model nor the true parameter $\theta$ to the LLM. These are only observable to the researcher. To test the model's ability to learn preferences from data, we vary the size (i.e., $h$) of the simulated choice data provided in the prompt. For each $h$, the above process yields recommendation data $R = \{(q_j, y_j)\}_{j=1}^{25}$ for each simulation instance, where $q_j$ is a price vector and $y_j$ is a recommendation vector. Due to the inherent randomness of LLM responses, we generate 30 recommendations for the same budget sets. To ensure independence across recommendations, we invoke a fresh API call for each recommendation task. As such, we treat this recommendation dataset as 30 samples of recommendation data for each parameter $\theta$.\footnote{The choice of 30 is due to the cost of API usage.} This recommendation dataset depends on the given learning set $D(\theta)$.


\subsection{Measurements}

We evaluate recommendation quality using both parametric and non-parametric methods as follows. Note that we can obtain the value of each measure for each recommendation data.

\vskip+1em

\noindent \textbf{Non-parametric measurements.} To  measure recommendation quality non-parametrically, we employ three distinct metrics. First, we consider the average normalized vector distance between the recommended allocation $y_j$ and the optimal allocation $x_j$ for the budget set under which $y_j$ is chosen:
\begin{align*}
\text{AVD}(R) = \frac{1}{25} \sum_{j=1}^{25} \frac{\| y_j - x_j \|}{\| x_j \|}.
\end{align*}

Second, we compute the average normalized difference in the risk neutrality measure. For recommendation and choice vectors, $y_j$ and $x_j$, risk neutrality, $y_j^{\text{RN}}$ and $x_j^{\text{RN}}$, is defined as the proportion of the allocation to the cheaper asset. It equals one if the entire budget is invested in the cheaper asset (i.e., maximizing expected value), and not smaller than $1/2$ if the choice does not violate the first-order stochastic dominance \citep{Choietal:2007:AER}. Consequently, the average normalized difference in the risk neutrality measure is defined as
\begin{align*}
\text{ARN}_{RN}(R) = \frac{1}{25} \sum_{j=1}^{25} \frac{| y_j^{\text{RN}} - x_j^{\text{RN}}|}{| x_j^{\text{RN}} |}.
\end{align*}

Third, to capture welfare implications, we compute the average normalized utility loss:
\begin{align*}
\text{AWL}(R) = \frac{1}{25} \sum_{j=1}^{25} \frac{ | U(x_j; \theta) - U(y_j; \theta) | }{ | U(x_j; \theta) |}.
\end{align*}
This measures the ex-ante welfare reduction a user would suffer by following the LLM's advice compared to their true optimal choice.

Note that for all three measures, the value is zero if the LLM generates recommendations that are perfectly aligned with the underlying preference.

\vskip+1em

\noindent \textbf{Parametric recovery.} We recover the implied parameters $\widehat{\theta} = (\widehat{\beta}, \widehat{\rho})$ from the recommended allocations using the non-linear least squares (NLLS) method employed by \citet{Choietal:2007:AER}.\footnote{Obviously, one may apply other parametric recovery methods, such as that of \citet{halevyetal:2018:JPE}. However, we use the NLLS method, as it can be easily applied to other settings, such as time preferences with the $\beta$--$\delta$ discounting model \citep[e.g.,][]{Andreonietal:2012:AER} and other-regarding preference models \citep[e.g.,][]{AndreoniMiller2002}.} Following their approach, boundary observations are incorporated by replacing any zero component with a small consumption level such that the demand ratio equals $\nu$ or $1/\nu$, where $\nu = 10^{-3}$. Specifically, we identify $\widehat{\theta}$ by minimizing the squared distance between the log-ratios of the recommended allocations and the theoretical optimal allocations:
\begin{align*}
\min_{(\beta, \rho)} \sum_{j=1}^{25} \left( \ln \left(\frac{y_{A,j}}{y_{B,j}} \right) - f\left(\ln \left(\frac{q_{A,j}}{q_{B,j}} \right); \theta \right) \right)^2,
\end{align*}
where $f$ is the piecewise nonlinear function derived from the first-order conditions of maximizing the DA utility with the CRRA Bernoulli utility function.\footnote{The full specification of $f$ is provided in \autoref{sec:calculation}.} We then define the normalized learning error for each parameter:
\begin{align*}
&\text{NLE($\beta$):} \quad \mathcal{L}_\beta = \frac{|\hat{\beta} - \beta|}{|\beta| + \mathbf{1}\{ \beta = 0 \}} \\
&\text{NLE($\rho$):} \quad \mathcal{L}_\rho = \frac{|\hat{\rho} - \rho|}{|\rho|}.
\end{align*}
The denominator in NLE($\beta$) contains an indicator function, $\mathbf{1}\{ \beta = 0 \}$, in order to make it well-defined for $\beta = 0$, which corresponds to the standard expected utility case. Thus, the NLE can be interpreted as the absolute percentage error between the recovered parameter and the true parameter, except when $\beta = 0$. These metrics allow us to decompose the learning performance and identify asymmetries in each AI recommendation system's ability to infer risk aversion versus disappointment aversion. Like the non-parametric measures, the above measures take value $0$ if the LLM generates recommendations that are perfectly aligned with the underlying preference.


\subsection{Bootstrapping Method}

To statistically evaluate the significance of our results and construct meaningful means and confidence intervals for each measure, we employ a bootstrap approach. We rely on bootstrapping for two primary reasons. First, the data-generating process involves an LLM, whose error structure is unknown and likely non-normal. Thus, standard asymptotic assumptions for statistical inference may not hold. Second, our key measures (e.g., the average normalized vector distance and the average normalized learning errors) are functions of the random recommendation data, which makes the derivation of analytical standard errors intractable.

We implement the bootstrap procedure as follows. For each experimental condition (defined by the history size $h$ and the true preference parameter vector $\theta$), we collect a bootstrap resample of size $30$, which is independently and identically drawn from the recommendation sets of size $30$. This random selection is with replacement, and a recommendation set is dropped and redrawn if the estimated parameters $\beta$ and $\rho$ are too large ($5$ for both parameters).\footnote{For the recommendation samples generated by GPT, our NLLS method calculates that 1.86\% of the estimates are strictly greater than 5 for $\rho$, while no estimate is strictly greater than 5 for $\beta$. Likewise, for Gemini and Claude, there are no cases in which $\beta$ is estimated to be strictly greater than 5. Only $\rho$ is estimated to exceed 5, with proportions of 0.41\% and 2.56\%, respectively. For the ML methods, outliers are observed only for $\rho$ in the SVR case, where 7.58\% of the estimates exceed 5. Since these baselines are deterministic and produce a single recommendation set per condition, the drop-and-redraw rule does not apply; outlier estimates are retained in the bootstrap procedure.} For the given resample of size $30$, we then calculate the four measurements. We generate $B=10,000$ bootstrap samples of the measurements by repeating the above resampling with replacement procedure. For each bootstrap sample, we calculate the mean of the target metric. Finally, we obtain point estimates of the mean for each measure and construct $95\%$ confidence intervals using the $2.5$th and $97.5$th percentiles of the bootstrap distribution.\footnote{We note that our confidence intervals quantify the stochasticity of the LLM's text generation conditional on a fixed training history and a fixed set of evaluation budget problems.}


\section{Results}

\subsection{Non-Parametric Analysis: Recommendation Quality}

We begin by evaluating the quality of the LLM's recommendations with non-parametric measures. 

\vskip+1em

\noindent \textbf{Heterogeneity of learning.} Our first main result is that, on average, the quality of recommendations improves as the GPT is provided with larger histories of revealed preference data. \autoref{fig:non_parametric_results}-(a) illustrates the average normalized vector distance for a representative agent ($\beta=0.1, \rho=0.45$) across increasing history sizes ($h \in \{1, 5, 25, 125\}$).\footnote{History figures for all parameter values are reported in Appendix \ref{label:subsection:additional_GPT}.} Interestingly, even with only a few simulated choice observations ($h=5$), the average normalized vector distance (AVD) is below 0.05, indicating that the average relative Euclidean deviation from the optimal allocation is under 5\%. This result is consistent with the finding that LLMs can generate high-quality results on many tasks using only a few examples or instructions, without additional training or fine-tuning \citep{brown2020languagemodelsfewshot}. Notably, we do not observe further improvement in recommendation quality at large history sizes ($h=125$) compared to the case of a very small history size ($h = 5$), suggesting that simply adding more data does not guarantee convergence to the optimum for all parameterizations.

\begin{figure}[ht]
\centering
\subfigure[AVD changes over history for $(\beta, \rho) = (0.1, 0.45)$.]{
    \includegraphics[width=0.48\linewidth]{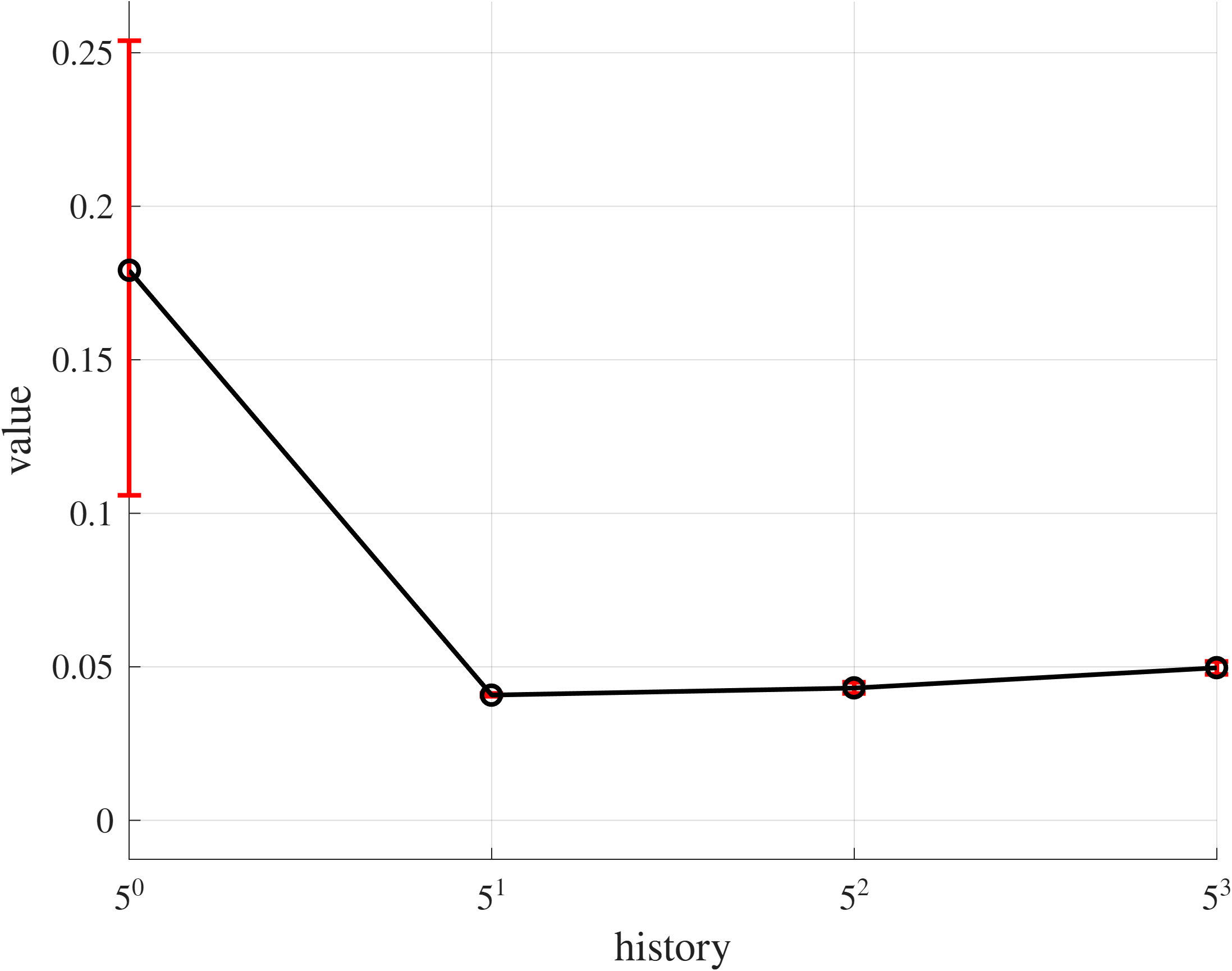}
    \label{fig:vec_distance}
}
\hfill 
\subfigure[Relative distance improvement.]{
    \includegraphics[width=0.48\linewidth]{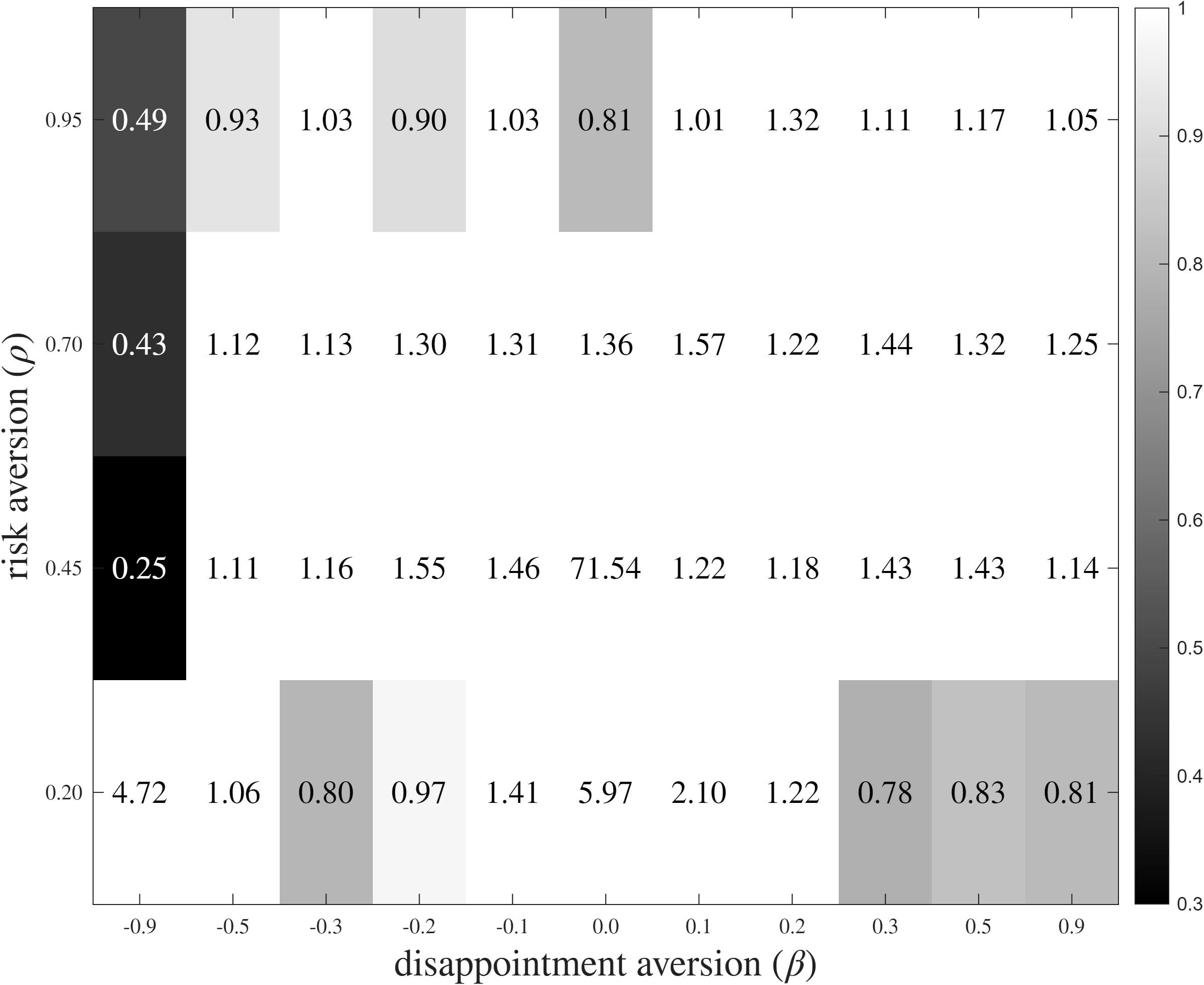}
    \label{fig:heatmap}
}    
\caption{Panel (a) shows the aggregate trend of the average normalized vector distance as the history size ($h$) increases. The bars represent 95\% bootstrap confidence intervals. Panel (b) displays the heatmap of the vector distance improvement ratio across the preference parameter space. White cells represent ratios at or above 1, and black cells represent ratios below 0.3. Panel (b) is generated based on Appendix \ref{label:subsection:additional_GPT}.}
\label{fig:non_parametric_results}
\end{figure}

\autoref{fig:non_parametric_results}-(b) presents a heatmap of the vector distance improvement ratio across the preference parameter space $(\beta,\rho)$. For each grid point representing a parameter vector $(\beta, \rho)$, the improvement ratio is defined by dividing the measure at $h=125$ by that at $h=5$. This ratio represents the relative distance between the recommendation with history $h = 125$ and the optimal choice, compared to the distance with history $h = 5$. Thus, darker cells indicate larger improvements (i.e., greater reductions in recommendation error), while lighter cells indicate little or no improvement. \autoref{fig:non_parametric_results}-(b) presents a heatmap of this ratio across the grid of risk aversion ($\rho$) and disappointment aversion ($\beta$) parameters. 

The results show that, across most of the preference parameter space, the improvement ratio remains close to 1 (white cells), implying little change in recommendation error as the history length increases from $h=5$ to $h=125$. The main exception is the extreme elation-seeking boundary (i.e., $\beta = -0.9$), where darker cells indicate sizable reductions in error with additional history. For the remainder of the parameter space, including most of the elation-seeking region (i.e., $\beta \in \{ -0.5, -0.3, -0.2, -0.1 \}$) and the entire disappointment-averse region ($\beta > 0$), the heatmap is almost uniformly white across a wide range of risk aversion levels. Overall, these patterns suggest that additional choice history yields limited gains in alignment across most of the preference space, even when the model is provided with a long history of past choices.

Taken together, these findings indicate that additional history helps only for a narrow subset of preference types, and that the LLM's ability to learn preferences from revealed-choice data is otherwise limited. Other non-parametric measures display the same qualitative patterns across the parameter space. In the next subsection, we turn to the DA model to provide a parametric decomposition of these results and to pinpoint the source of the remaining misalignment.


\subsection{Parametric Analysis: Preference Learning and Decomposition}

To diagnose the source of the patterns observed in the non-parametric analysis, we utilize the NLLS estimates to recover the implied preference parameters $(\hat{\beta}, \hat{\rho})$ from the LLM's recommendations. By comparing these estimates to the true generating parameters $(\beta^0, \rho^0)$, we can decompose the learning performance into its two constituent dimensions: risk aversion and disappointment aversion.

\vskip+1em

\noindent \textbf{Asymmetric learning.} We find that the LLM exhibits a distinct asymmetry in its learning capabilities: it learns risk aversion ($\rho$) significantly better than it learns disappointment aversion ($\beta$). To quantify this, we examine the recommendation error improvement ratio for each parameter across the preference grid. \autoref{fig:parametric_heatmaps} displays these ratios for the two parameters.

As shown in \autoref{fig:parametric_heatmaps}-(a), the heatmap for $\rho$ exhibits a clear pattern: at higher levels of risk aversion ($\rho \in \{0.70, 0.95\}$), the cells are predominantly dark, indicating that the estimation error decreases substantially as the history size increases from $h = 5$ to $h = 125$. This suggests that the LLM is capable of identifying and adapting to the curvature of the utility function when risk aversion is sufficiently high. However, at lower levels of risk aversion ($\rho \in \{0.20, 0.45\}$), especially for the case of $\rho = 0.20$, the ratios frequently exceed 1, indicating that additional data does not improve.

\begin{figure}[ht]
\centering
\subfigure[$\rho$ estimation error improvement ratio.]{
\includegraphics[width=0.48\linewidth]{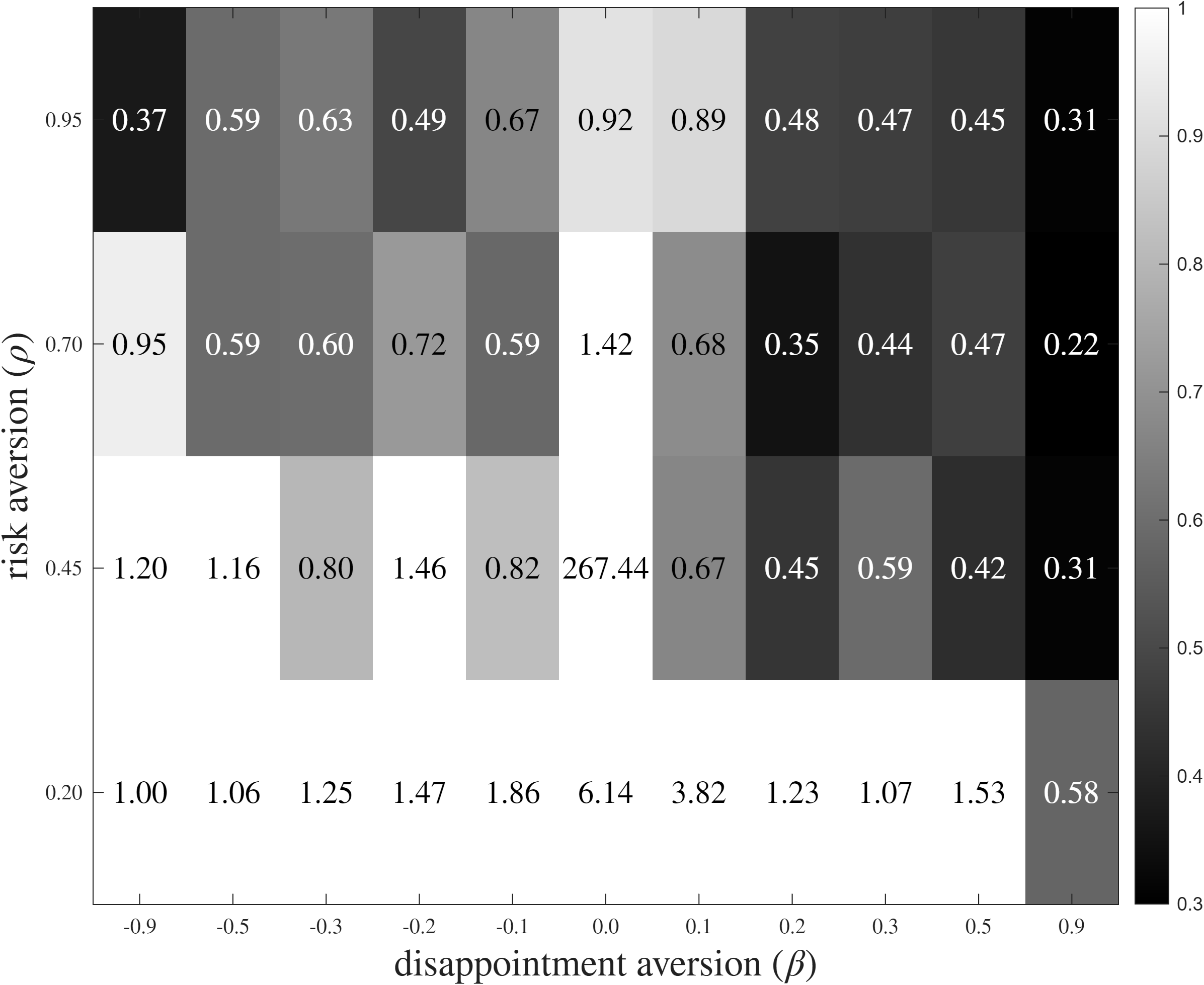}
\label{fig:rho_heatmap}
}
\hfill
\subfigure[$\beta$ estimation error improvement ratio.]{
\includegraphics[width=0.48\linewidth]{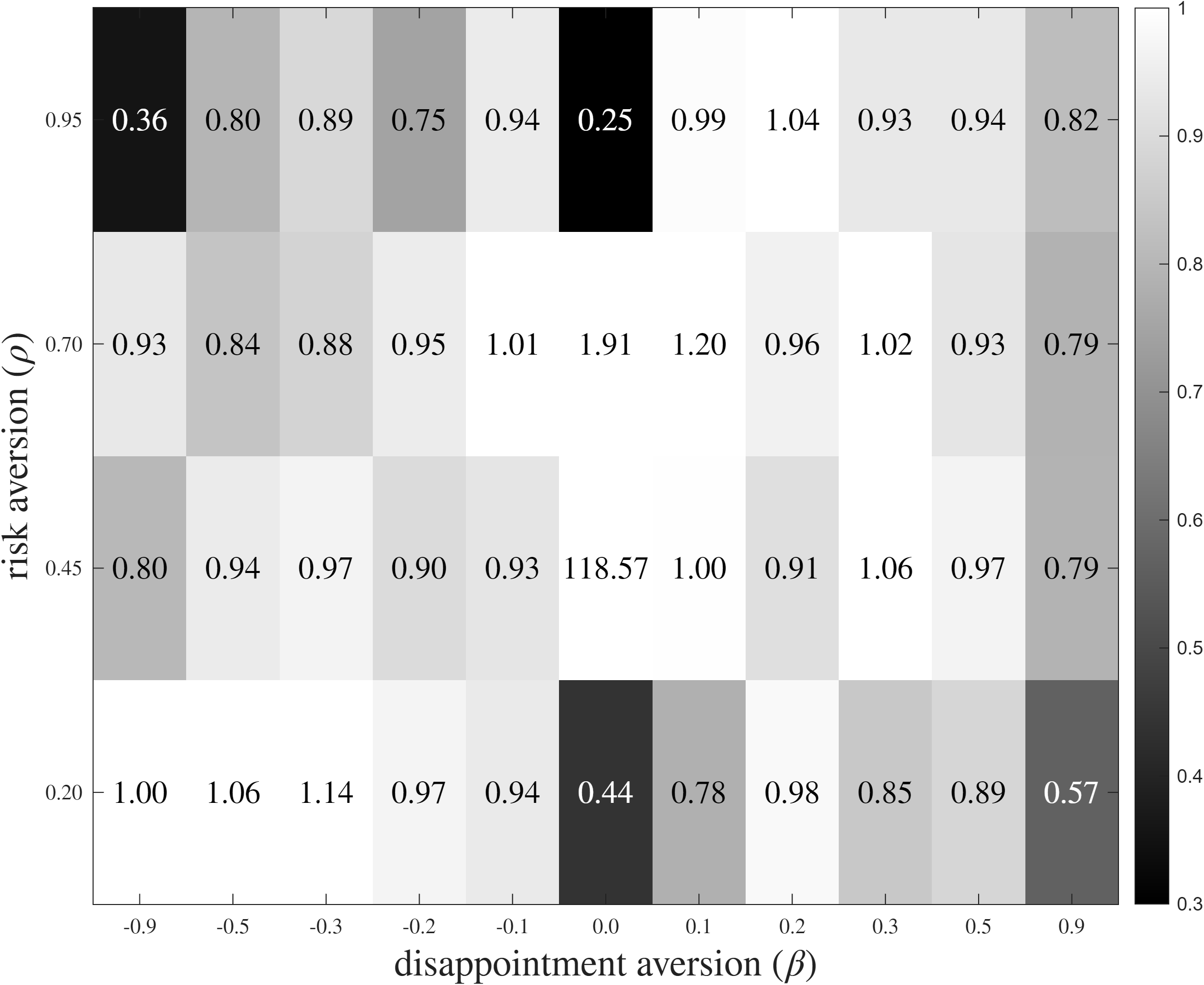}
\label{fig:beta_heatmap}
}
\caption{The heatmaps display the estimation error improvement ratio for risk aversion (Panel a) and disappointment aversion (Panel b). Darker cells indicate effective learning (reduced error), while lighter cells indicate little or no improvement. Values above 1 indicate that the estimation error increased as the history size grew from $h=5$ to $h=125$.}
\label{fig:parametric_heatmaps}
\end{figure}

In contrast, \autoref{fig:parametric_heatmaps}-(b) reveals that the heatmap for $\beta$ remains largely light across most of the grid, indicating that the estimation error for disappointment aversion does not improve significantly for most parameter combinations. A partial exception appears at $\beta = 0.9$, where the ratios fall below 1 across all levels of $\rho$, suggesting some learning of extreme disappointment aversion. Nevertheless, across the most of the parameter space, the LLM fails to recover $\beta$ even with a large history of observations.

\vskip+1em 

\noindent \textbf{Distributional evidence.} To further illustrate this phenomenon, \autoref{fig:cdfs_GPT} plots the cumulative distribution functions of the estimated parameters for a representative agent with parameters $\beta = 0.1$ and $\rho = 0.45$. On one hand, in \autoref{fig:cdfs_GPT}-(a), we find that as the history size increases (moving from the solid blue line ($h = 1$) to the solid black line ($h = 125$)), the distribution of $\hat{\rho}$ tightens and shifts toward the true value, which is indicated by the vertical red line in the figure. This confirms that with sufficient data, the LLM converges toward the correct level of risk aversion. On the other hand, in \autoref{fig:cdfs_GPT}-(b), the distribution of $\hat{\beta}$ shows almost no movement toward the true value of 0.1, even at $h = 125$. Instead, the estimates are anchored at zero, which corresponds to standard EUT behavior. This finding highlights the model's persistent inability to infer disappointment aversion from choice data.

\begin{figure}[ht]
    \centering
    \subfigure[CDF of Estimated Risk Aversion ($\hat{\rho}$).]{
        \includegraphics[width=0.48\linewidth]{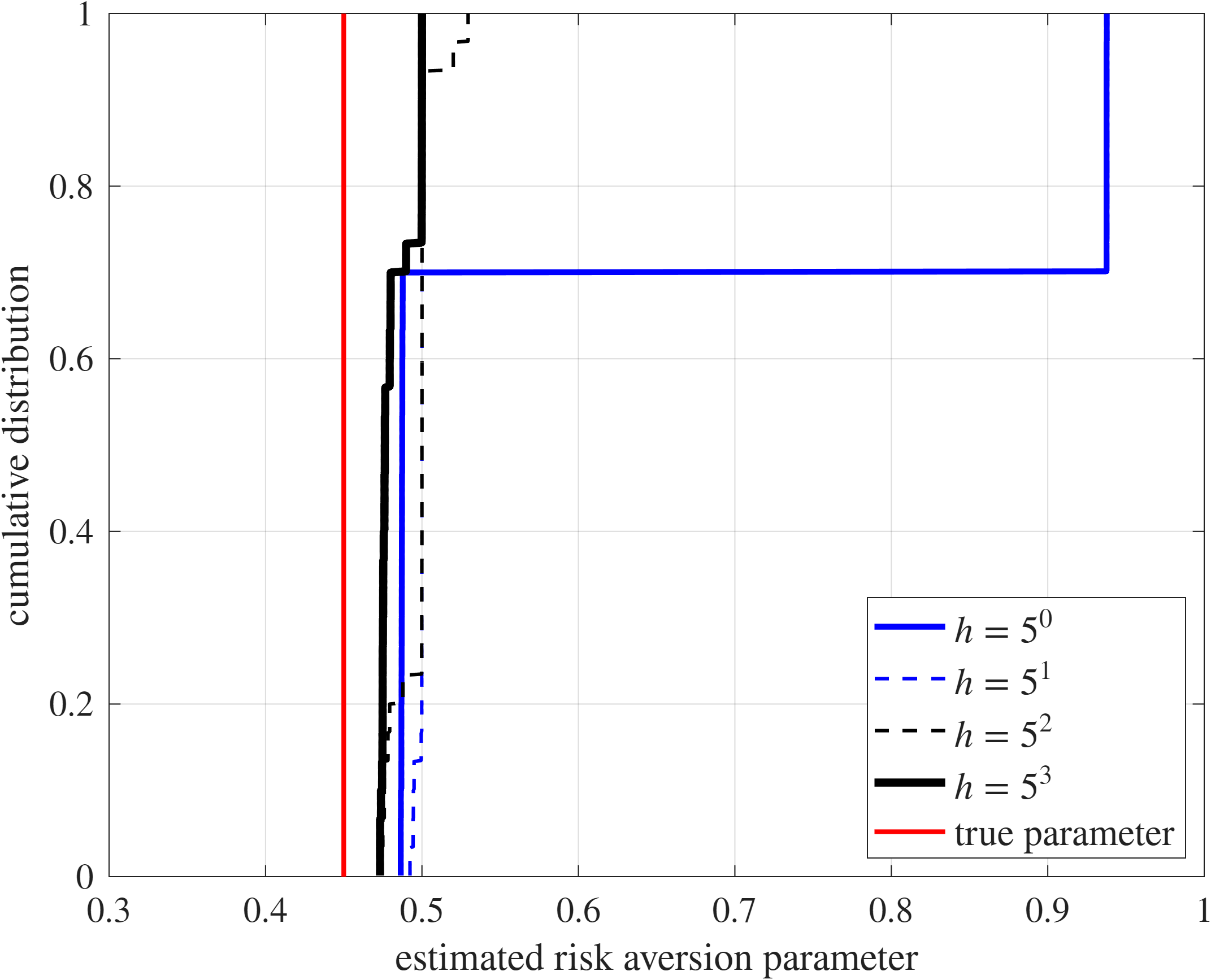}
        \label{fig:cdf_rho}
    }
    \hfill
    \subfigure[CDF of Estimated Disappointment Aversion ($\hat{\beta}$).]{
        \includegraphics[width=0.48\linewidth]{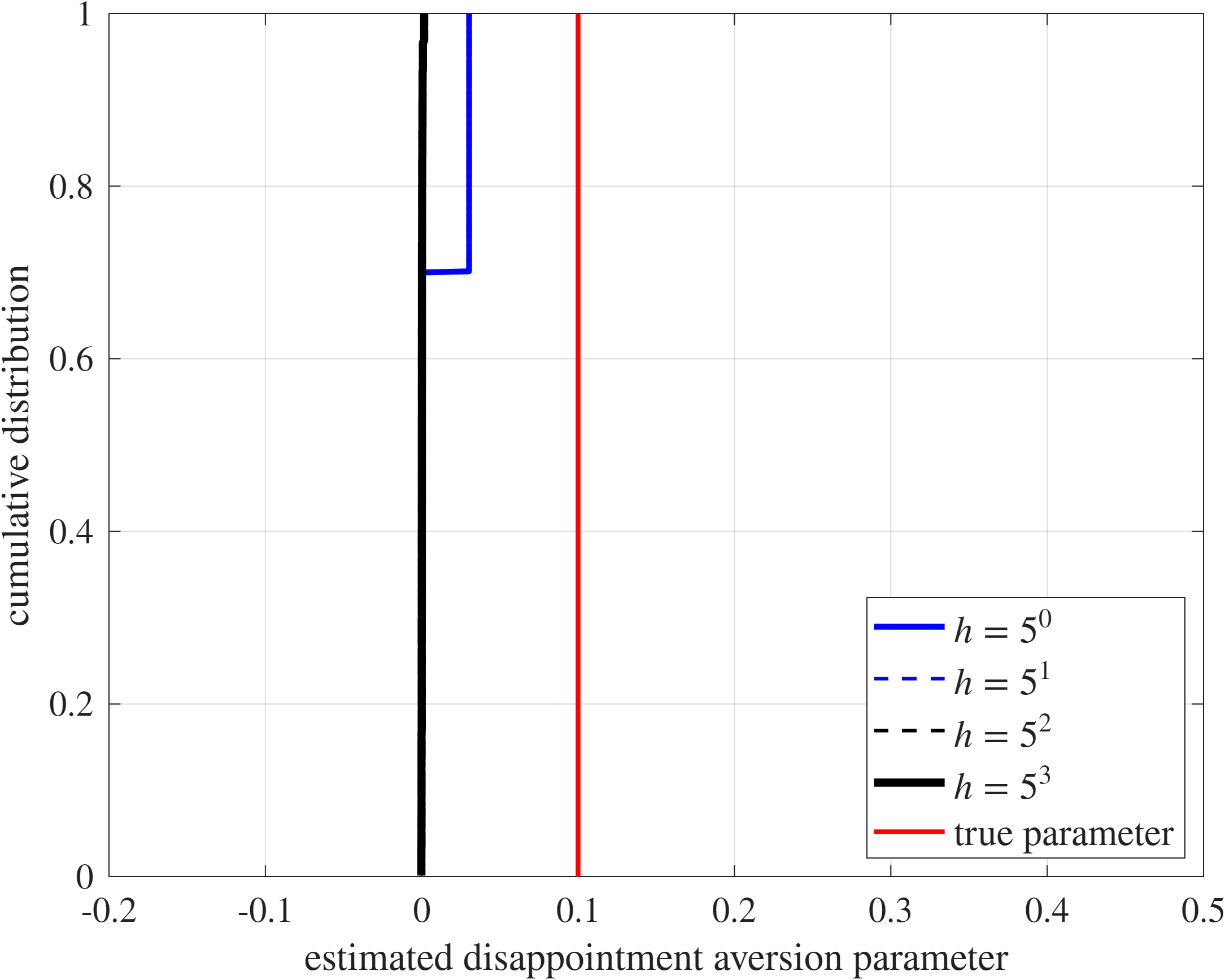}
        \label{fig:cdf_beta}
    }
    \caption{The figures plot the CDFs of the recovered parameters for a representative agent ($\beta = 0.1$ and $\rho = 0.45$) across different history sizes. Panel (a) displays the cumulative distribution of estimated risk aversion ($\hat{\rho}$), and Panel (b) displays the cumulative distribution of estimated disappointment aversion ($\hat{\beta}$). The vertical red lines indicate the true parameter values.}
    \label{fig:cdfs_GPT}
\end{figure}



\section{Discussion}

\noindent \textbf{Comparison across LLMs.} Our analysis reveals substantial differences in preference learning across the three LLMs. We gather results for Gemini and Claude in Appendix \ref{label:subsection:gemini} and Appendix \ref{label:subsection:claude}, respectively. GPT learns risk aversion reasonably well but largely fails to recover disappointment aversion. Gemini displays a distinct learning pattern: where it does improve, learning is concentrated in the high disappointment aversion region rather than the elation-seeking region where GPT improves, and its parametric recovery of $\rho$ is weakest near the expected utility benchmark (i.e., $\beta = 0$), strengthening as preferences diverge from it in either direction. Claude demonstrates the strongest performance among the three LLMs, exhibiting broadly effective learning across most of the preference parameter space for all non-parametric measures. Interestingly, Claude does not exhibit the stark asymmetry between risk aversion and disappointment aversion recovery that characterizes GPT: its parametric heatmaps show predominantly dark cells for both $\rho$ and $\beta$, indicating successful recovery of both preference dimensions as history size grows.

These cross-model differences suggest that the capacity for preference learning from revealed-choice data depends on model-specific factors rather than being a generic property of LLMs, and that the asymmetric learning patterns observed in GPT and Gemini reflect model-specific characteristics rather than an inherent limitation of the task.

\vskip+1em 

\noindent \textbf{Comparison to machine learning models.} We benchmark the three LLMs against three standard machine learning methods, k-nearest neighbors (kNN), random forest (RF), and support vector regression (SVR), whose detailed construction and results are reported in Appendices \ref{label:subsection:ML} --\ref{label:subsection:SVR}.\footnote{We refer to \citet{james2013islr} for a detailed discussion of these machine learning methods.} All three ML models exhibit uniformly dark heatmaps across the entire preference parameter space for both non-parametric and parametric measures, indicating that recommendation quality improves consistently as the history size increases from $h = 5$ to $h = 125$. In particular, SVR achieves the strongest convergence, with welfare loss improvement ratios falling below 0.05 for most parameter combinations. Parametric recovery is also broadly successful: all three models recover both $\beta$ and $\rho$ effectively, though RF exhibits some difficulty near the expected utility benchmark and kNN shows isolated clusters of light cells. Moreover, none of the ML models display the stark asymmetry between risk aversion and disappointment aversion recovery that characterizes GPT. 

The above contrast highlights that GPT's failure to learn disappointment aversion is not an inherent limitation of learning from revealed-choice data in this environment, but rather reflects a model-specific deficiency.

\vskip+1em 

\noindent \textbf{Illustrative examples of LLM learning failure.} Figures \ref{fig:NLLS_GPT1}--\ref{fig:NLLS_GPT3} in Appendix \ref{label:subsection:additional_GPT} illustrate GPT's recommendations for $\rho = 0.2$ across different values of $\beta$ and $h$. Three qualitative patterns emerge. First, with minimal data ($h = 1$), GPT extrapolates from a single observed choice and produces recommendations that violate FOSD severely as recommendations allocate nearly the entire budget to the expensive asset over a wide range of relative prices. This suggests that GPT does not impose normative constraints, but instead imitates the local pattern in the data it receives. Second, for elation-seeking preferences ($\beta < 0$), a small number of observations ($h = 5$) is sufficient to reproduce the corner-solution pattern, whereas for disappointment-averse preferences ($\beta > 0$), the same history size yields dispersed recommendations with no stable demand shape. Third, as the history grows ($h = 25$ to $125$), GPT's recommended share in asset $A$ for disappointment-averse agents becomes anchored near $0.5$ (i.e., hedging) across a wide range of price ratios, rather than converging to the true S-shaped demand implied by the DA model.

\vskip+1em 

\noindent \textbf{Extension to other settings.} The SRE framework can be applied to test LLMs' learning capability in other preference domains, such as time preference. 
To illustrate, consider the convex time budget (CTB) design of \citet{Andreonietal:2012:AER}, where a subject allocates a budget between a sooner and a later payment under a linear budget constraint. This is structurally equivalent to our portfolio choice setting, with the relative price of future consumption playing the role of the asset price ratio. The quasi-hyperbolic discounting model provides a natural two-dimensional analog to the DA model, and one can construct a parameter grid over the present bias parameter $\beta$ and the exponential discounting parameter $\delta$, spanning empirically reasonable values estimated in the literature \citep{cohen2020measuring, imai2021meta}. The SRE procedure then applies directly: generate optimal CTB allocations from known preference parameter vectors $(\beta, \delta)$, provide choice histories to the LLM, and evaluate recommendations using the same non-parametric metrics and parametric decompositions. 

Beyond gauging LLMs learning capacity in other preference dimensions, such applications would also test whether the asymmetric learning pattern documented in the current paper generalizes. Just as GPT learns risk aversion but not disappointment aversion, it may learn long-run patience but not present bias, suggesting that LLMs are systematically anchored to standard economic benchmarks. We leave this as a future research exercise.

We also note that the SRE framework does not require the underlying data-generating process to be fully rational or parametric, and can therefore be applied to settings that do not  admit canonical preference representations. In our implementation, we use a structured utility representation to generate revealed-choice data, as it allows for transparent  evaluation and interpretation. However, this structure is not essential. The framework only requires access to observed choice behavior from the decision maker of interest. For example, a firm seeking to evaluate whether an LLM can align with a client’s preferences need not assume that the client’s behavior is rationalizable by a utility-maximizing model. Instead, it can treat the client’s past choices as the object to be learned and use SRE to assess whether the LLM produces recommendations that are consistent with those observed choices. The framework thus also applies to settings with non-parametric, noisy, or even systematically biased behavior, as long as evaluation is anchored by observed choice data.

\vskip+1em 

\noindent \textbf{Practical guidelines for evaluating preference learning.}
Our findings suggest the following practical guidelines for researchers and practitioners who use LLMs as preference-based recommendation systems:

\begin{itemize}
    \item \textbf{Evaluate out of sample and with welfare-relevant metrics.}
    Treat the observed choice history as training data and assess performance on new, unseen decision problems. In addition to descriptive notions of ``reasonable'' advice, report welfare-relevant outcomes (e.g., welfare loss under the benchmark model) to quantify the economic cost of misalignment.

    \item \textbf{Report both non-parametric alignment and parametric recovery.}
    Non-parametric measures (e.g., normalized distance between recommended and optimal allocations, welfare loss) capture how close recommendations are to the benchmark without committing to a particular parametric structure. Parametric recovery diagnostics (e.g., how well key preference parameters are recovered) help distinguish surface-level imitation from genuine learning of preference primitives.

    \item \textbf{Map heterogeneity across the preference space and highlight worst-case regions.}
    Average improvements can mask persistent failures for specific preference types. We therefore recommend reporting results over a broad parameter space and emphasizing where the model performs poorly (including economically relevant worst-case regions), rather than relying solely on aggregate summaries.

    \item \textbf{Compare LLMs to simple baselines, and add safety checks for real-world use.}
To understand whether poor performance reflects a fundamental identification problem or a limitation of the LLM, compare the LLM to simple and transparent benchmarks trained on the same revealed-choice data (e.g., a standard regression-based predictor or a structural estimator). If these baselines succeed while the LLM does not, the evidence points to a model-specific limitation; if both fail, the revealed-choice data or the decision environment may not be informative enough to pin down preferences. For practical use, implement simple diagnostics to detect unreliable inference (e.g., recommendations that change markedly under minor prompt variations or weak/unstable parameter recovery across histories) and adopt fail-safe rules that either request additional elicitation or fall back to conservative, constraint-respecting recommendations when confidence is low.
\end{itemize}



\newpage

\bibliographystyle{ecta}
\bibliography{references}


\newpage

\appendix
\counterwithin{figure}{section}
\counterwithin{table}{section}

\section*{Online Appendix}


\section{Prompts}
\label{sec:prompts}

\subsection{Recommendation without Data}

The following is the prompt used for the GPT recommendation without data:

\begin{itemize}

    \item System role: I want you to act as a recommendation system for our valuable customers. One of our customers will be given 25 rounds of decision-making tasks, and you will be responsible for providing recommendations for the customer. You should use your best judgment to come up with solutions that the customer likes most. You must provide your answers in every round. If you do not provide an answer, I will assume you are making a random choice and will implement it for the customer.


    \item User role: In each round, the customer is endowed with a budget of 100 dollars for allocation between asset A and asset B. One of the 25 rounds is randomly selected for payment. In the selected paying round, the customer has a 50\% probability that the realized payoff is determined by the allocation to asset A and a 50\% probability that it is determined by the allocation to asset B.

    The following table displays the prices of two assets. The first column represents the round number, with a total of 25 rounds. The second column reports the dollar price of one unit of asset A. The third column reports the dollar price of one unit of asset B.

    \underline{[Return Table]}

    What is your recommendation for investments in each round for the two assets? 

    IMPORTANT: Please provide your answer in the following EXACT format:

    Round 1: I recommend investing M1 units in asset A and N1 units in asset B.
    
    Round 2: I recommend investing M2 units in asset A and N2 units in asset B.
    
    Round 3: I recommend investing M3 units in asset A and N3 units in asset B.
    ... (continue for all 25 rounds)

    Remember:
    
    - Provide recommendations for ALL 25 rounds
    
    - Note that in each round, total spending must be 100 dollars because the customer is endowed with a budget of 100 dollars.
    
    - Use the exact format shown above.
    
    - Please do not comment on anything else; just answer with the recommendation for each round.
    
    - When answering the recommended unit, please respond using only numbers.
    
    - Number the rounds from 1 to 25
    
\end{itemize}


\subsection{Recommendation with Sample Choice Data} 

The following is the prompt used for the GPT personalized recommendation with sample choice data: 

\begin{itemize}

    \item System role: I want you to act as a recommendation system for our valuable customers. One of our customers will be given 25 rounds of decision-making tasks, and you will be responsible for providing recommendations for the customer. You should use your best judgment to come up with solutions that the customer likes most. You must provide your answers in every round. If you do not provide an answer, I will assume you are making a random choice and will implement it for the customer.


    \item User role: In each round, the customer is endowed with a budget of 100 dollars for allocation between asset A and asset B. One of the 25 rounds is randomly selected for payment. In the selected paying round, the customer has a 50\% probability that the realized payoff is determined by the allocation to asset A and a 50\% probability that it is determined by the allocation to asset B.

    To help you understand the customer's preferences, we asked the customer to participate in 25 rounds of the same tasks. The following data table summarizes the customer's choices.
    
    \underline{[DATA]} 
    
    The above table displays the prices of the two assets. The first column represents the round number, with a total of 25 rounds. The second column reports the dollar price of one unit of asset A, and the third column reports the dollar price of one unit of asset B. The fourth column shows the customer's allocated units of asset A. The last column shows the customer's allocated units of asset B. Note that in each round, total spending is fixed at 100 dollars because the budget size is 100 dollars.

    The following table displays the prices of two assets. The first column represents the round number, with a total of 25 rounds. The second column reports the dollar price of one unit of asset A. The third column reports the dollar price of one unit of asset B.

    \underline{[Return Table]}

    What is your recommendation for investments in each round for the two assets? 

    IMPORTANT: Please provide your answer in the following EXACT format:

    Round 1: I recommend investing M1 units in asset A and N1 units in asset B.
    
    Round 2: I recommend investing M2 units in asset A and N2 units in asset B.
    
    Round 3: I recommend investing M3 units in asset A and N3 units in asset B.
    ... (continue for all 25 rounds)

    Remember:
    
    - Provide recommendations for ALL 25 rounds
    
    - Note that in each round, total spending must be 100 dollars because the customer is endowed with a budget of 100 dollars.
    
    - Use the exact format shown above.
    
    - Please do not comment on anything else; just answer with the recommendation for each round.
    
    - When answering the recommended unit, please respond using only numbers.
    
    - Number the rounds from 1 to 25
    
\end{itemize}

The number of rounds of recommendation, fixed at 25, remains constant. However, the sample choice data (\underline{[DATA]} in the prompt) and the related prompt (\underline{25} in the prompt) vary based on the entity generating sample choice data and the data's size. The size of the sample choice data can take values in the set $\{1, 5, 25, 125\}$.


We consistently employ 25 return vectors throughout the experiments, ensuring that \underline{[RETURN TABLE]} remains constant and does not vary.


\newpage

\subsection{Sample Screenshots}
\label{subsection:screenshots}

\noindent \textbf{GPT personalized recommendation.} The following screenshots illustrate the prompts used to generate the GPT recommendation dataset and the corresponding responses. In each API call, GPT receives 25 sample choice pairs and returns recommendations for 25 rounds in a single answer. 

\autoref{fig:screenshot1}, for each call of GPT API, we provide prompts for the system role. In the prompt, we provide a set of choices in table form (highlighted in the first red box in the figure). Then, we give a set of 25 pairs of returns from two assets in table form (highlighted in the second red box in the figure).

\begin{figure}[ht]
\centering
\includegraphics[width = 0.82 \textwidth]{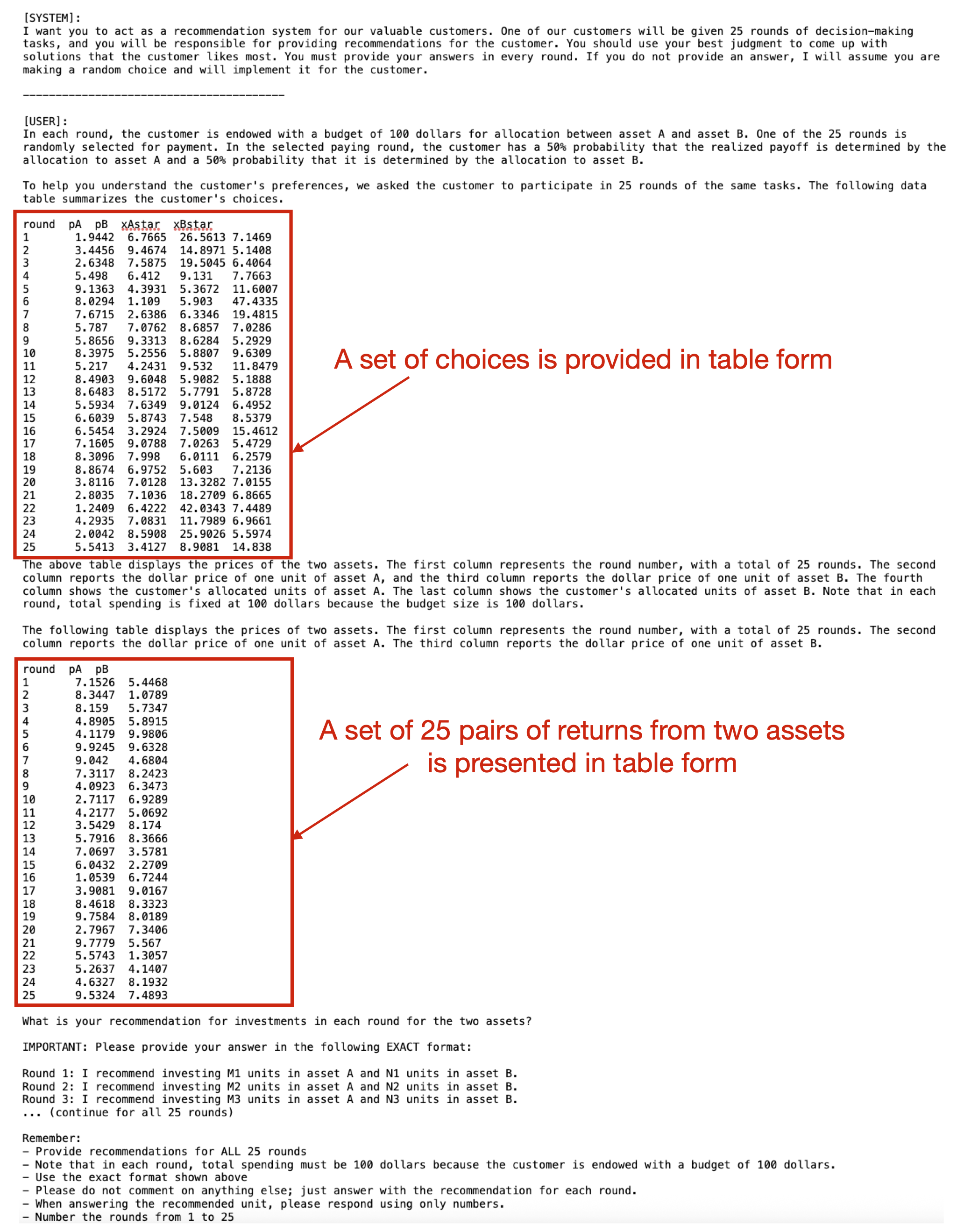}
\caption{Screenshot of the prompts for generating the recommendation dataset by GPT.}
\label{fig:screenshot1}
\end{figure}

\autoref{fig:screenshot2} is the subsequent answers (recommendations) provided by GPT following \autoref{fig:screenshot1}. GPT provides recommendations for 25 rounds in a single answer.

\begin{figure}[ht]
\centering
\includegraphics[width = 0.82 \textwidth]{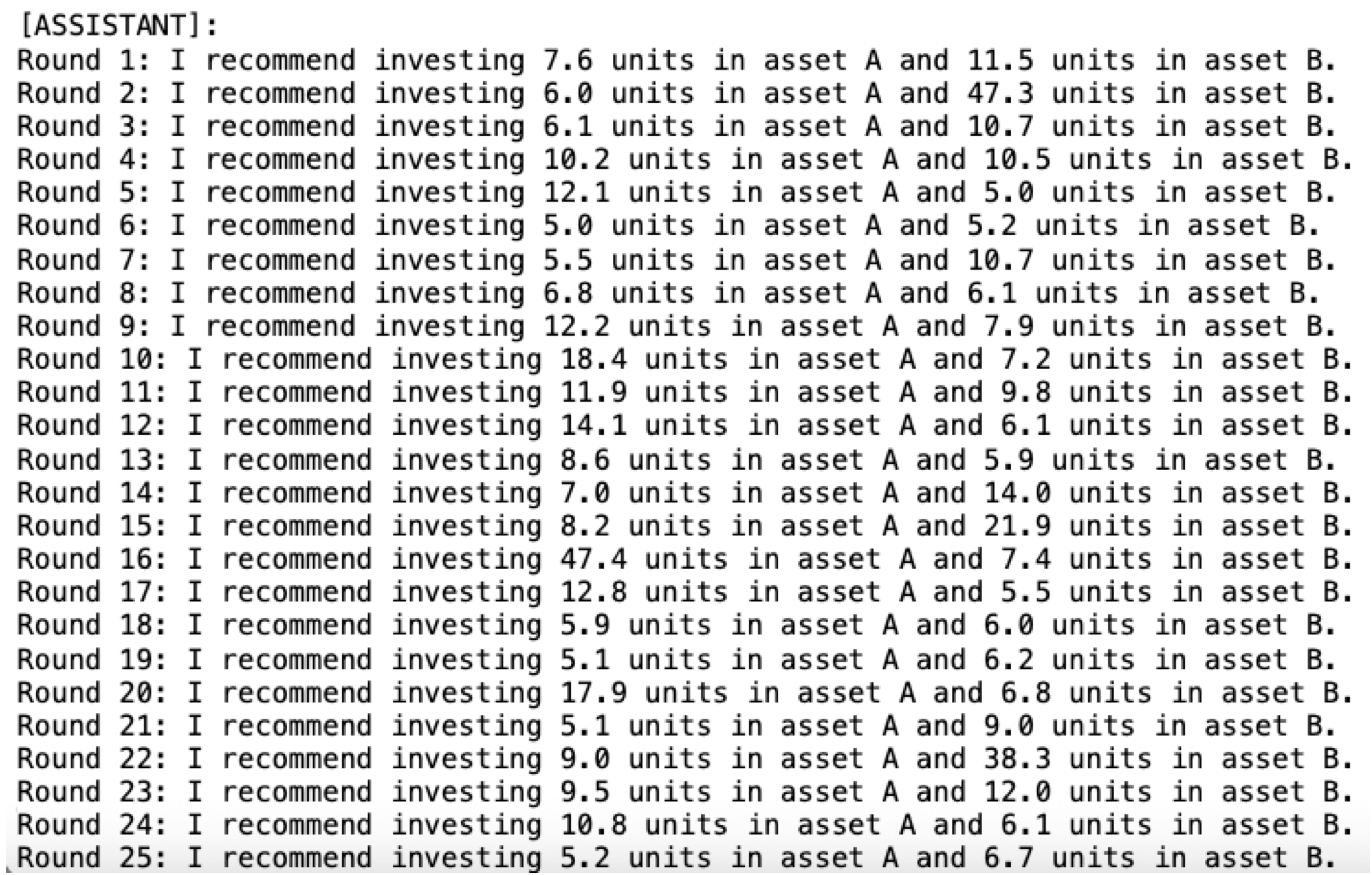}
\caption{Screenshot of responded recommendation dataset by GPT. }
\label{fig:screenshot2}
\end{figure}


\newpage

\section{NLLS Method}
\label{sec:calculation}

\noindent \textbf{NLLS.} By carefully following \citet{Choietal:2007:AER}, our NLLS method considers the following minimization problem:
\begin{align*}
\text{minimize}_{(\beta, \rho)} \sum_{i=1}^{25} \left( \ln\!\left(\frac{y_{A}^{i}}{y_{B}^{i}}\right) - f\!\left[\,\ln\!\left(\frac{q^{\,i}_{A}}{q^{\,i}_{B}}\right);\,\beta,\rho,\omega\right] \right)^2 \text{ with $\omega = 10^{-3}$},
\end{align*}
where $f$ takes the following for $\beta > 0$ (disappointment aversion):
\begin{align*}
& f\!\left[\, x ;\,\beta,\rho,\omega\right]\\
&=
\begin{cases}
- \ln \omega,
& \text{if } x \leq - \ln (1+\beta) + \rho \ln \omega, \\[6pt]
-\dfrac{1}{\rho}\Big[ x + \ln (1+\beta) \Big],
& \text{if } - \ln (1+\beta) + \rho \ln \omega < x < - \ln (1+\beta), \\[8pt]
0,
& \text{if } -\ln (1+\beta) \leq x \leq \ln (1+\beta), \\[8pt]
-\dfrac{1}{\rho}\Big[ x - \ln (1+\beta) \Big],
& \text{if } \ln (1+\beta) < x < \ln (1+\beta) - \rho \ln \omega, \\[8pt]
\ln \omega,
& \text{if } x \geq \ln (1+\beta) - \rho \ln \omega,
\end{cases}
\end{align*}
and $f$ takes the following for $\beta \leq 0$ (elation seeking):
\begin{align*}
& f\!\left[\, x ;\,\beta,\rho,\omega\right]\\
&=
\begin{cases}
-\ln \omega,
& \text{if } x \leq - \ln (1+\beta) + \rho \ln \omega \text{  and  } x < 0, \\[6pt]
-\dfrac{1}{\rho}\Big[ x + \ln (1+\beta) \Big],
& \text{if } - \ln (1+\beta) + \rho \ln \omega < x < 0, \\[8pt]
-\dfrac{1}{\rho}\Big[ x - \ln (1+\beta) \Big],
& \text{if } 0  < x < \ln (1+\beta) - \rho \ln \omega, \\[8pt]
\ln \omega,
& \text{if } x \geq \ln (1+\beta) - \rho \ln \omega \text{  and  } x > 0.
\end{cases}
\end{align*}

\autoref{fig:f_function} illustrates the graph of $f$ for values of the disappointment aversion parameter $\beta > -1$. \autoref{fig:f_function}-(a) corresponds to the disappointment aversion region (i.e., $\beta > 0$), and \autoref{fig:f_function}-(b) corresponds to the elation-seeking region (i.e., $\beta \leq 0$). The red lines have a common slope of $-\frac{1}{\rho}$ in both cases. Note that the blue flat region exists only in the disappointment region, which represents perfect hedging behavior when prices are not sufficiently different (i.e., when the price ratio is close to one). $\omega$ is chosen to be a very small number, as in \citet{Choietal:2007:AER}. In the MATLAB implementation, we solve the minimization problem separately for the two cases distinguished by $\beta$, and then select the values of $\beta$ and $\rho$ that yield the smaller sum of squared residuals (SSR).

\begin{figure}[ht]
\centering
\subfigure[disappointment aversion: $\beta > 0$]{
\begin{tikzpicture}[>=stealth, scale=0.7]
  \draw[->, thick] (0.5,0) -- (8.2,0) node[below right] {$\ln\!\left(\frac{p_A}{p_B}\right)$};
  \draw[->, thick] (4,-3) -- (4,3.2) node[above left] {$\ln\!\left(\frac{x_A}{x_B}\right)$};

  \node[left] at (0.8,0){};

  \draw (4,0.15)--(4,-0.15);
  \node[below left] at (4,-0.15) {$0$};

  \draw[line width=1pt] (1.0,2.0) -- (2.2,2.0);
  \node[above left] at (1.5,2.0) {\small (c)};

  \draw[dashed] (1.0,2.0) -- (0.5,2.0);
  \node[above left] at (0.5,2.0) {$-\ln \omega$};

  \draw[line width=1.5pt, red] (2.2,2.0) -- (3.0,0.0);
  \node[right] at (2.3,1.4) {\small (d)};

  \draw[line width=1.5pt, blue] (3.0,0.0) -- (5.0,0.0);
  \node[above right] at (3.1,0.1) {\small (e) (width = $|2 \ln(1+\beta)|$)};

  \draw[line width=1.5pt, red] (5.0,0.0) -- (5.8,-2.0);
  \node[right] at (5.25,-1.0) {\small (b)};

  \draw[line width=1pt] (5.8,-2.0) -- (7.2,-2.0);
  \node[below] at (6.5,-2.0) {\small (a)};

  \draw[dashed] (7.2,-2.0) -- (7.5,-2.0);
  \node[right] at (7.5,-2.0) {$\ln \omega$};
  
\end{tikzpicture}
}
\hfill
\subfigure[elation seeking: $\beta \leq 0$]{
\begin{tikzpicture}[>=stealth, scale=0.7]
  \draw[->, thick] (0.5,0) -- (8.2,0) node[below right] {$\ln\!\left(\frac{p_A}{p_B}\right)$};
  \draw[->, thick] (4,-3) -- (4,3.2) node[above left] {$\ln\!\left(\frac{x_A}{x_B}\right)$};

  \node[left] at (0.8,0){};

  \draw (4,0.15)--(4,-0.15);
  \node[below left] at (4,-0.15) {$0$};

  \draw[line width=1pt] (1.0,2.0) -- (3.6,2.0);
  \node[above left] at (1.5,2.0) {\small (c)};

  \draw[dashed] (1.0,2.0) -- (0.5,2.0);
  \node[above left] at (0.5,2.0) {$-\ln \omega$};

  \draw[line width=1.5pt, red] (3.6,2.0) -- (4.0,1.0);
  \node[left] at (3.75,1.4) {\small (d)};

  \draw[line width=1.5pt, red] (4.0,-1.0) -- (4.4,-2.0);
  \node[right] at (4.25,-1.0) {\small (b)};

  \draw[line width=1pt] (4.4,-2.0) -- (7.2,-2.0);
  \node[below] at (6.5,-2.0) {\small (a)};

  \draw[dashed] (7.2,-2.0) -- (7.5,-2.0);
  \node[right] at (7.5,-2.0) {$\ln \omega$};
  
\end{tikzpicture}
}
\caption{Illustration of $f$ by values of $\beta$}
\label{fig:f_function}
\end{figure}

\,

\clearpage


\newpage

\section{Additional Tables and Figures}
\label{sec:additional_tables_and_figures}

\noindent \textbf{Model labels.} We use the following labels for the data collected using LLMs and benchmark machine learning models.

\begin{table}[ht]
\centering
\footnotesize
\begin{tabular}{|l|p{12cm}|}
\hline 
\textbf{Name} & \textbf{Description} \\ 
\hline 
GPT & GPT-5 prompted to generate recommendations as a recommendation system. \\
\hline 
Gemini & Gemini 2.5 prompted to generate recommendations as a recommendation system. \\
\hline 
Claude & Claude-Opus-4.1 prompted to generate recommendations as a recommendation system. \\
\hline 
kNN & k-Nearest Neighbors benchmark model implemented using reference code. \\
\hline
RF & Random Forest benchmark model implemented using reference code. \\
\hline 
SVR & Support Vector Regression benchmark model implemented using reference code. \\
\hline
\end{tabular}
\caption{Model labels used in the dataset}
\label{tab:model_label}
\end{table}
\newpage
\subsection{Additional Figures for GPT}
\label{label:subsection:additional_GPT}

\noindent \textbf{Illustrative examples of LLM learning failure.} We present examples of GPT's recommendations that illustrate the various ways in which the model fails to align with the optimal allocations. In the following figures, hollow circles denote the optimal allocations implied by the DA model, red dots denote GPT's recommendations, the black curve is the theoretical demand function, and the red curve is the NLLS fit to the recommendations. 

Figure \ref{fig:NLLS_GPT1} displays the case $(\beta, \rho) = (-0.1, 0.2)$ at two history sizes. In Panel (a), GPT's recommendations cluster near $\frac{x_A}{x_A+x_B} \approx 1$ regardless of the price ratio, including the region where asset $A$ is much more expensive. This is a severe FOSD violation.\footnote{When $h = 1$, the single sample choice given to GPT is $(p_A, p_B) = (1.944, 6.7665)$ and $(x_A, x_B) = (51.229, 0.059)$.} In Panel (b), the FOSD violations disappear and the recommendations polarize toward $0$ and $1$, capturing the elation seeker's corner solution pattern. However, the NLLS fit yields $\widehat{\beta} > 0$, rationalizing the recommendations through a hedging region rather than the corner-solution pattern implied by elation seeking, illustrating how NLLS can assign structurally mismatched parameters to heuristic recommendation patterns.

\begin{figure}[ht]
\centering
\includegraphics[width=0.45\linewidth]{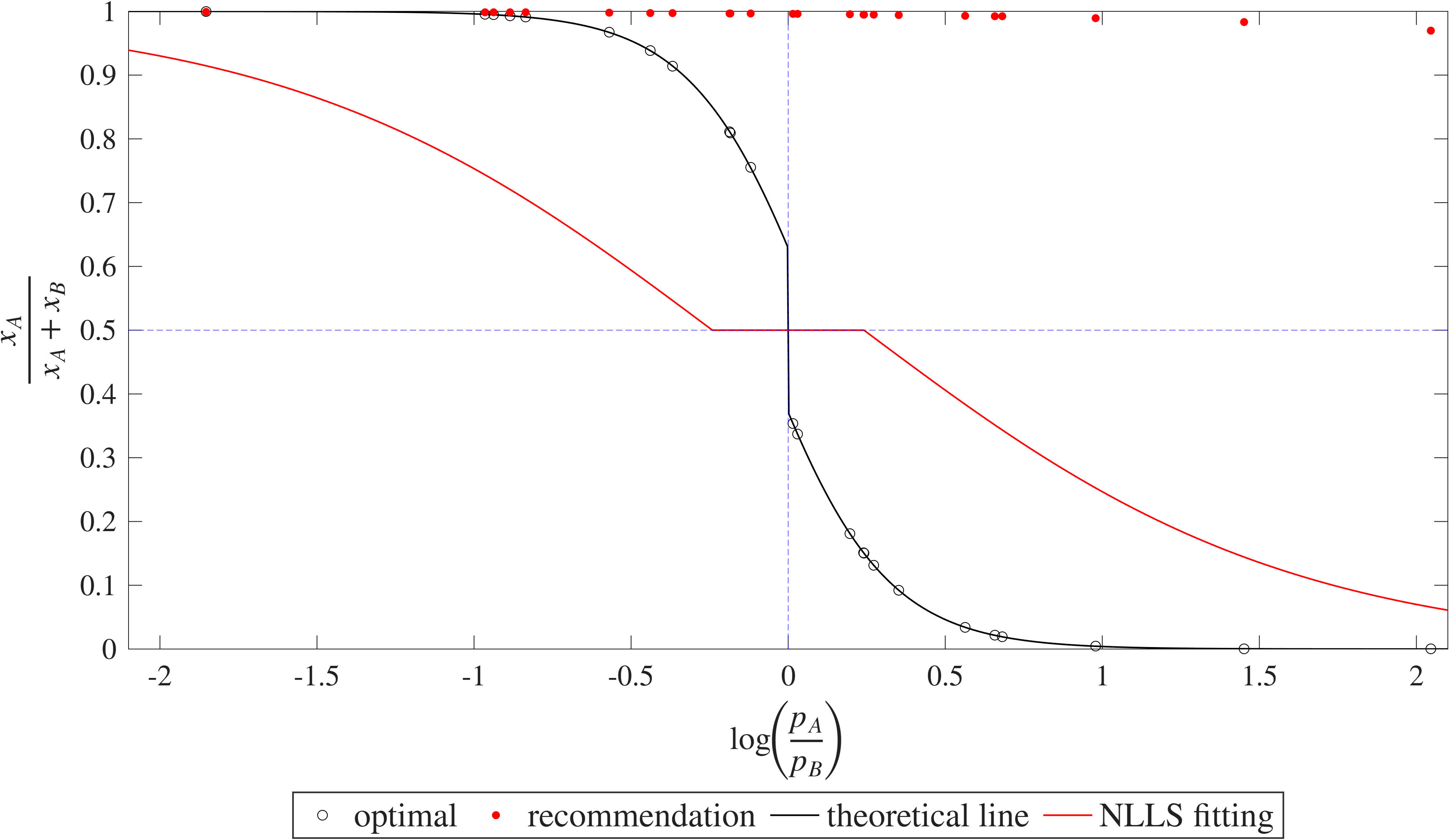}
\includegraphics[width=0.45\linewidth]{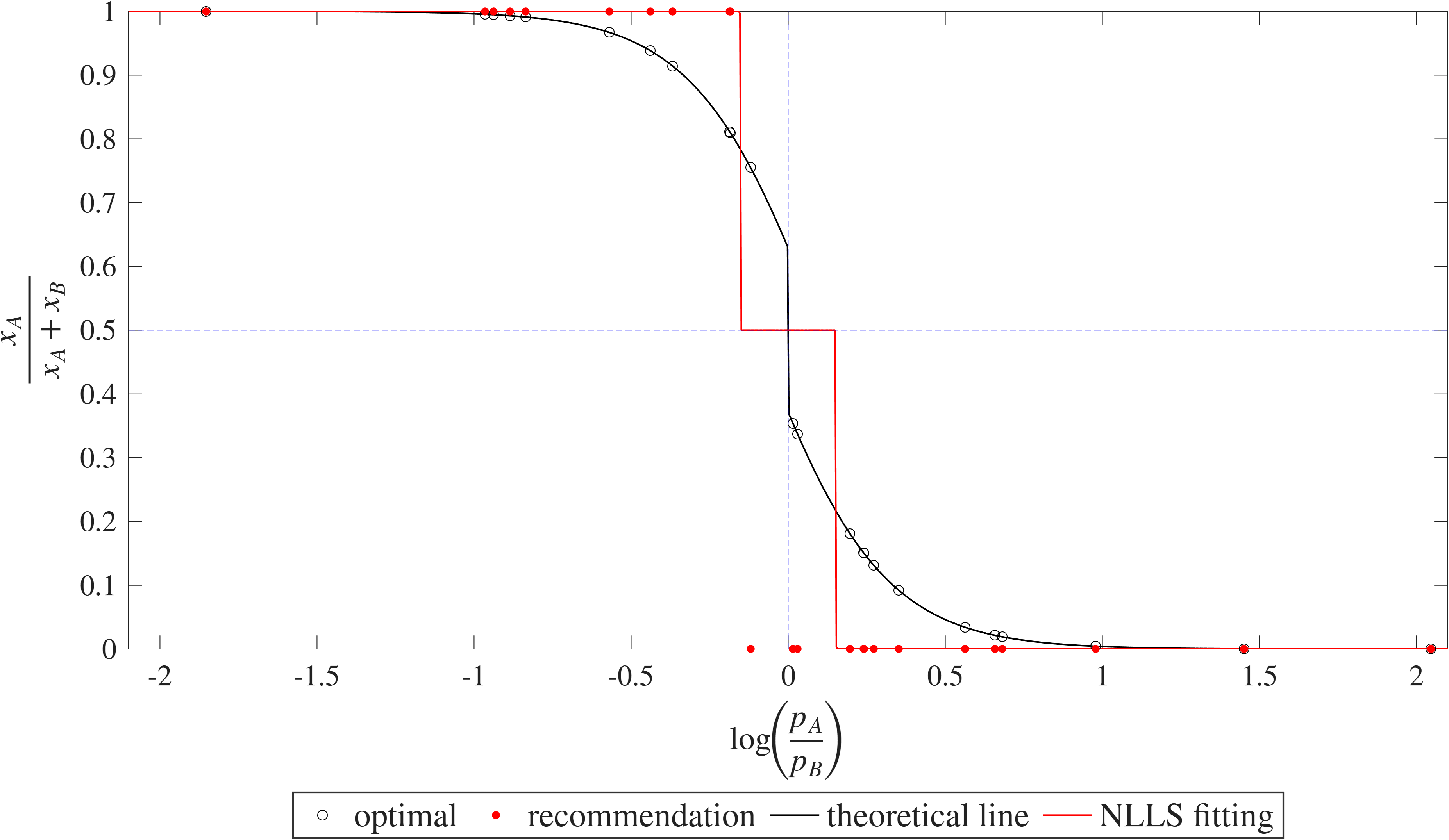}
\caption{Sample recommendations for $(\beta,\rho)=(-0.1,0.2)$. Panel (a): $h=1$. Panel (b): $h=5$.}
\label{fig:NLLS_GPT1}
\end{figure}

Figure \ref{fig:NLLS_GPT2} displays the case $(\beta, \rho) = (0.1, 0.2)$ at two history sizes. In Panel (a), GPT's recommendations appear as an mirror image of the optimal allocations. The NLLS fit is a flat line at 0.5. In Panel (b), recommendations remain anchored at 0.5. GPT's recommendations converge to a naive equal-allocation heuristic ($x_A \approx x_B$), rationalized by the NLLS fit as $\widehat{\rho} \approx \infty$ and $\widehat{\beta} \approx 0$.

\begin{figure}[ht]
\centering
\includegraphics[width=0.45\linewidth]{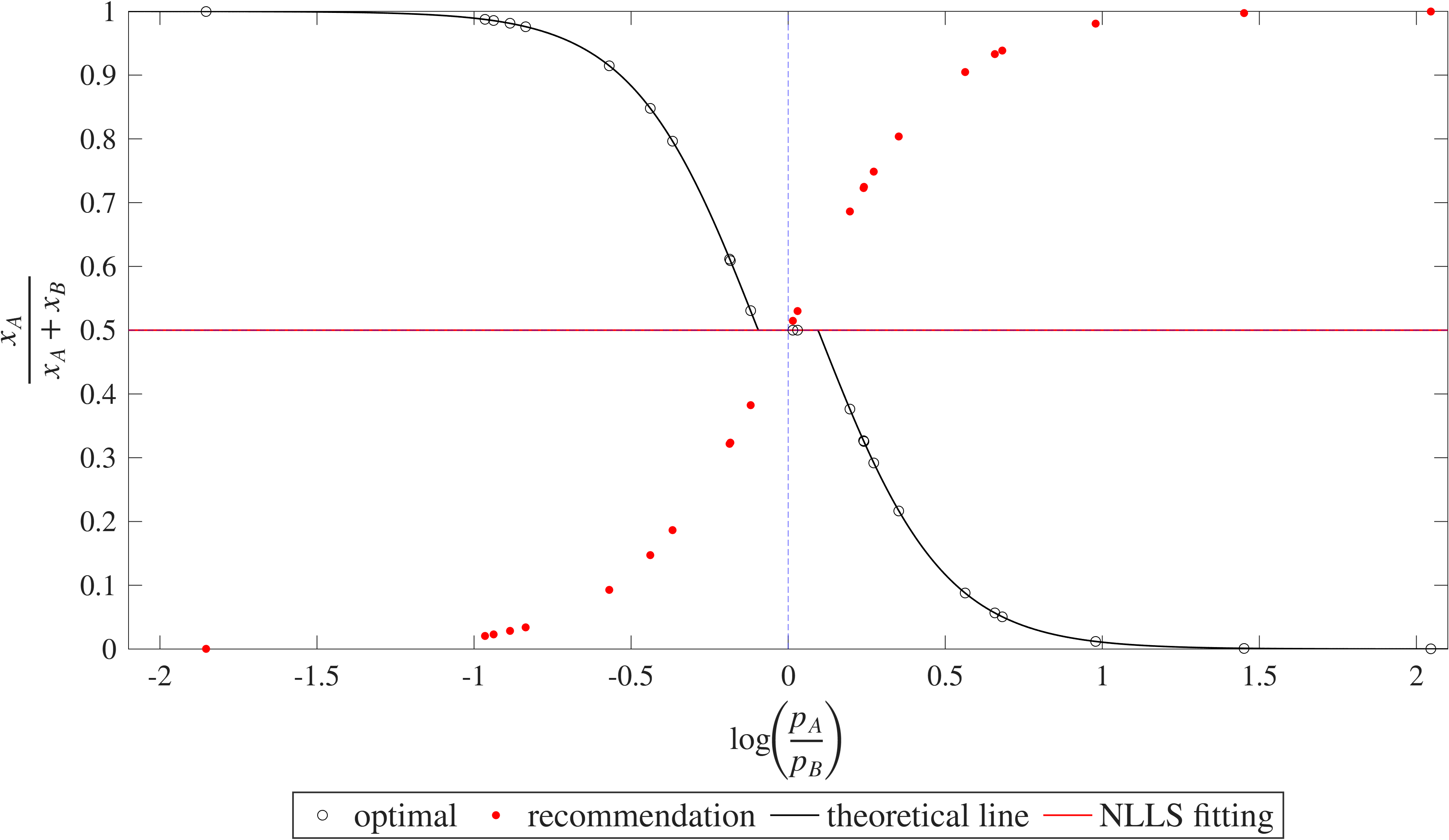}
\includegraphics[width=0.45\linewidth]{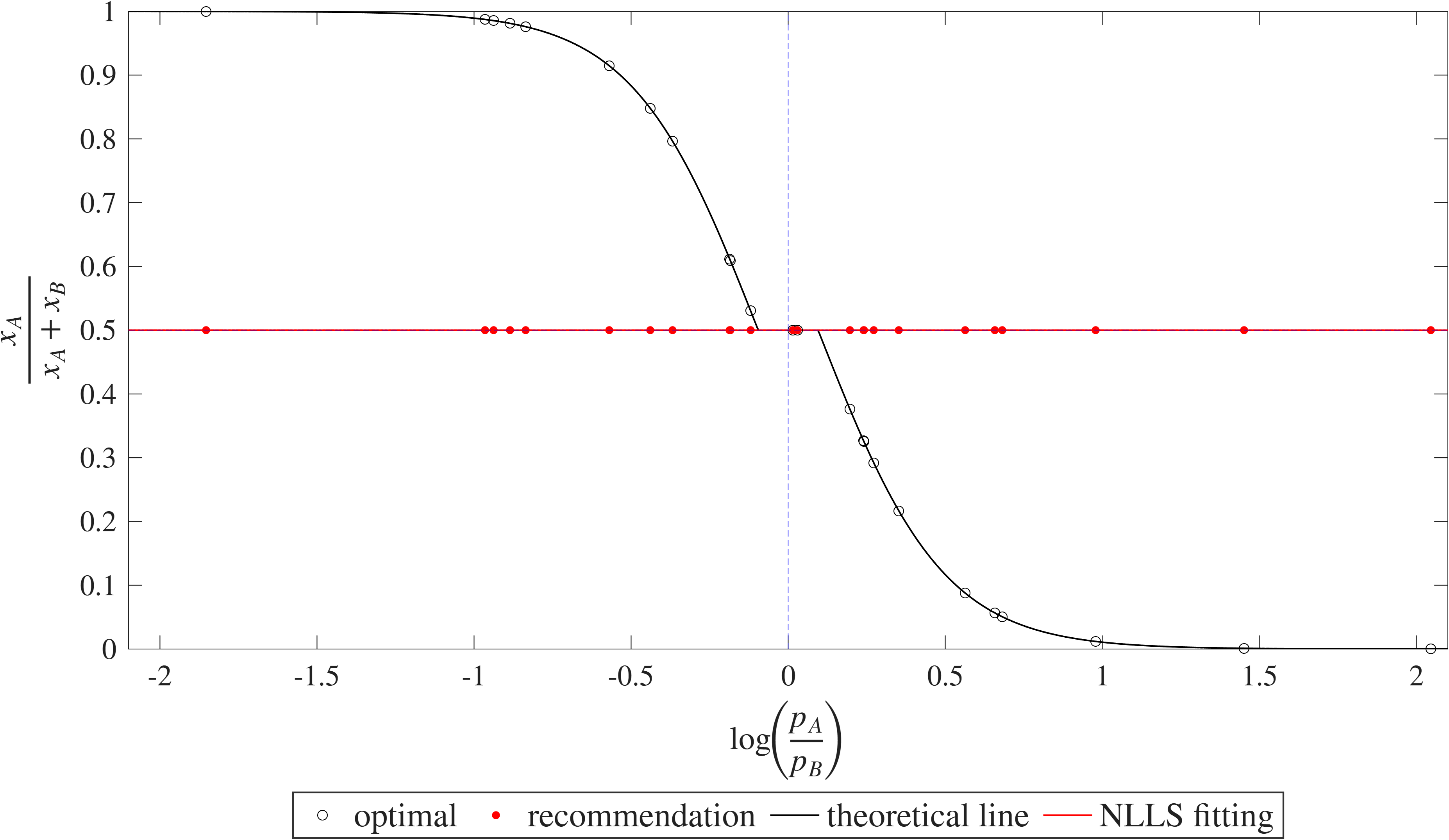}
\caption{Sample recommendations for $(\beta,\rho)=(0.1,0.2)$. Panel (a): $h=25$. Panel (b): $h=125$. }
\label{fig:NLLS_GPT2}
\end{figure}

Figure \ref{fig:NLLS_GPT3} displays the case $(\beta, \rho) = (0.2, 0.2)$ at two history sizes. In Panel (a), GPT's recommendations scatter widely above and below the optimal S-curve with no discernible structure. In Panel~(b) ($h = 125$), the recommendations converge to approximately 0.5 across almost the entire price range, mirroring the pattern in Panel (b) of Figure \ref{fig:NLLS_GPT2}. However, at extreme price ratios, GPT switches to corner solutions near 0 and 1, which distort the NLLS fit into a step function with a wide flat region, which is rationalized by a high-$\beta$ model rather than the true low-$\beta$ specification.

\begin{figure}[ht]
\centering
\includegraphics[width=0.45\linewidth]{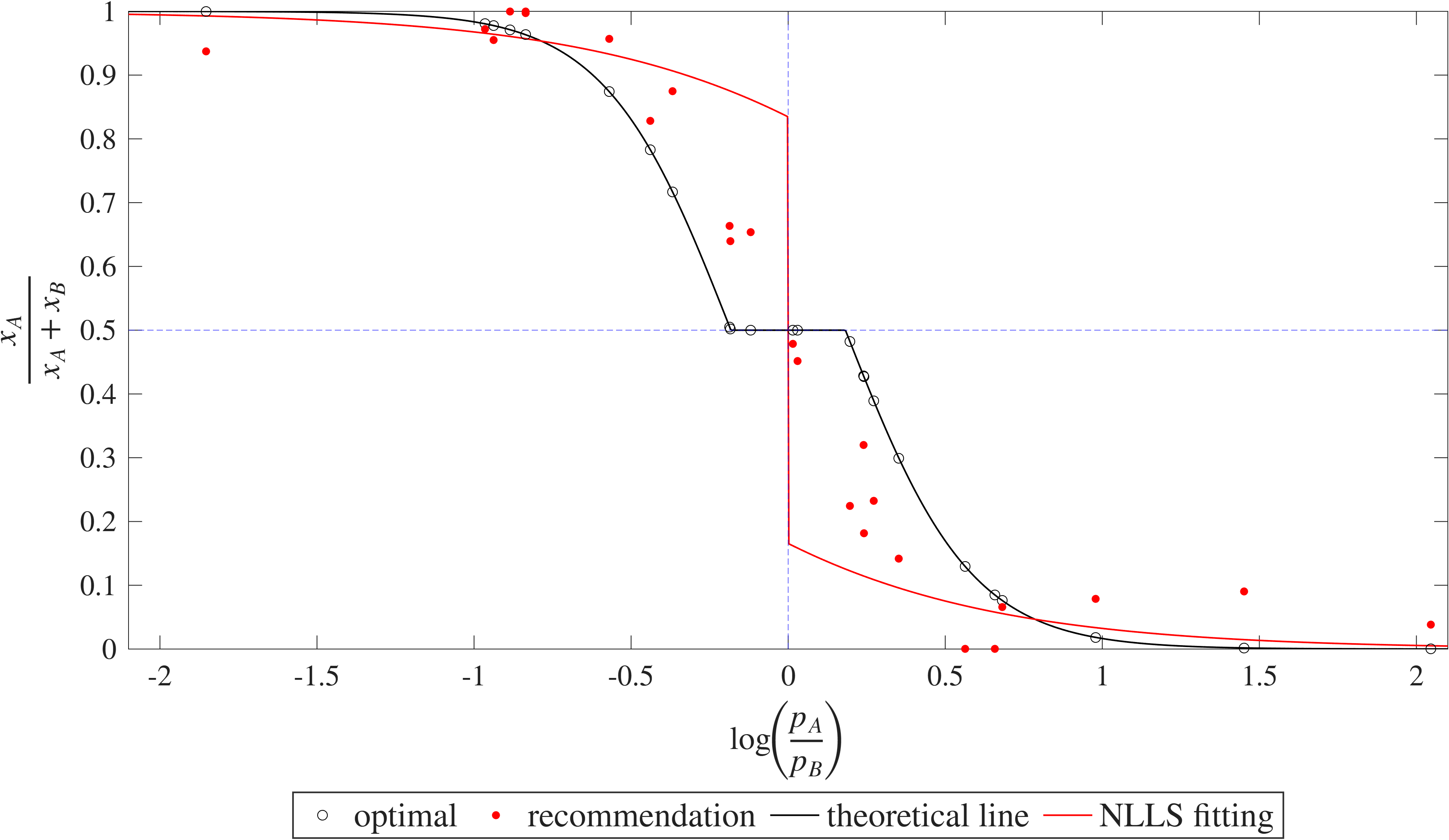}
\includegraphics[width=0.45\linewidth]{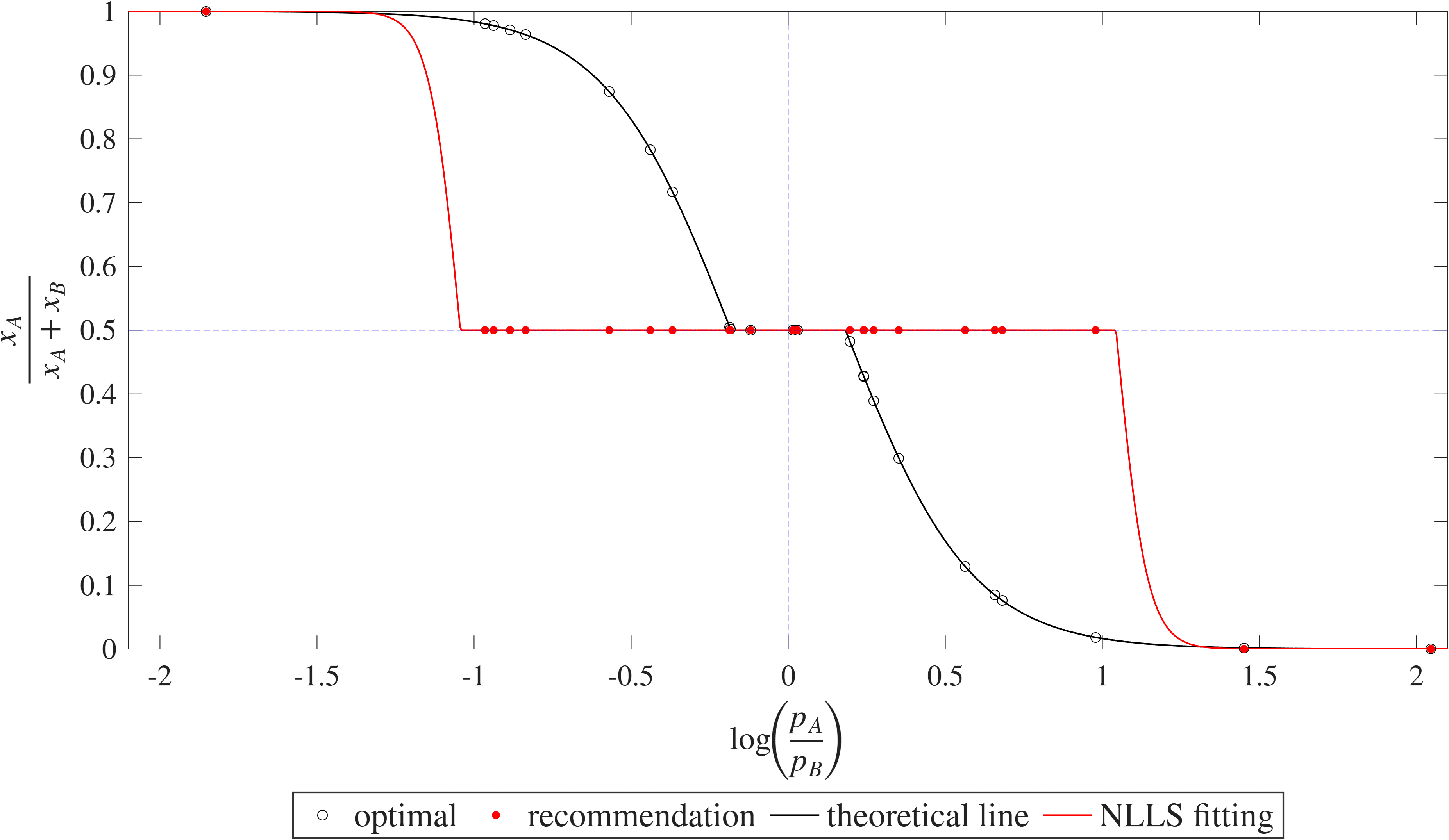}
\caption{Sample recommendations for $(\beta,\rho)=(0.2,0.2)$. Panel (a): $h=5$. Panel (b): $h=125$.}
\label{fig:NLLS_GPT3}
\end{figure}



\vskip+1em

\noindent \textbf{Additional heatmaps.} \autoref{fig:GPT_additional_heatmaps} confirms that the other non-parametric measures exhibit qualitatively similar patterns. For both the risk neutrality measure (Panel a) and the welfare loss measure (Panel b), the improvement ratios remain above or close to 1 across most of the preference parameter space, indicating little to no improvement as the history size increases. As with the vector distance, the most notable improvements are concentrated in the elation-seeking region with high risk aversion, where the ratios fall well below 1. In contrast, the disappointment-averse region ($\beta > 0$) shows ratios consistently above 1 for both measures. Overall, these results are consistent with the discussion of the normalized vector distance ratio in the main text.

\begin{figure}[ht]
\centering
\subfigure[RN improvement ratio]{
\includegraphics[width=0.48\linewidth]{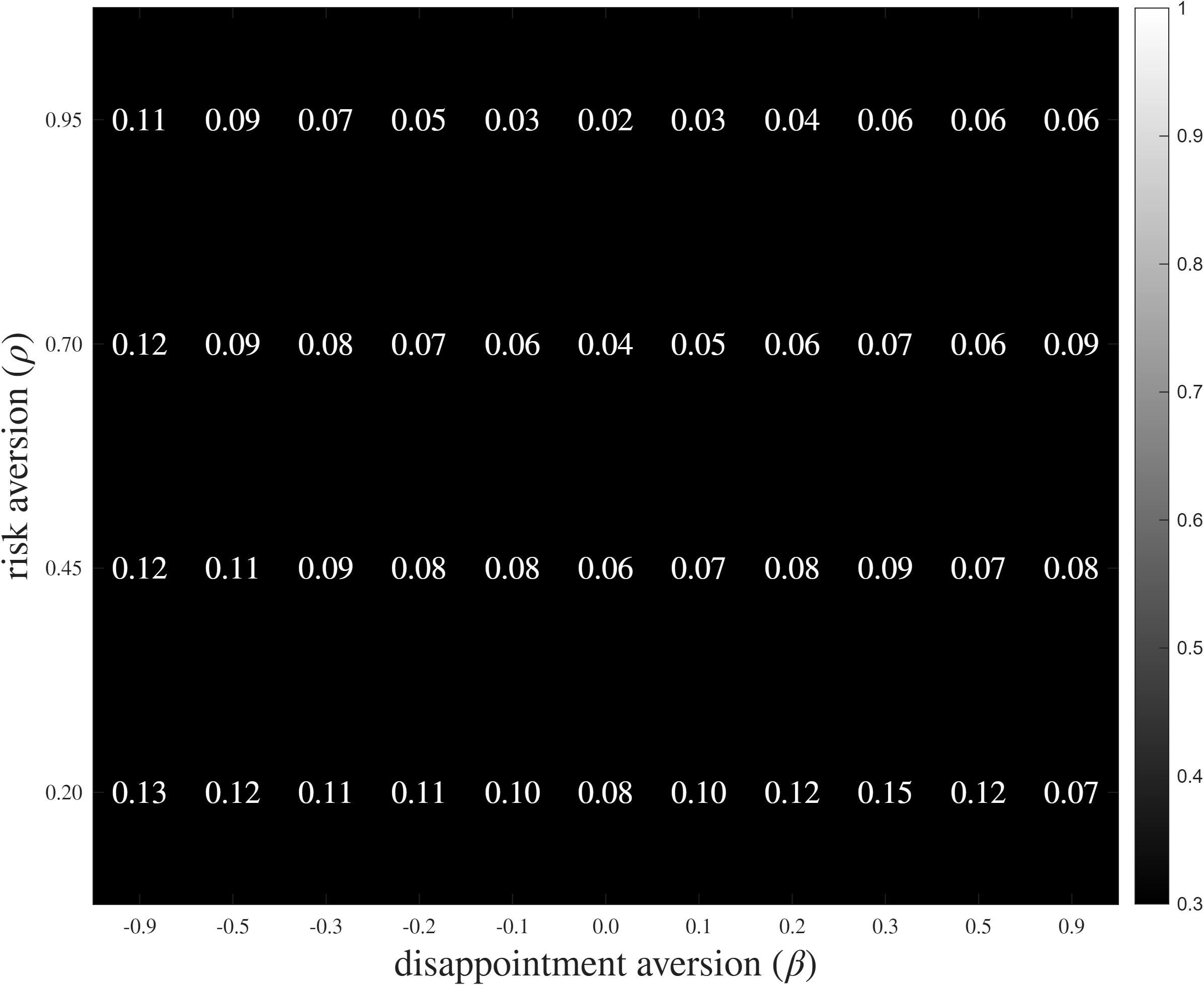}
}
\hfill
\subfigure[welfare improvement ratio]{
\includegraphics[width=0.48\linewidth]{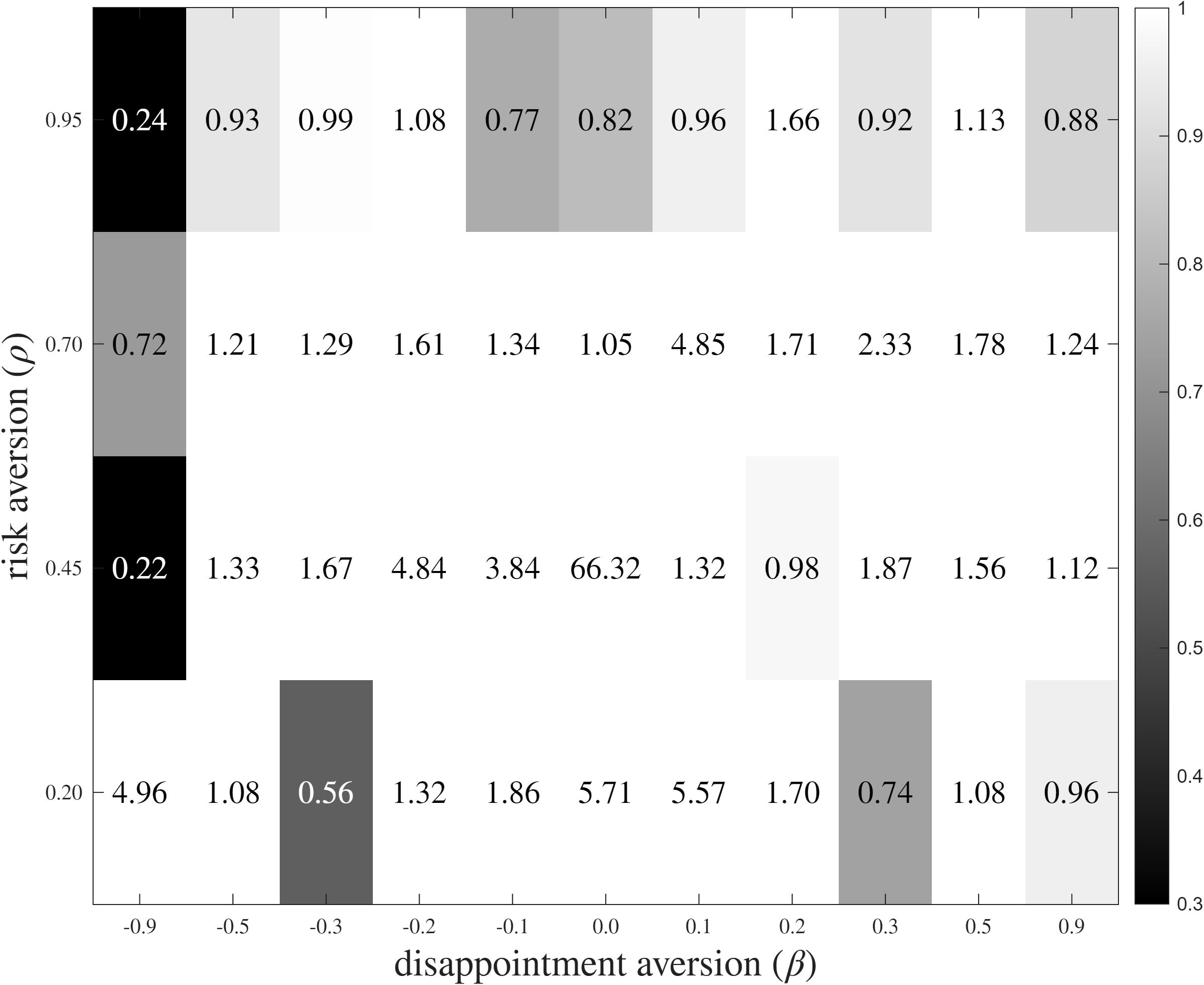}
}
\caption{The heatmaps display the improvement ratio (measure at $h=125$ divided by that at $h=5$) for the risk neutrality measure (Panel a) and the welfare loss measure (Panel b). Darker cells indicate effective learning (reduced error), while lighter cells indicate little or no improvement. Values above 1 indicate that recommendation quality worsened as the history size increased.}
\label{fig:GPT_additional_heatmaps}
\end{figure}

\vskip+1em

\newpage
\noindent \textbf{Detailed variation.} For all the measures, the changes for history is gathered below.

\begin{figure}[!htbp]
\centering
\includegraphics[width=1.2\linewidth,angle=90]{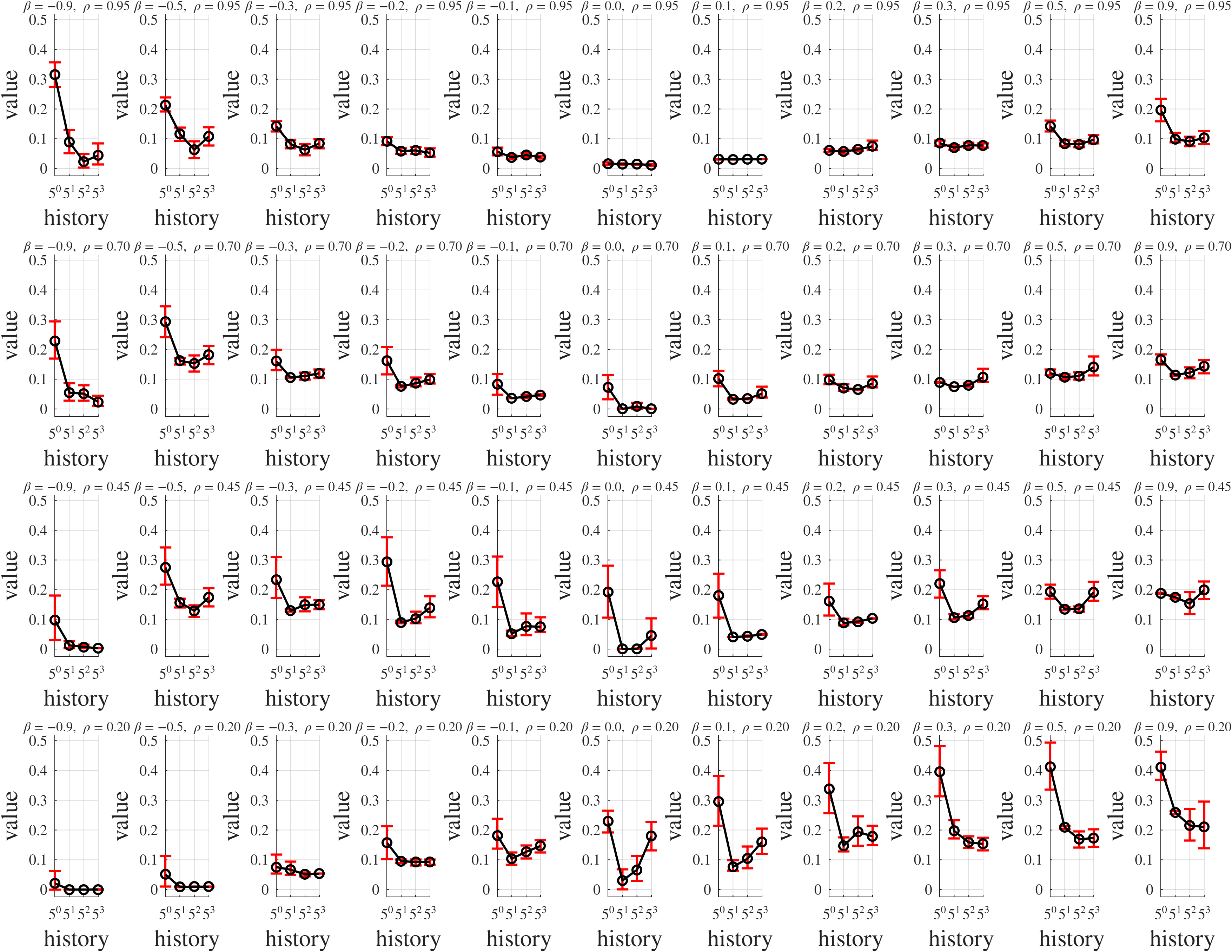}
\caption{Average normalized vector distance (AVD) by history size for each parameter combination ($\beta$,$\rho$) under GPT. Each panel plots the bootstrap mean of AVD across history sizes $h \in \{5^0, 5^1, 5^2, 5^3 \}$, with bars representing 95\% bootstrap confidence intervals. Rows correspond to increasing values of $\rho$ (from bottom to top), and columns correspond to increasing values of $\beta$ (from left to right).}
\label{fig:GPT_additional_graphs1}
\end{figure}

\begin{figure}[!htbp]
\centering
\includegraphics[width=1.2\linewidth,angle=90]{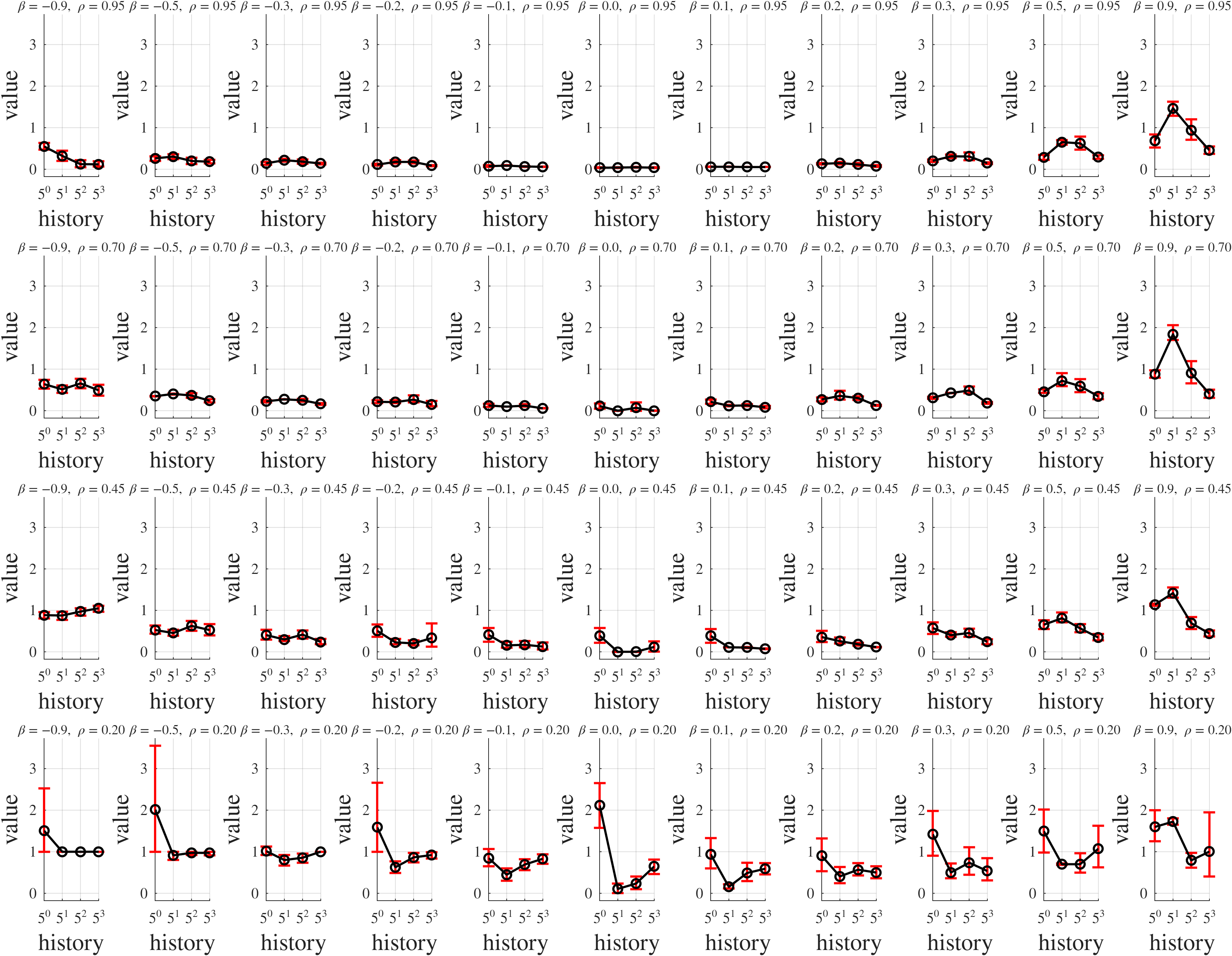}
\caption{Normalized learning error for risk aversion (NLE($\rho$)) by history size for each parameter combination ($\beta$,$\rho$) under GPT. Each panel plots the bootstrap mean of NLE($\rho$) across history sizes $h \in \{5^0, 5^1, 5^2, 5^3 \}$, with bars representing 95\% bootstrap confidence intervals. Rows correspond to increasing values of $\rho$ (from bottom to top), and columns correspond to increasing values of $\beta$ (from left to right).}
\label{fig:GPT_additional_graphs2}
\end{figure}

\begin{figure}[!htbp]
\centering
\includegraphics[width=1.2\linewidth,angle=90]{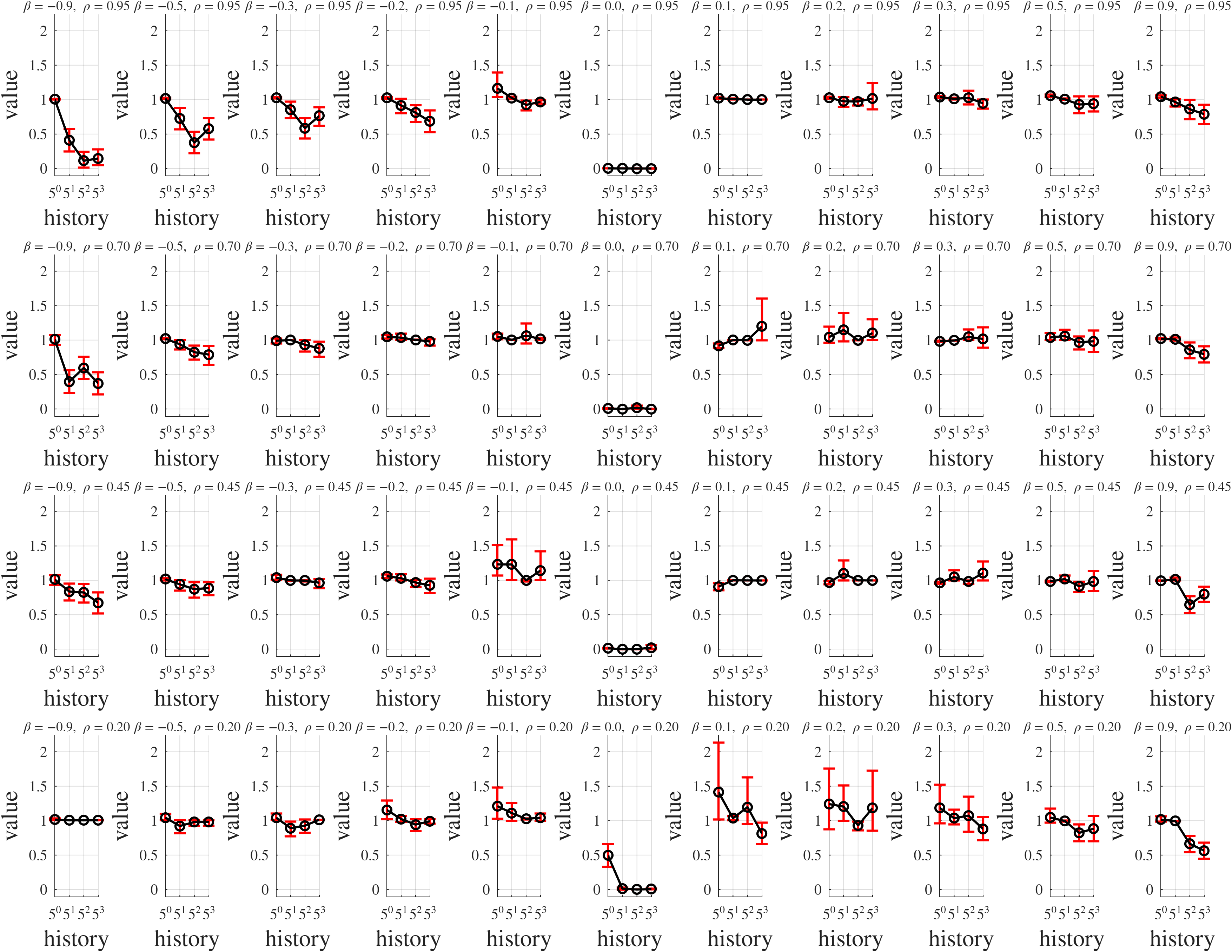}
\caption{Normalized learning error for disappointment aversion (NLE($\beta$)) by history size for each parameter combination ($\beta$,$\rho$) under GPT. Each panel plots the bootstrap mean of NLE($\beta$) across history sizes $h \in \{5^0, 5^1, 5^2, 5^3 \}$, with bars representing 95\% bootstrap confidence intervals. Rows correspond to increasing values of $\rho$ (from bottom to top), and columns correspond to increasing values of $\beta$ (from left to right).}
\label{fig:GPT_additional_graphs3}
\end{figure}

\, \,

\newpage

\subsection{Additional Figures for Gemini}
\label{label:subsection:gemini}

\noindent \textbf{Additional heatmaps.} \autoref{fig:gemini_additional_heatmaps1} presents the non-parametric improvement ratios for Gemini. Across all three measures, the vast majority of cells remain light with ratios above 1, indicating that Gemini's recommendation quality generally does not improve, and in many cases worsens, as the history size increases from $h=5$ to $h=125$. The most consistent improvement is observed at $\beta = 0.9$, where darker cells appear across most levels of risk aversion for all three measures. At $\beta = 0.5$, improvement is more limited, appearing primarily at lower levels of risk aversion but not at $\rho = 0.95$. This pattern is notably different from GPT, where the most pronounced improvement was observed in the elation-seeking region.

\begin{figure}[ht]
\centering
\subfigure[vector distance improvement ratio]{
\includegraphics[width=0.48\linewidth]{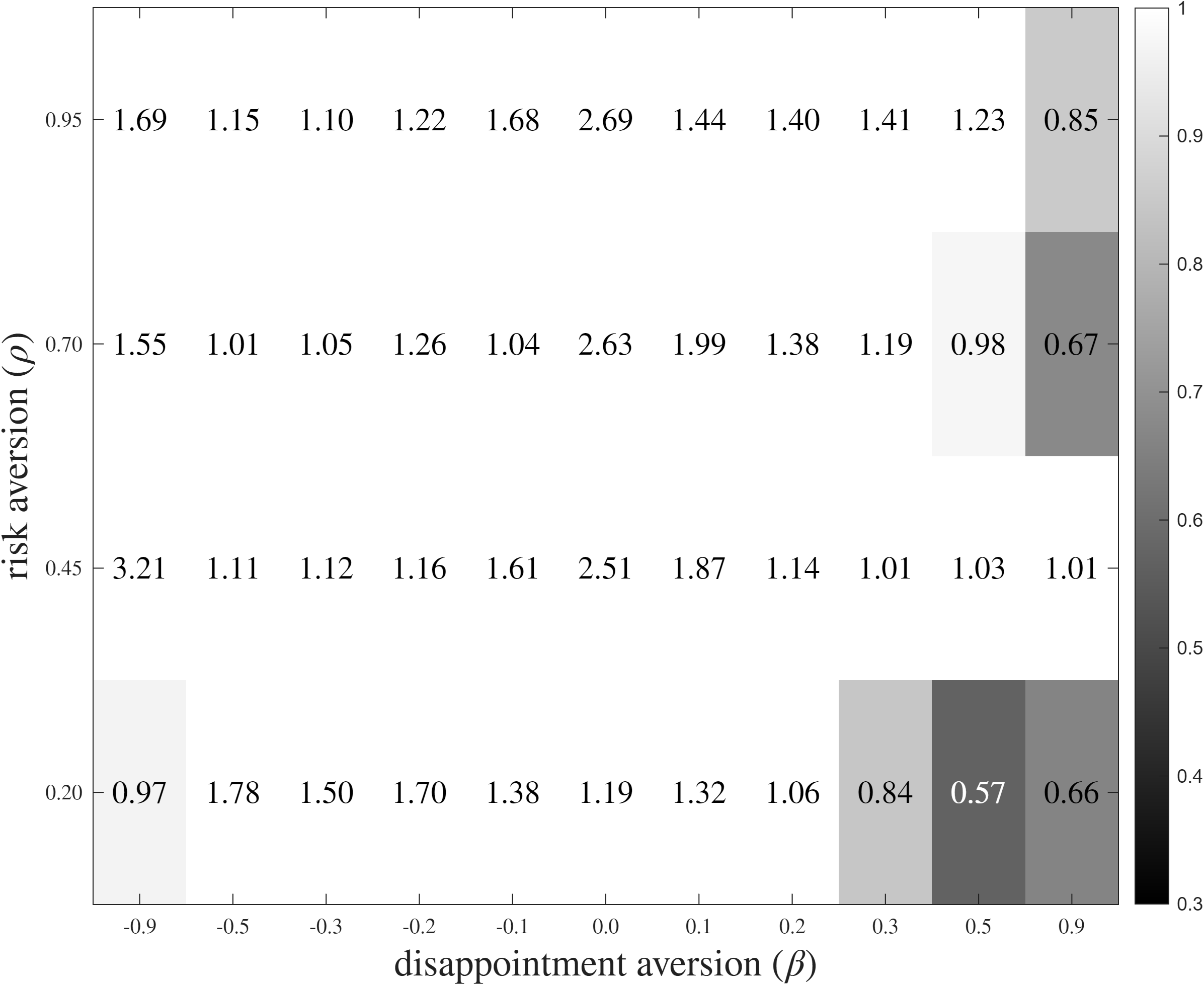}
}
\hfill
\subfigure[RN improvement ratio]{
\includegraphics[width=0.48\linewidth]{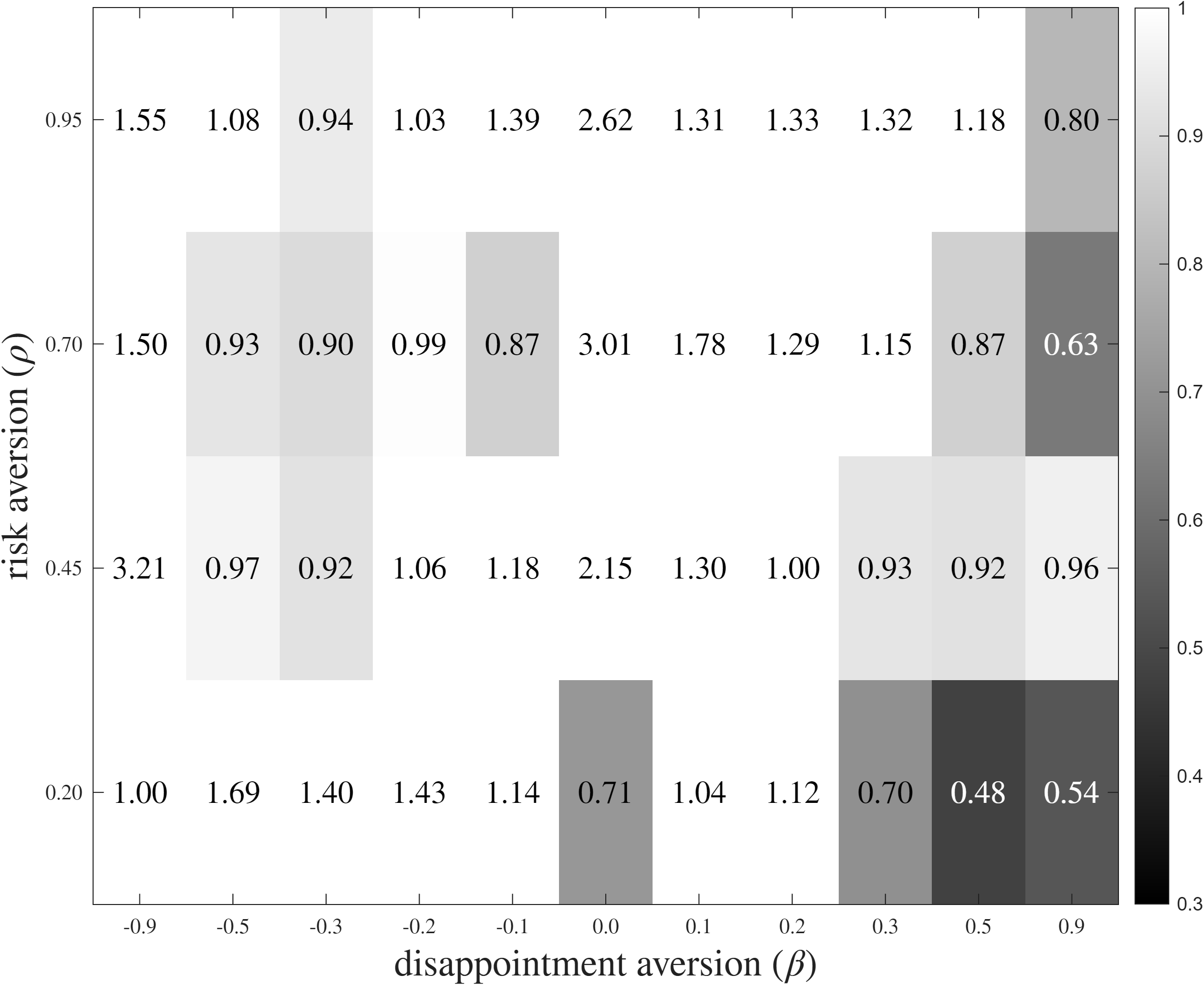}
}
\hfill
\subfigure[welfare improvement ratio]{
\includegraphics[width=0.48\linewidth]{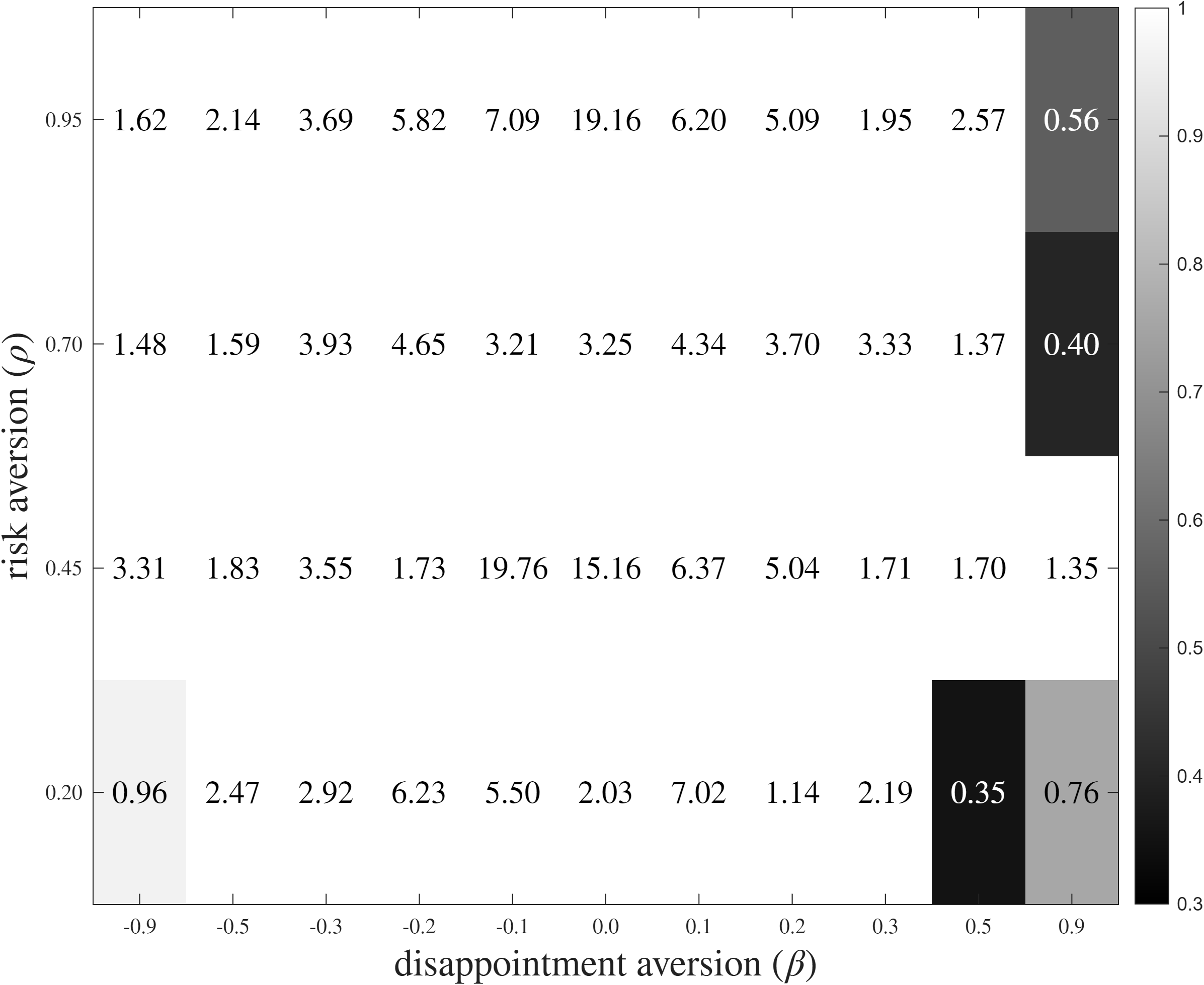}
}
\caption{The heatmaps display the improvement ratio (measure at $h=125$ divided by that at $h=5$) for (i) the vector distance measure (Panel a), (ii) the risk neutrality measure (Panel b), and (iii) the welfare loss measure (Panel c). Darker cells indicate effective learning (reduced error), while lighter cells indicate little or no improvement. Values above 1 indicate that recommendation quality worsened as the history size increased.}
\label{fig:gemini_additional_heatmaps1}
\end{figure}

\autoref{fig:gemini_additional_heatmaps2} presents the parametric estimation error improvement ratios for Gemini. In Panel (a), the heatmap for $\rho$ shows that learning is weakest near $\beta = 0$ and improves as preferences diverge from the expected utility benchmark in either direction, with the strongest improvement at high disappointment aversion ($\beta \in \{0.5, 0.9\}$). Panel (b) displays a broadly similar but asymmetric pattern for $\beta$: improvement is strong and consistent in the high disappointment aversion region, while in the elation-seeking region it appears only at intermediate values ($\beta \in \{-0.5, -0.3\}$) and vanishes at $\beta = -0.9$.

\begin{figure}[ht]
\centering
\subfigure[$\rho$ estimation error improvement ratio]{
\includegraphics[width=0.48\linewidth]{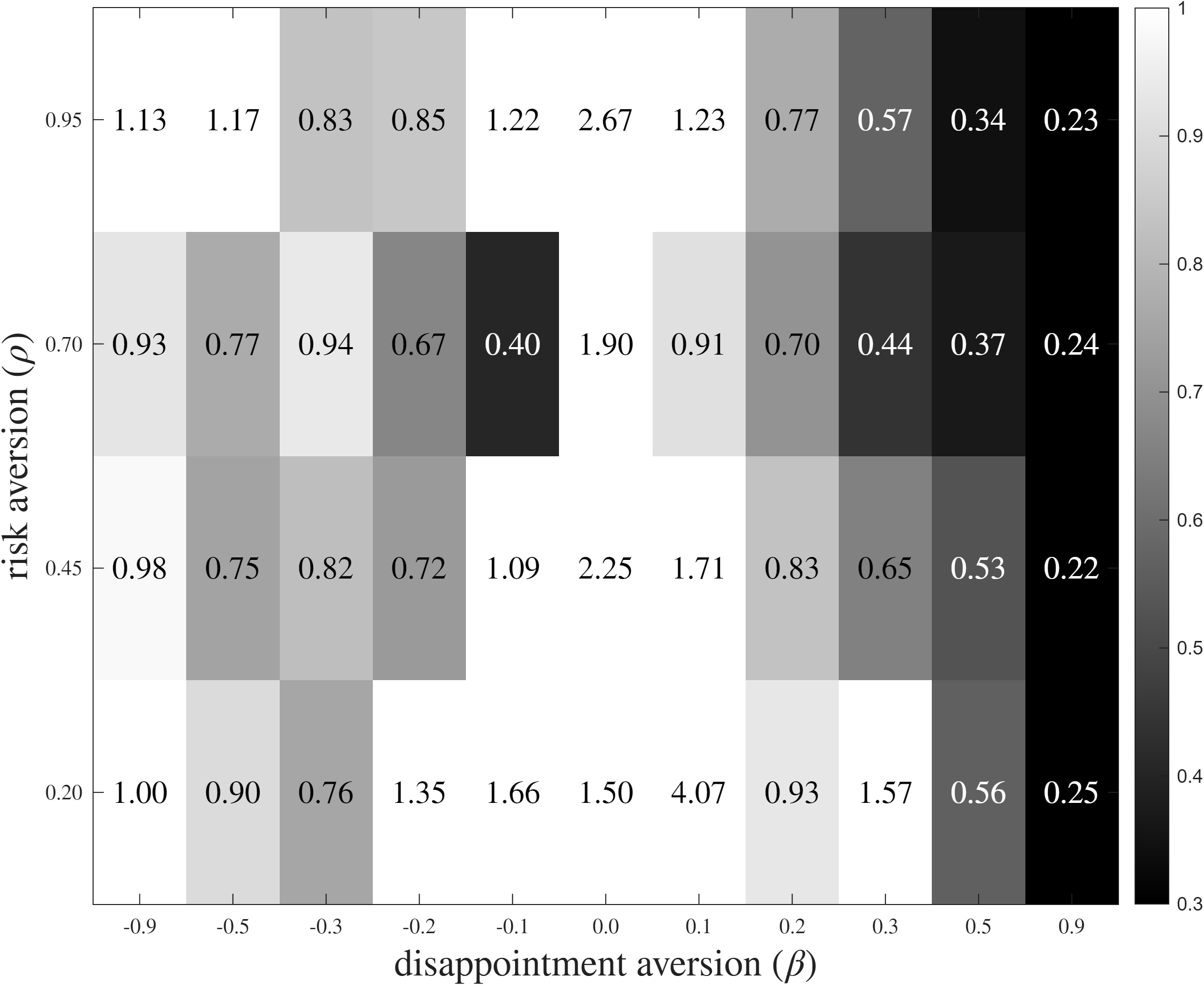}
}
\hfill
\subfigure[$\beta$ estimation error improvement ratio]{
\includegraphics[width=0.48\linewidth]{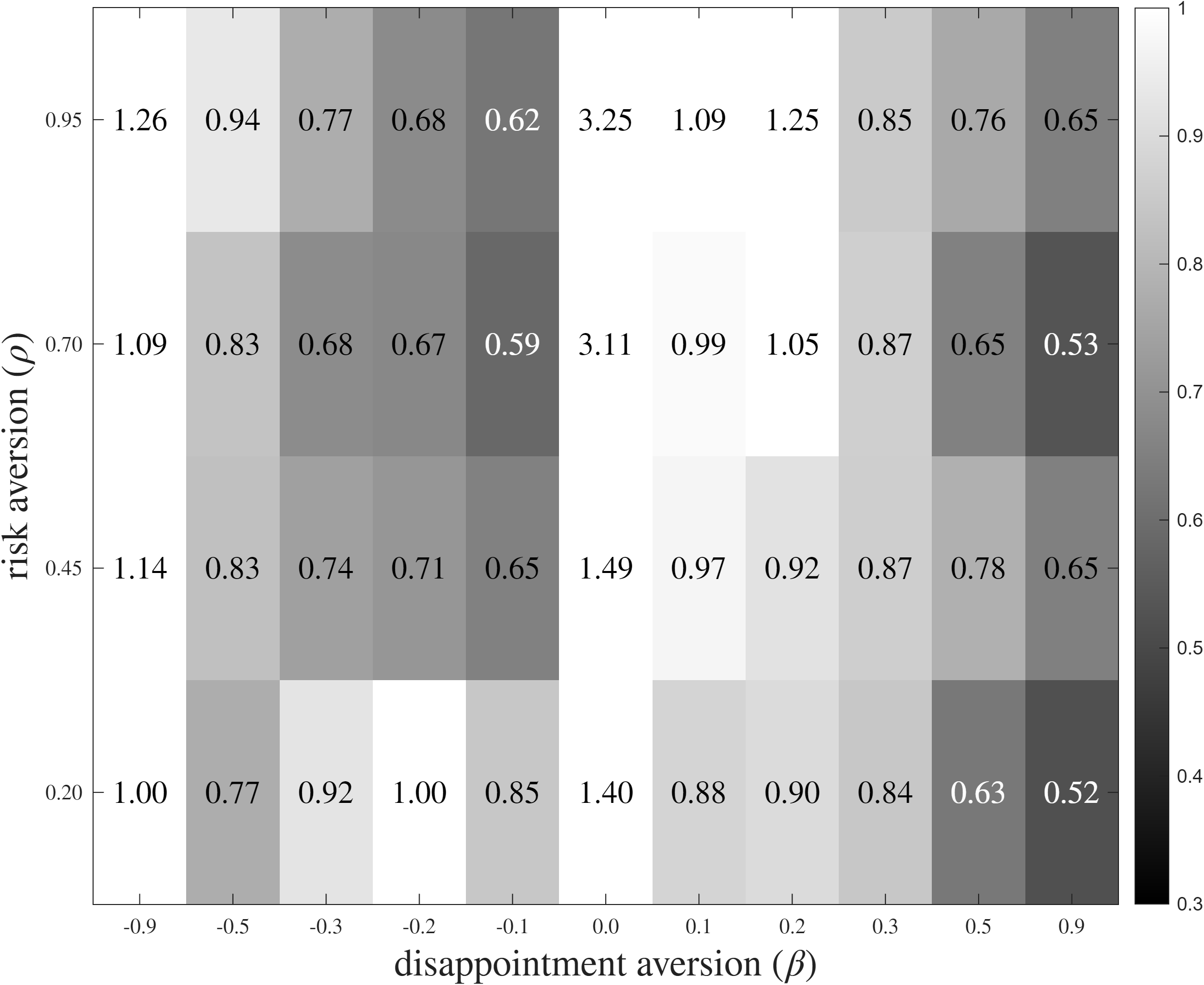}
}
\caption{The heatmaps display the estimation error improvement ratio (measure at $h = 125$ divided by that at $h = 5$) for risk aversion (Panel a) and disappointment aversion (Panel b). Darker cells indicate effective learning (reduced error), while lighter cells indicate little or no improvement. Values above 1 indicate that the estimation error increased as the history size grew from $h = 5$ to $h = 125$.}
\label{fig:gemini_additional_heatmaps2}
\end{figure}


\newpage

\noindent \textbf{Detailed variation.} For all the measures, the changes for history is gathered below.

\begin{figure}[!htbp]
\centering
\includegraphics[width=1.2\linewidth,angle=90]{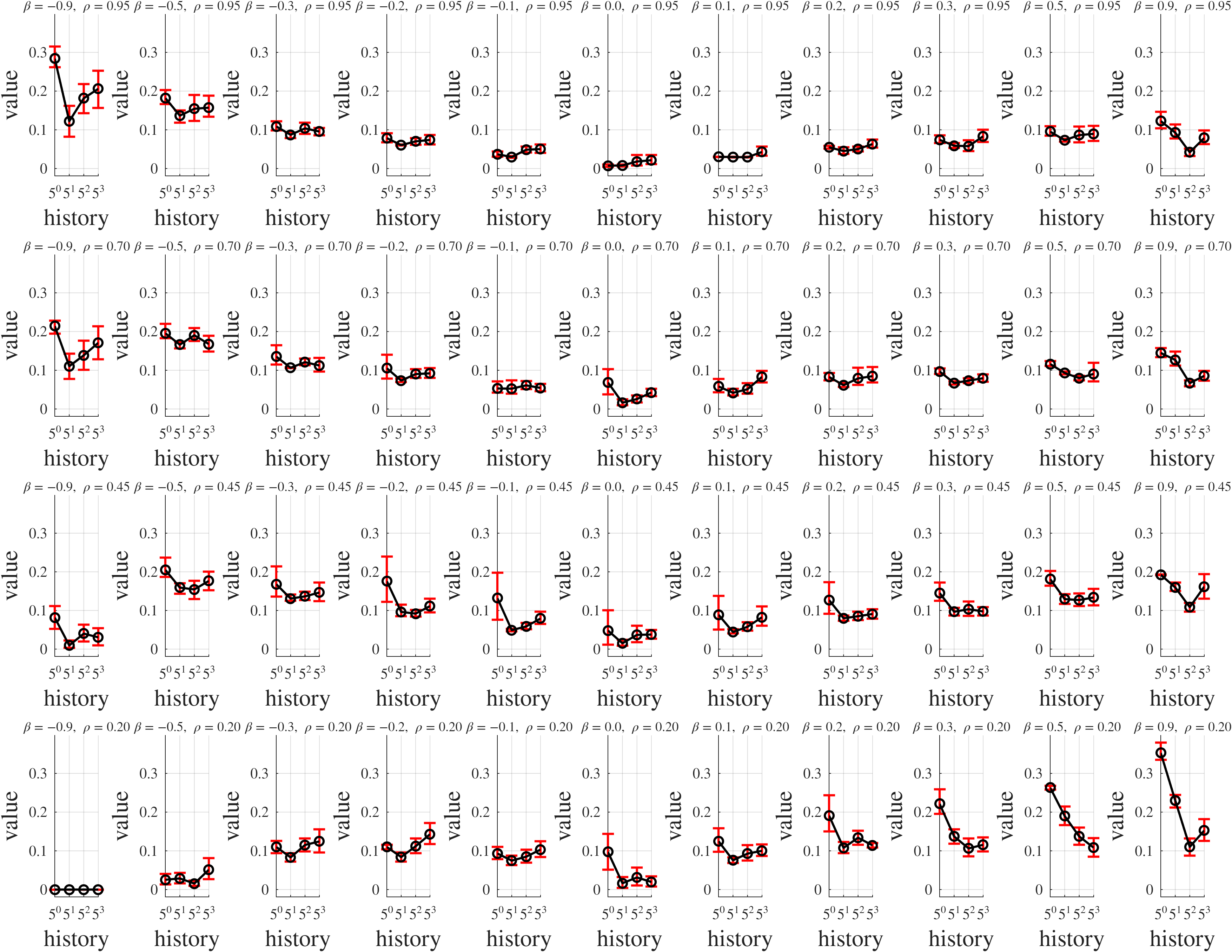}
\caption{Average normalized vector distance (AVD) by history size for each parameter combination ($\beta$,$\rho$) under Gemini. Each panel plots the bootstrap mean of AVD across history sizes $h \in \{5^0, 5^1, 5^2, 5^3 \}$, with bars representing 95\% bootstrap confidence intervals. Rows correspond to increasing values of $\rho$ (from bottom to top), and columns correspond to increasing values of $\beta$ (from left to right).}
\label{fig:gemini_additional_graphs1}
\end{figure}

\begin{figure}[!htbp]
\centering
\includegraphics[width=1.2\linewidth,angle=90]{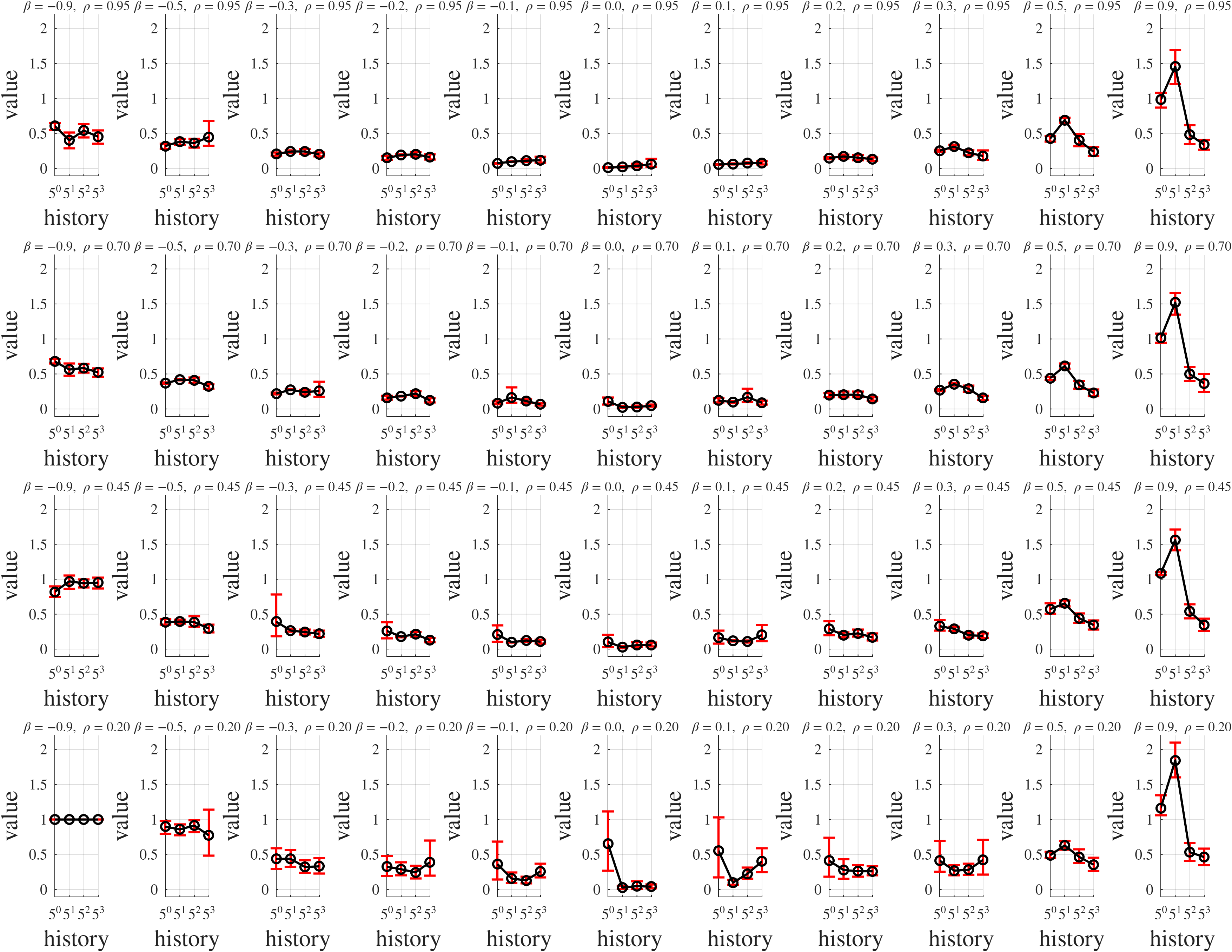}
\caption{Normalized learning error for risk aversion (NLE($\rho$)) by history size for each parameter combination ($\beta$,$\rho$) under Gemini. Each panel plots the bootstrap mean of NLE($\rho$) across history sizes $h \in \{5^0, 5^1, 5^2, 5^3 \}$, with bars representing 95\% bootstrap confidence intervals. Rows correspond to increasing values of $\rho$ (from bottom to top), and columns correspond to increasing values of $\beta$ (from left to right).}
\label{fig:gemini_additional_graphs2}
\end{figure}

\begin{figure}[!htbp]
\centering
\includegraphics[width=1.2\linewidth,angle=90]{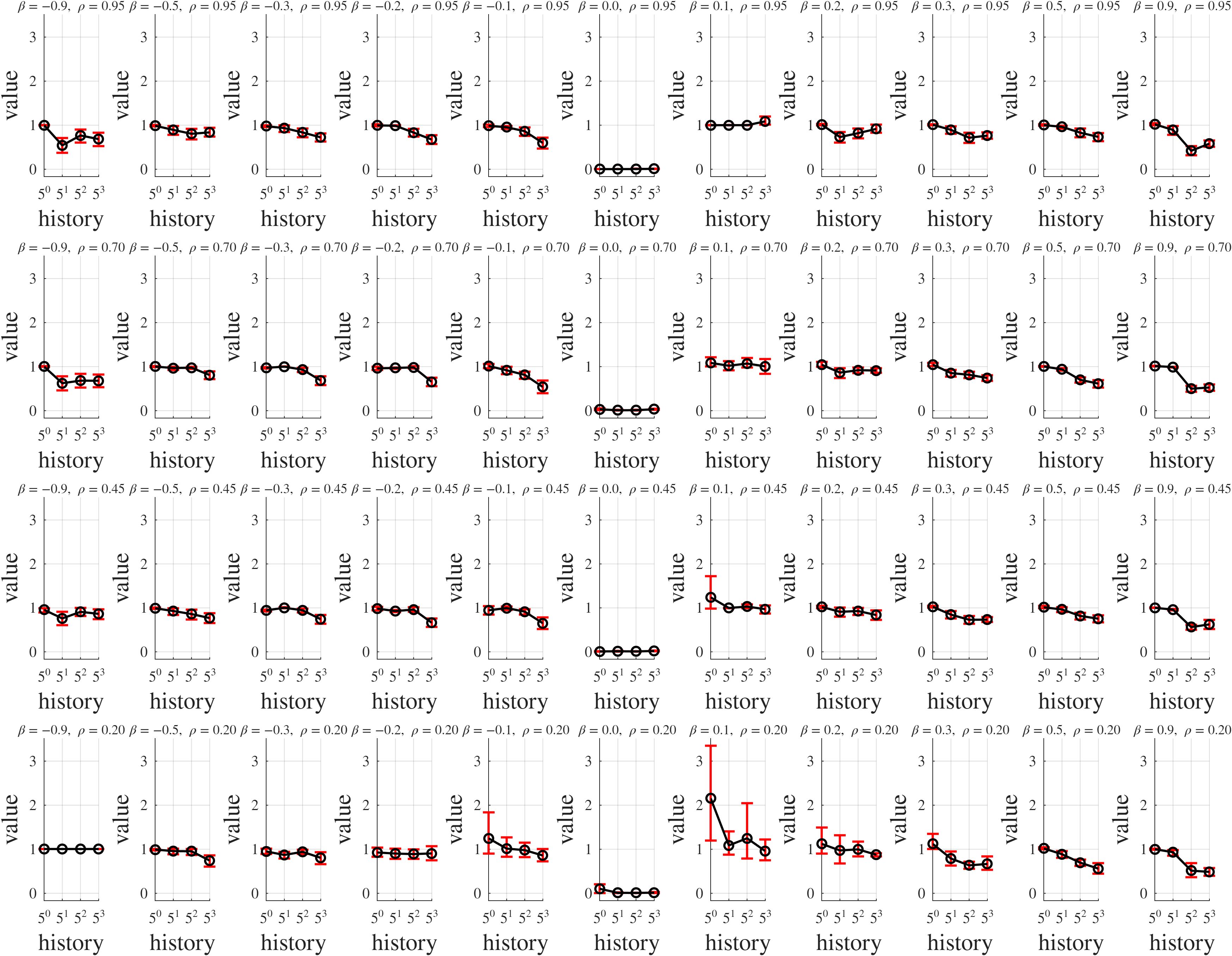}
\caption{Normalized learning error for disappointment aversion (NLE($\beta$)) by history size for each parameter combination ($\beta$,$\rho$) under Gemini. Each panel plots the bootstrap mean of NLE($\beta$) across history sizes $h \in \{5^0, 5^1, 5^2, 5^3 \}$, with bars representing 95\% bootstrap confidence intervals. Rows correspond to increasing values of $\rho$ (from bottom to top), and columns correspond to increasing values of $\beta$ (from left to right).}
\label{fig:gemini_additional_graphs3}
\end{figure}


\newpage

\subsection{Additional Figures for Claude}
\label{label:subsection:claude}

\noindent \textbf{Additional heatmaps.}  \autoref{fig:Claude_additional_heatmaps1} presents the corresponding non-parametric improvement ratios for Claude. In contrast to GPT, Claude exhibits substantially darker cells across most of the preference parameter space for all three measures, indicating that recommendation quality improves broadly as the history size increases from $h = 5$ to $h = 125$. These results suggest that Claude is considerably more effective than GPT at learning preferences from revealed-choice data. 

\begin{figure}[ht]
\centering
\subfigure[vector distance improvement ratio]{
\includegraphics[width=0.48\linewidth]{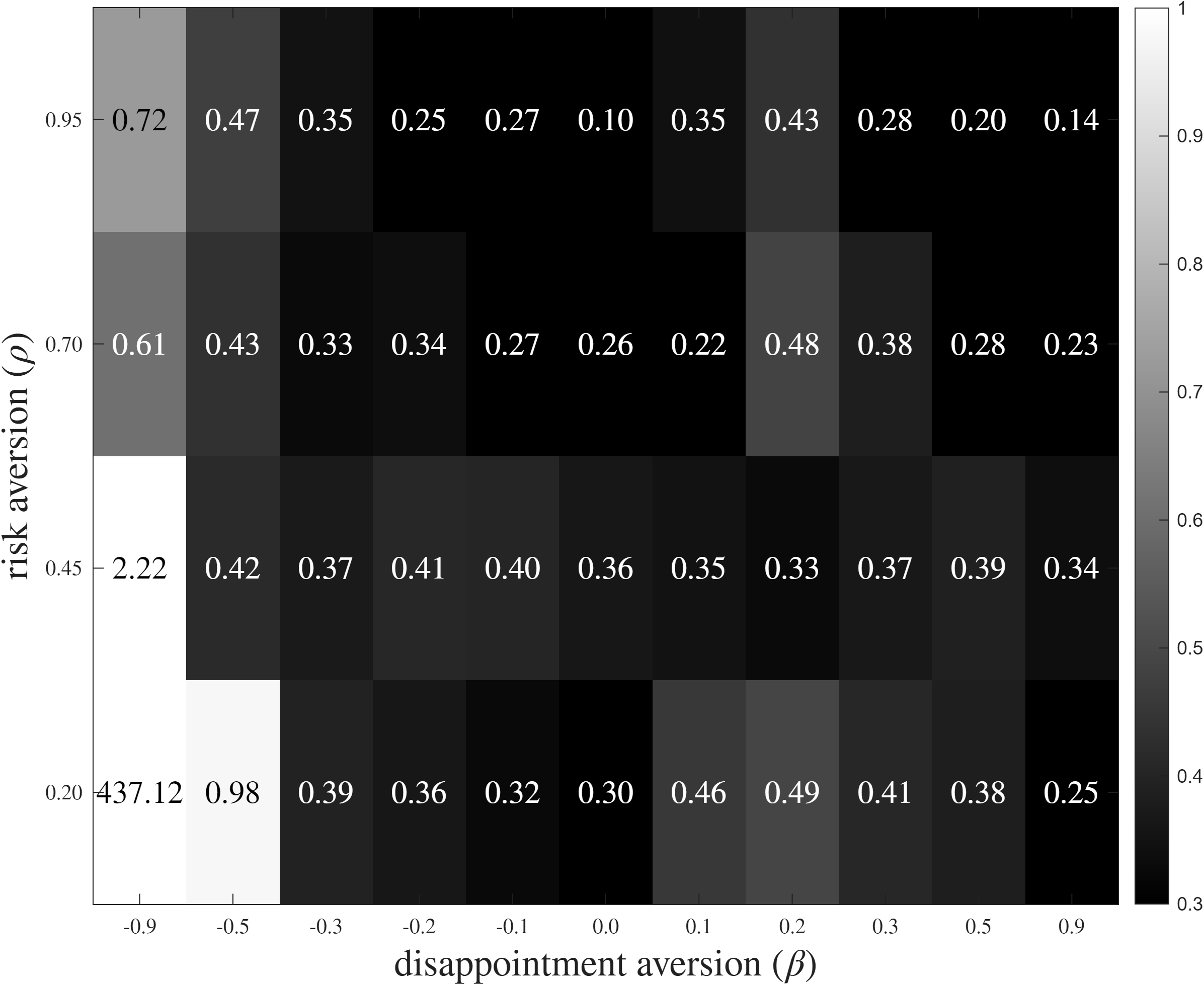}
}
\hfill
\subfigure[RN improvement ratio]{
\includegraphics[width=0.48\linewidth]{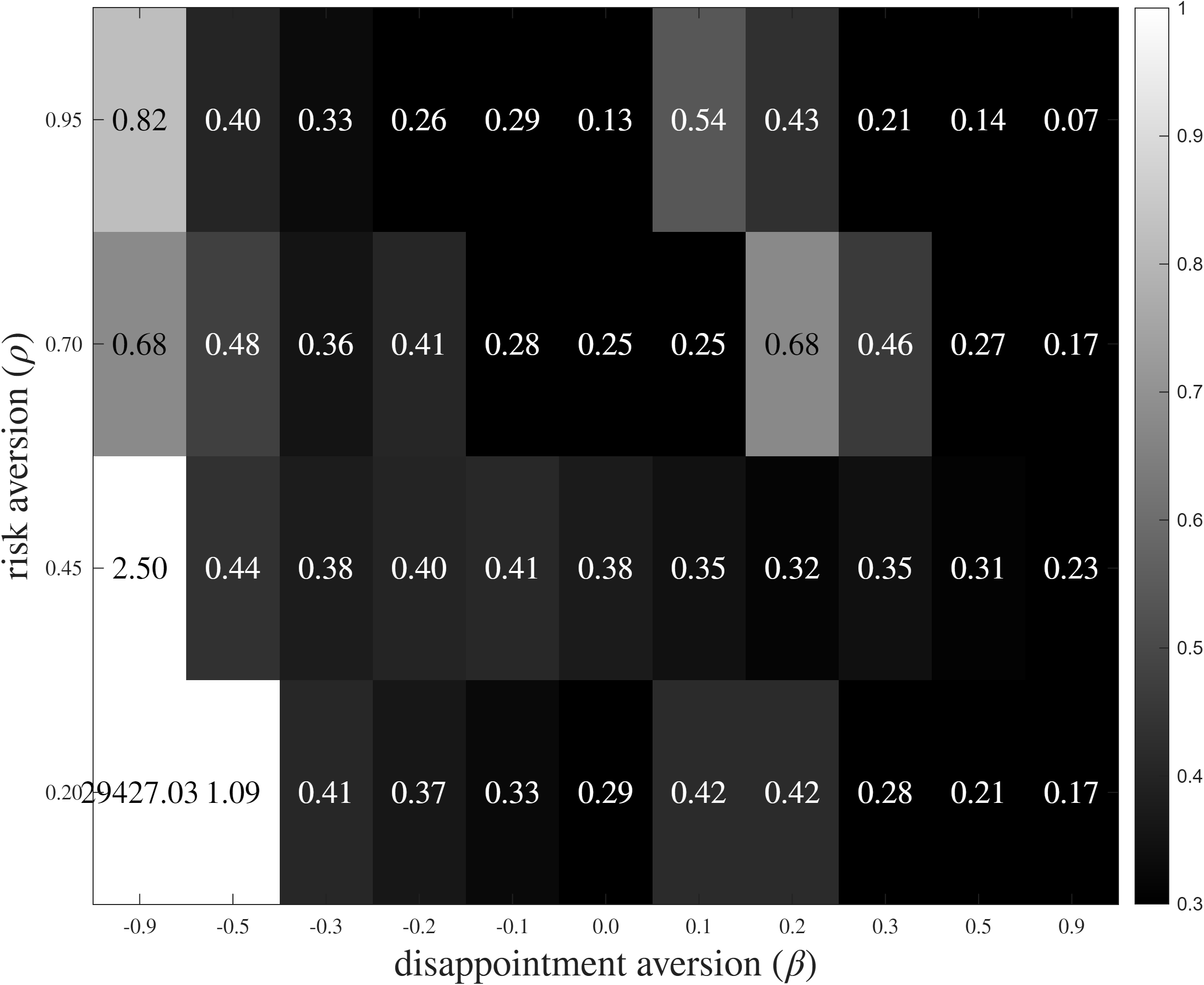}
}
\hfill
\subfigure[welfare improvement ratio]{
\includegraphics[width=0.48\linewidth]{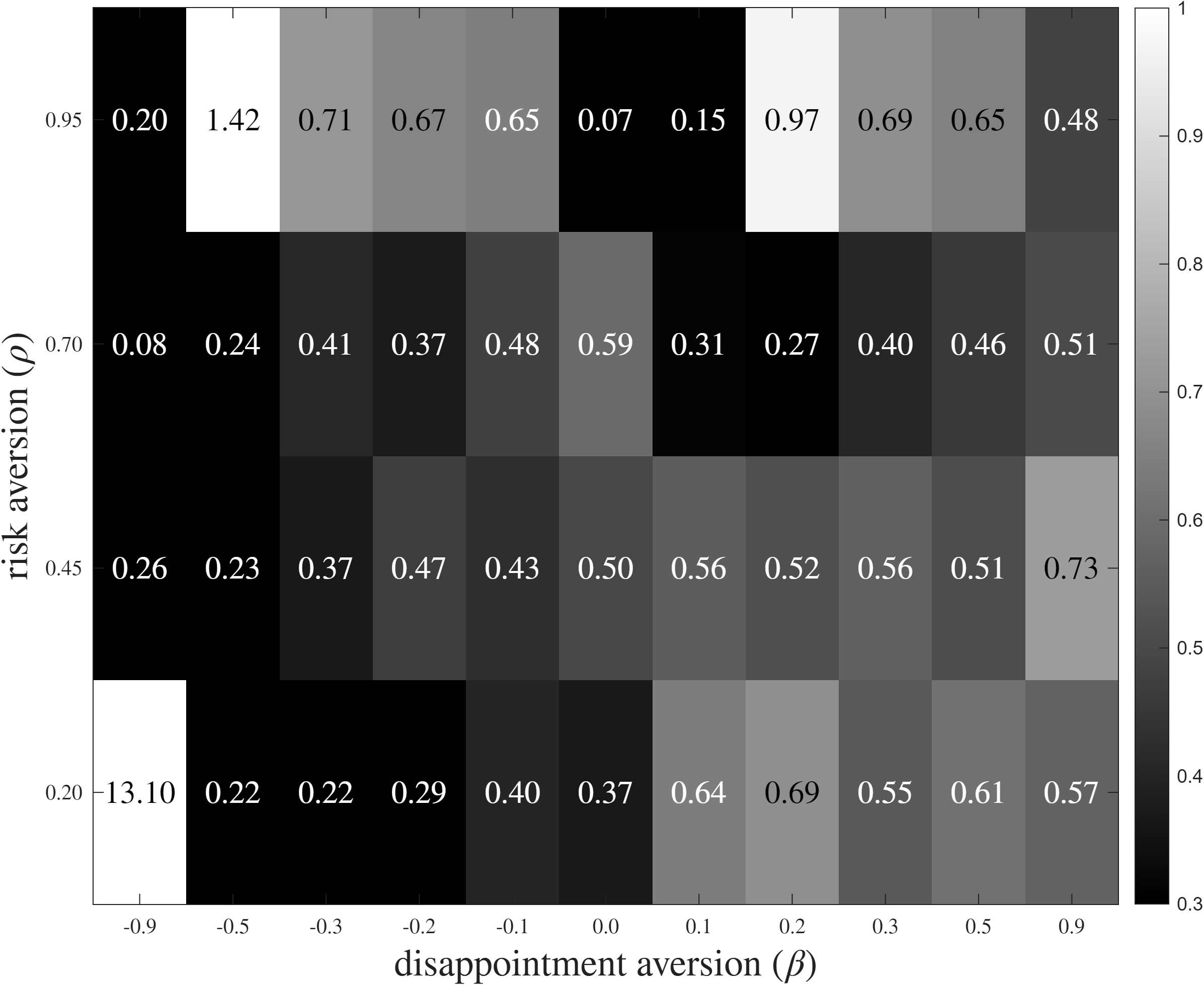}
}
\caption{The heatmaps display the improvement ratio (measure at $h=125$ divided by that at $h=5$) for (i) the vector distance measure (Panel a), (ii) the risk neutrality measure (Panel b), and (iii) the welfare loss measure (Panel c). Darker cells indicate effective learning (reduced error), while lighter cells indicate little or no improvement. Values above 1 indicate that recommendation quality worsened as the history size increased.}
\label{fig:Claude_additional_heatmaps1}
\end{figure}

\autoref{fig:claude_additional_heatmaps2} presents the parametric estimation error improvement ratios for Claude. Unlike GPT, which exhibits a stark asymmetry between the two parameters, Claude shows broadly effective learning for both $\rho$ and $\beta$ across most of the preference parameter space. In Panel (a), the majority of cells are dark, indicating substantial reductions in the estimation error for risk aversion. Notably, Panel (b) also displays predominantly dark cells, suggesting that Claude successfully recovers disappointment aversion as the history size increases, a dimension that GPT largely fails to learn.

\begin{figure}[ht]
\centering
\subfigure[$\rho$ estimation error improvement ratio]{
\includegraphics[width=0.48\linewidth]{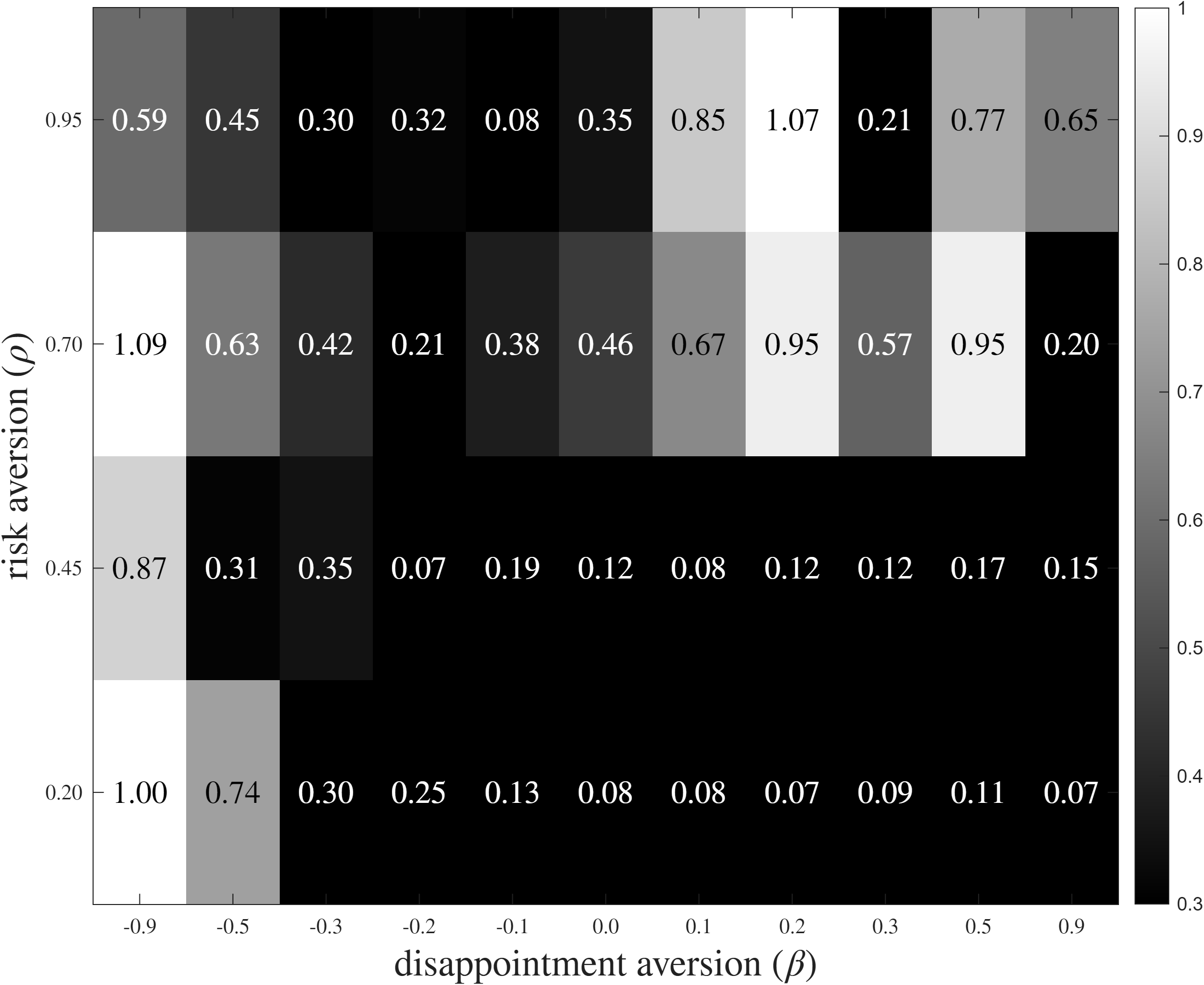}
}
\hfill
\subfigure[$\beta$ estimation error improvement ratio]{
\includegraphics[width=0.48\linewidth]{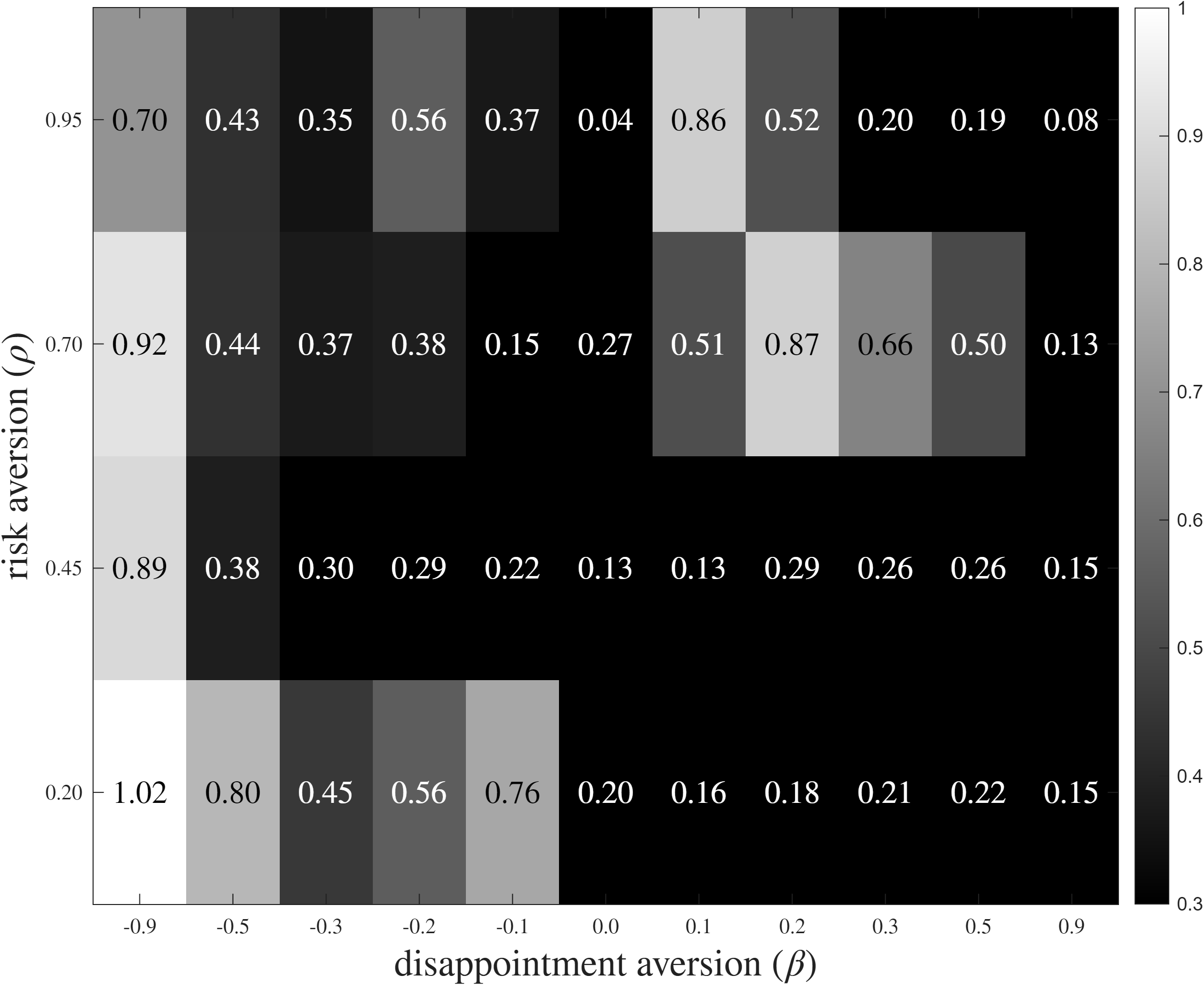}
}
\caption{The heatmaps display the estimation error improvement ratio (measure at $h = 125$ divided by that at $h = 5$) for risk aversion (Panel a) and disappointment aversion (Panel b). Darker cells indicate effective learning (reduced error), while lighter cells indicate little or no improvement. Values above 1 indicate that the estimation error increased as the history size grew from $h = 5$ to $h = 125$.}
\label{fig:claude_additional_heatmaps2}
\end{figure}


\newpage

\noindent \textbf{Detailed variation.} For all the measures, the changes for history is gathered below.

\begin{figure}[!htbp]
\centering
\includegraphics[width=1.2\linewidth,angle=90]{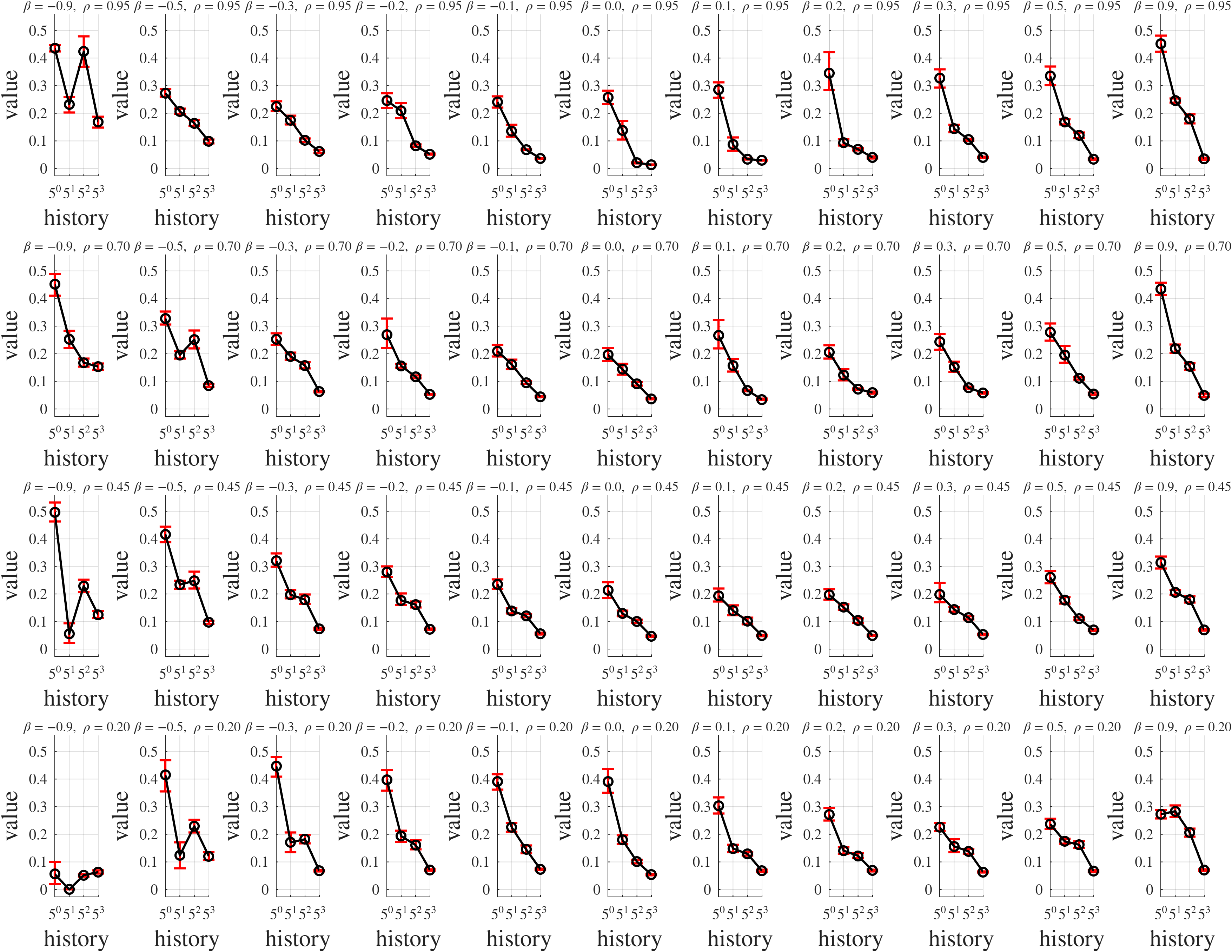}
\caption{Average normalized vector distance (AVD) by history size for each parameter combination ($\beta$,$\rho$) under Claude. Each panel plots the bootstrap mean of AVD across history sizes $h \in \{5^0, 5^1, 5^2, 5^3 \}$, with bars representing 95\% bootstrap confidence intervals. Rows correspond to increasing values of $\rho$ (from bottom to top), and columns correspond to increasing values of $\beta$ (from left to right).}
\label{fig:claude_additional_graphs1}
\end{figure}

\begin{figure}[!htbp]
\centering
\includegraphics[width=1.2\linewidth,angle=90]{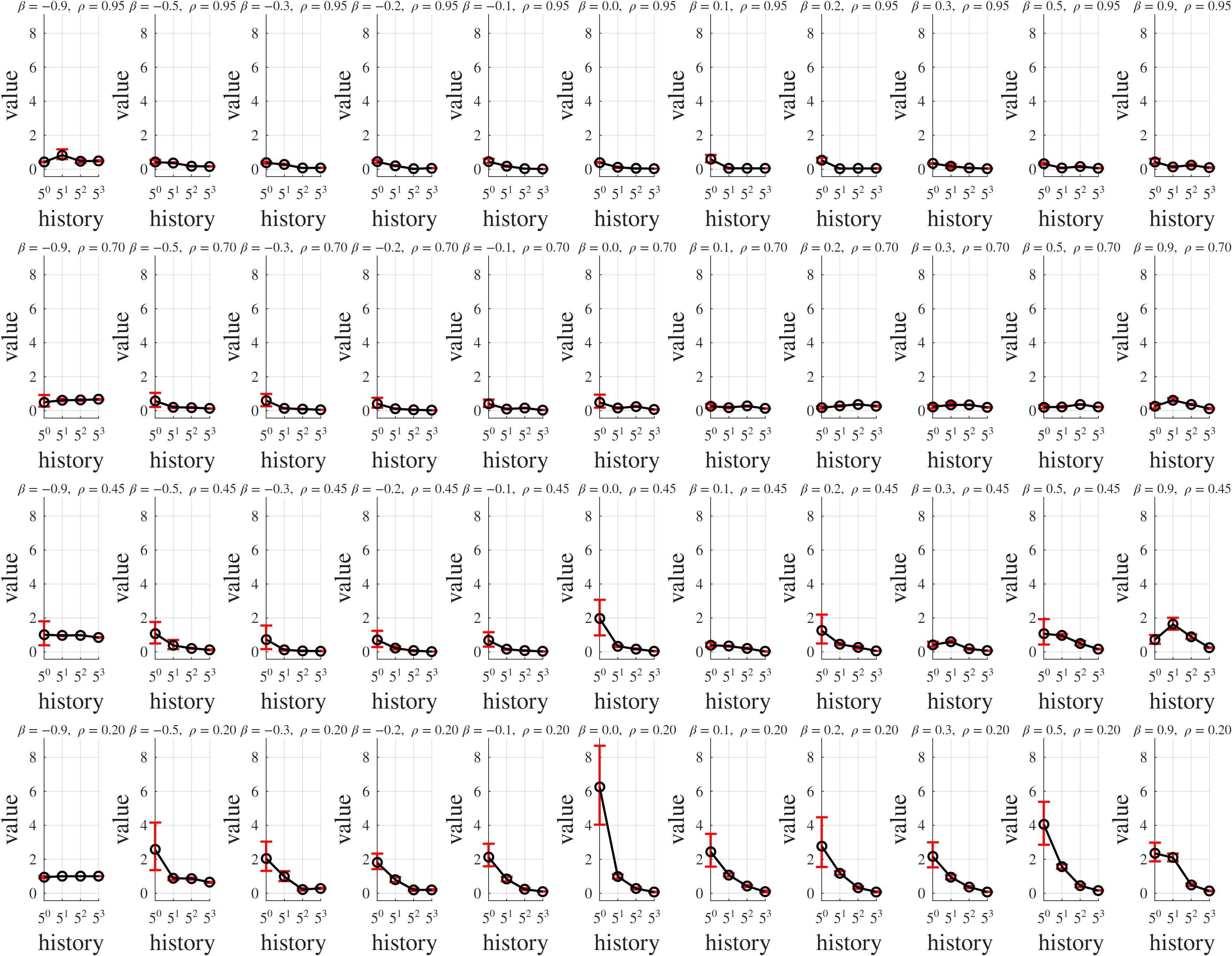}
\caption{Normalized learning error for risk aversion (NLE($\rho$)) by history size for each parameter combination ($\beta$,$\rho$) under Claude. Each panel plots the bootstrap mean of NLE($\rho$) across history sizes $h \in \{5^0, 5^1, 5^2, 5^3 \}$, with bars representing 95\% bootstrap confidence intervals. Rows correspond to increasing values of $\rho$ (from bottom to top), and columns correspond to increasing values of $\beta$ (from left to right).
}
\label{fig:claude_additional_graphs2}
\end{figure}

\begin{figure}[!htbp]
\centering
\includegraphics[width=1.2\linewidth,angle=90]{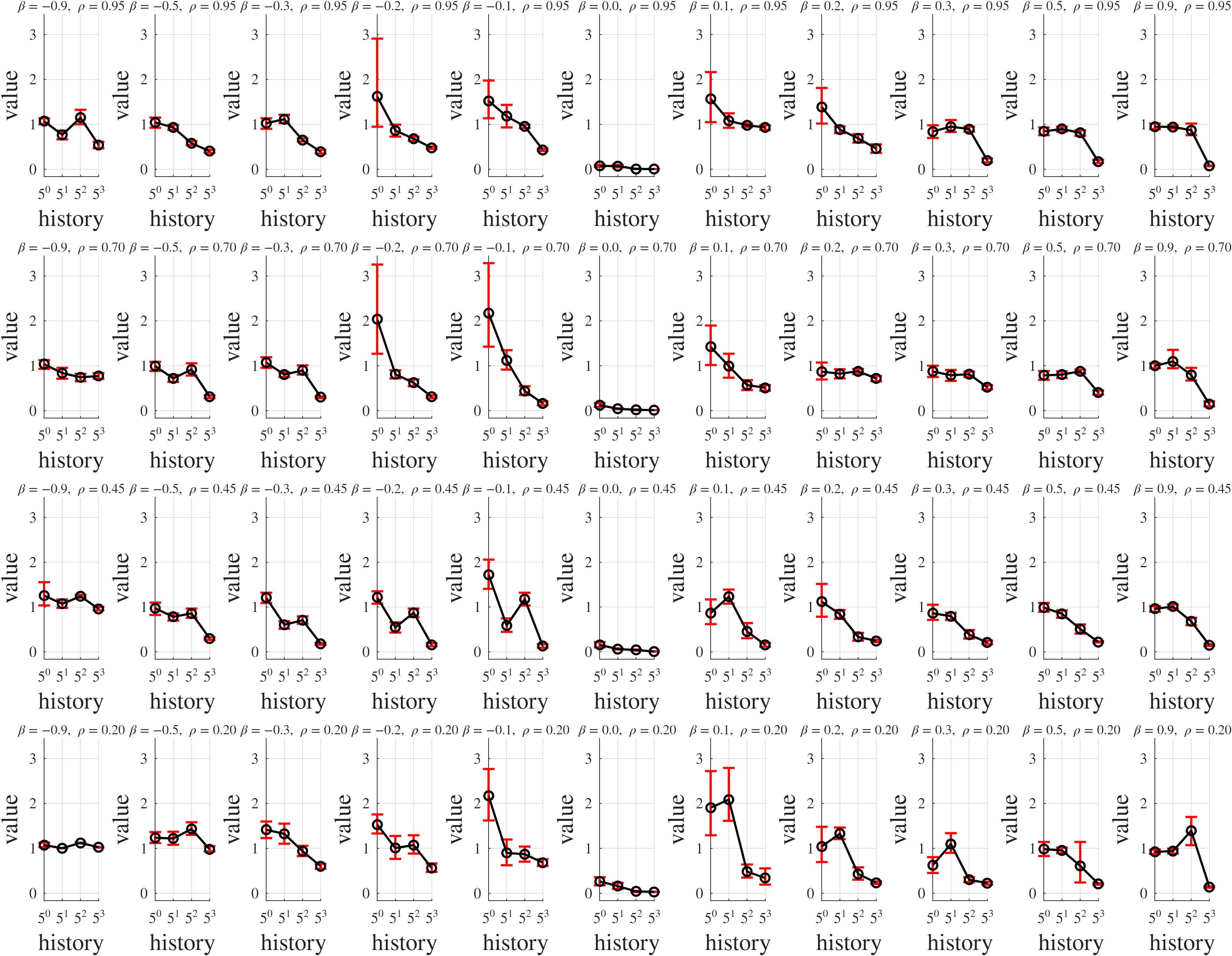}
\caption{Normalized learning error for disappointment aversion (NLE($\beta$)) by history size for each parameter combination ($\beta$,$\rho$) under Claude. Each panel plots the bootstrap mean of NLE($\beta$) across history sizes $h \in \{5^0, 5^1, 5^2, 5^3 \}$, with bars representing 95\% bootstrap confidence intervals. Rows correspond to increasing values of $\rho$ (from bottom to top), and columns correspond to increasing values of $\beta$ (from left to right).}
\label{fig:claude_additional_graphs3}
\end{figure}


\newpage

\subsection{Machine Learning Models}
\label{label:subsection:ML}

\noindent \textbf{kNN model construction.} {We applied a $k$-nearest neighbors (kNN) model.}
The training data were extracted from the learning dataset corresponding to each preference parameter vector $(\beta, \rho)$ for $h = 5, 25, 125$.

At the prediction step, the $k$ training observations with the shortest Euclidean distance to the new price vector were identified, and the average of these neighbors' $x_A^*$ values was returned as the prediction. The optimal purchase quantity $x_B^*$ for asset B was then derived from the budget constraint $p_A x_A^{*} + p_B x_B^{*} = 100$.

We performed hyperparameter tuning exclusively within the training dataset using cross-validation (CV): Leave-One-Out CV for $h=5$ and 5-fold CV for $h=25$ and $h=125$. The value of $k$ was searched over $k \in \{1,2,3,4,5\}$; however, for $h=5$, it was restricted to $k \in \{1,2,3,4\}$ because each training fold contains only four observations. The final model was selected as the combination that minimized the mean squared error of the asset allocation ratio $\frac{x_A^*}{x_A^*+x_B^*}$. The recommendation set was used only for evaluating model performance on out-of-sample data.


\vskip+1em

\noindent \textbf{RF model construction.} We applied a random forest (RF) regression model. The training data were extracted from the learning dataset corresponding to each preference parameter vector $(\beta, \rho)$, taking the first $h$ observations where $h \in \{5, 25, 125\}$. RF is an ensemble technique that independently trains multiple decision trees and uses the average of their predictions as the final output, mitigating overfitting of individual trees and enhancing prediction stability.

The hyperparameter search range was adjusted according to the size of the training data. Grid search was performed across all combinations of the number of trees, maximum tree depth, the minimum sample size required for internal node splits, and the minimum sample size required for leaf nodes. We performed hyperparameter tuning exclusively within the training dataset using cross-validation (CV): Leave-One-Out CV for $h=5$ and 5-fold CV for $h=25$ and $h=125$. The final model was selected as the combination that minimized the mean squared error of the asset allocation ratio $\frac{x_A^*}{x_A^*+x_B^*}$. The recommendation set was used only for evaluating model performance on out-of-sample data. The optimal purchase quantity $x_B^*$ for asset B was derived from the budget constraint $p_A x_A^{*}+p_B x_B^{*}=100$.


\vskip+1em

\noindent \textbf{SVR model construction.} We applied a support vector regression (SVR) model based on a radial basis function kernel. The training data were extracted from the learning dataset corresponding to each parameter combination $(\beta, \rho)$, taking the first $h$ observations where $h \in \{5, 25, 125\}$. SVR performs robust prediction in the presence of noise in the training data by learning a regression hyperplane that minimizes the $\epsilon$-insensitive loss function in a high-dimensional feature space.

The search range for hyperparameters was adjusted according to the size of the training data. Grid search was performed across all combinations of the regularization parameter, tolerance, and kernel width. We performed hyperparameter tuning exclusively within the training dataset using cross-validation (CV): Leave-One-Out CV for $h=5$ and 5-fold CV for $h=25$ and $h=125$. The final model was selected as the combination that minimized the mean squared error of the asset allocation ratio $\frac{x_A^*}{x_A^*+x_B^*}$. The recommendation set was used only for evaluating model performance on out-of-sample data. The optimal purchase quantity $x_B^*$ for asset B was derived from the budget constraint $p_A x_A^{*} + p_B x_B^{*} = 100$. 


\vskip+1em

\noindent \textbf{Target Variable and Post-Prediction Adjustment.} All the three machine learning models were trained to predict the raw quantity $x_A^*$ directly, using the observed price vectors $(p_A, p_B)$ as input features. The quantity $x_B^*$ is then derived from the budget constraint: $x_B^* = \frac{100-p_A x_A^*}{p_B}$. Since the models do not inherently enforce non-negativity or budget constraint, a post-prediction adjustment was applied to all predicted values. Specifically, for each prediction $x_A^*$, if $x_A^* < 0$, it was set to $0$ and $x_B^*$ was recalculated as $100/p_B$; if the resulting $x_B^* < 0$ (i.e., $p_A x_A^* > 100$), then $x_B^*$ was set to $0$ and $x_A^*$ was recalculated as $100/p_A$.


\newpage

\subsection{Additional Figures for kNN}
\label{label:subsection:kNN}

\noindent \textbf{Additional heatmaps.} \autoref{fig:kNN_additional_heatmaps1} presents the non-parametric improvement ratios for kNN. In contrast to the LLMs, the heatmaps are nearly uniformly dark across the entire preference parameter space for all three measures, indicating that kNN's recommendation quality improves consistently as the history size increases from $h=5$ to $h=125$. 

\begin{figure}[ht]
\centering
\subfigure[vector distance improvement ratio]{
\includegraphics[width=0.48\linewidth]{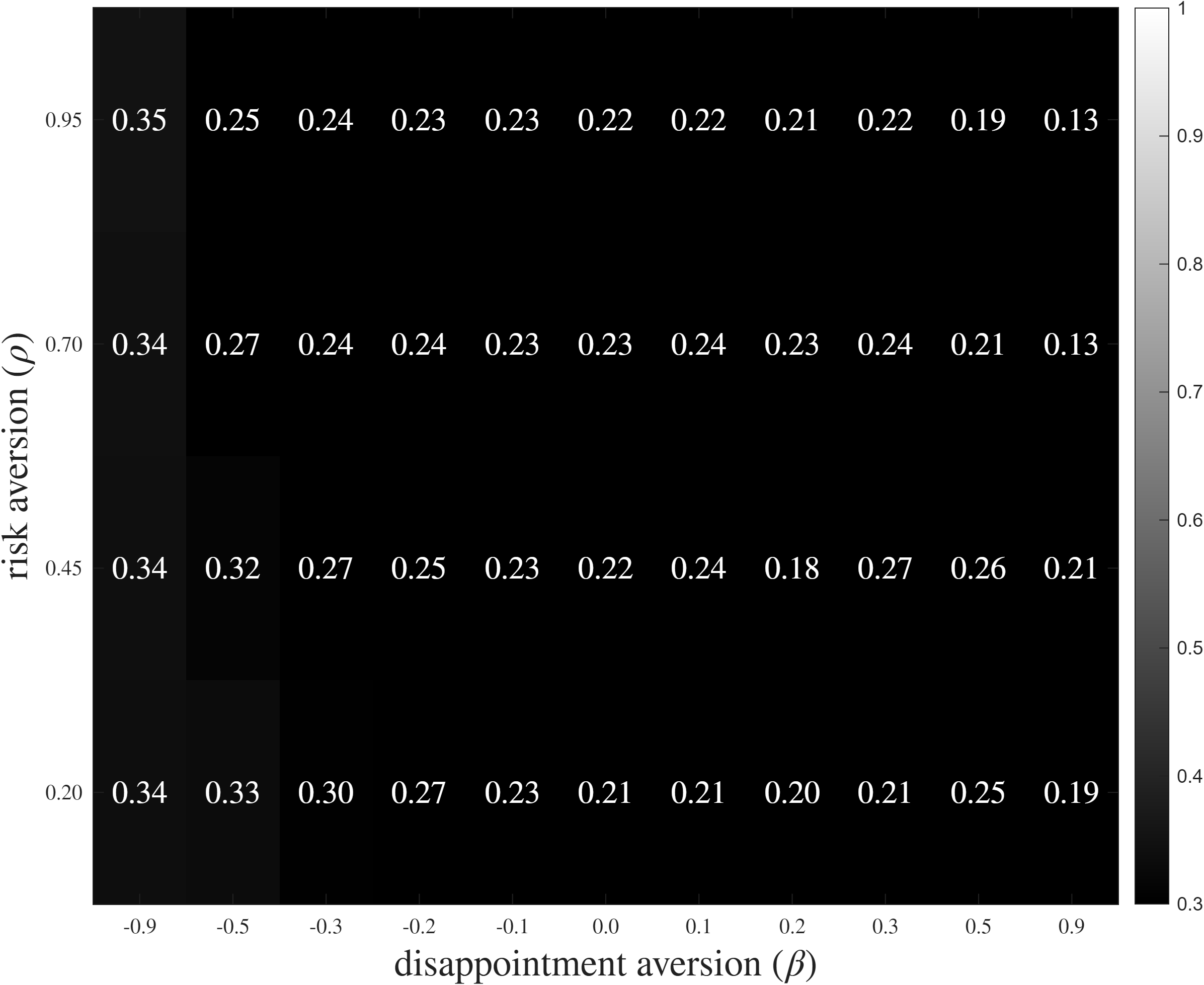}
}
\hfill
\subfigure[RN improvement ratio]{
\includegraphics[width=0.48\linewidth]{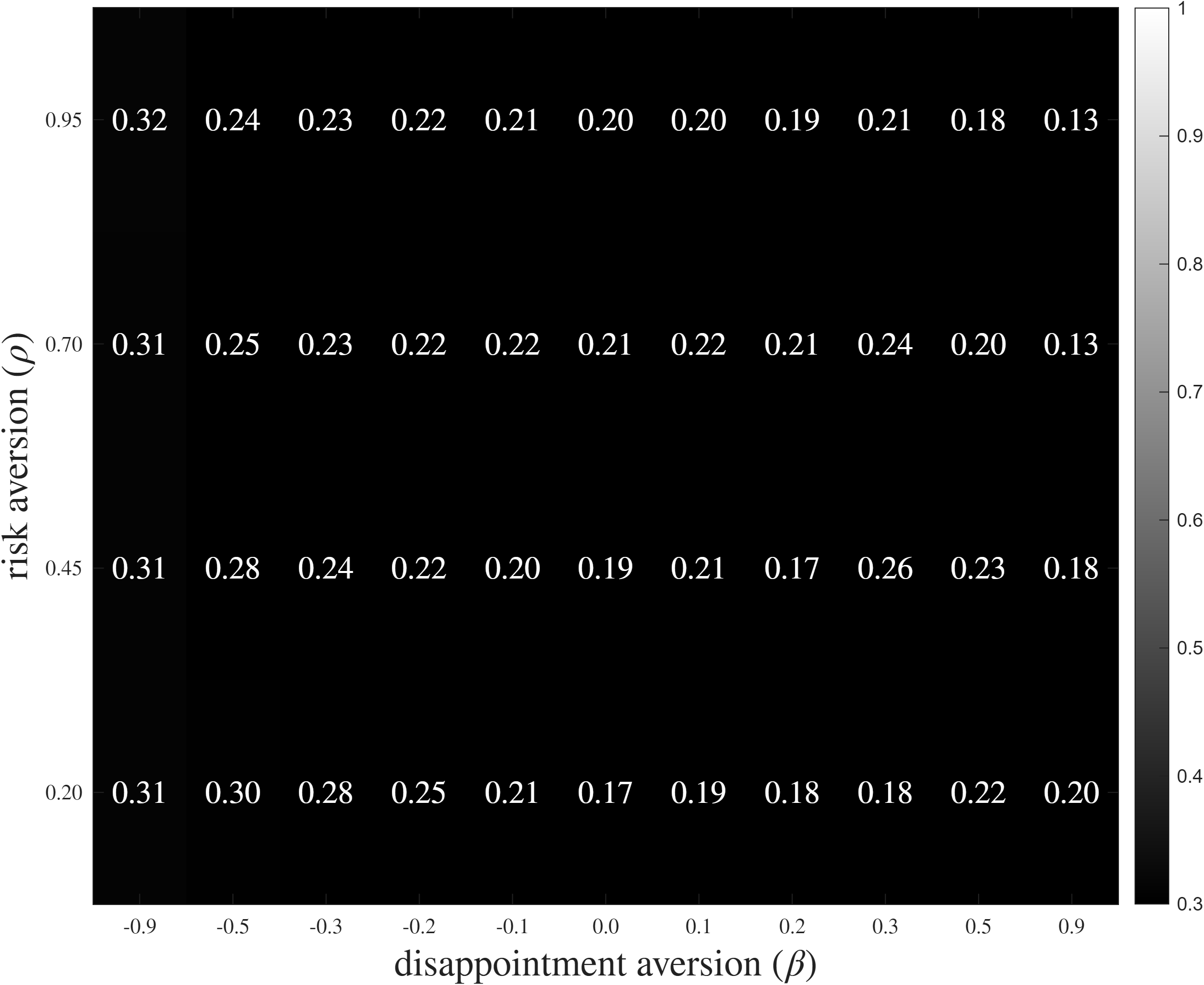}
}
\hfill
\subfigure[welfare improvement ratio]{
\includegraphics[width=0.48\linewidth]{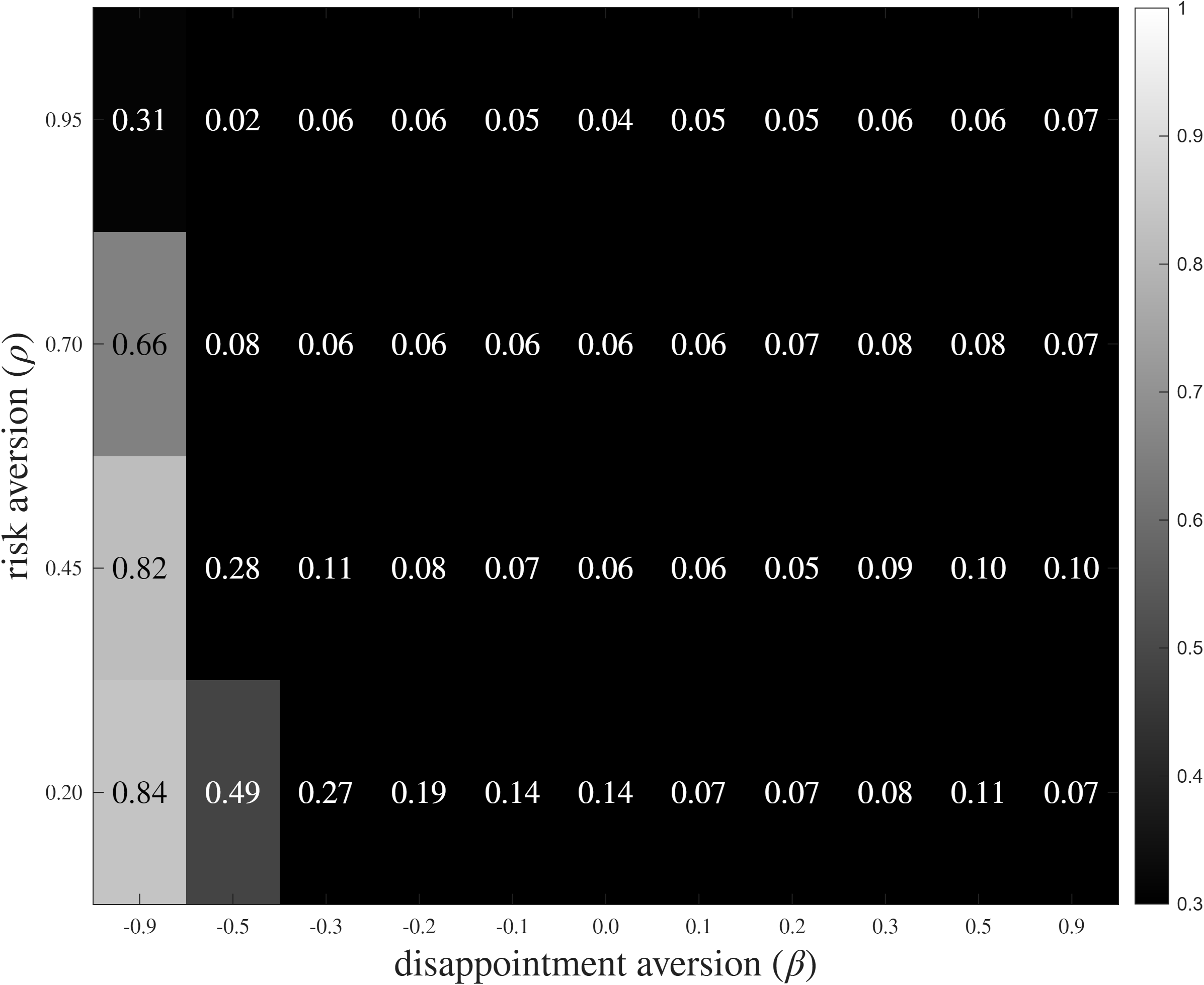}
}
\caption{The heatmaps display the improvement ratio (measure at $h=125$ divided by that at $h=5$) for (i) the vector distance measure (Panel a), (ii) the risk neutrality measure (Panel b), and (iii) the welfare loss measure (Panel c). Darker cells indicate effective learning (reduced error), while lighter cells indicate little or no improvement. Values above 1 indicate that recommendation quality worsened as the history size increased.}
\label{fig:kNN_additional_heatmaps1}
\end{figure}

\autoref{fig:kNN_additional_heatmaps2} presents the parametric estimation error improvement ratios for kNN. Panel (a) shows that $\rho$ is well recovered across most of the parameter space, as indicated by the predominance of dark cells. However, several clusters of lighter cells appear in some regions, suggesting that the improvement in estimation accuracy is limited there. Panel (b) displays a broadly similar pattern for $\beta$, with most cells dark but with a few localized areas of lighter shading. These patterns indicate that increasing the history size generally improves parameter recovery, although the magnitude of improvement varies across the $(\beta,\rho)$ parameter space.

\begin{figure}[ht]
\centering
\subfigure[$\rho$ estimation error improvement ratio]{
\includegraphics[width=0.48\linewidth]{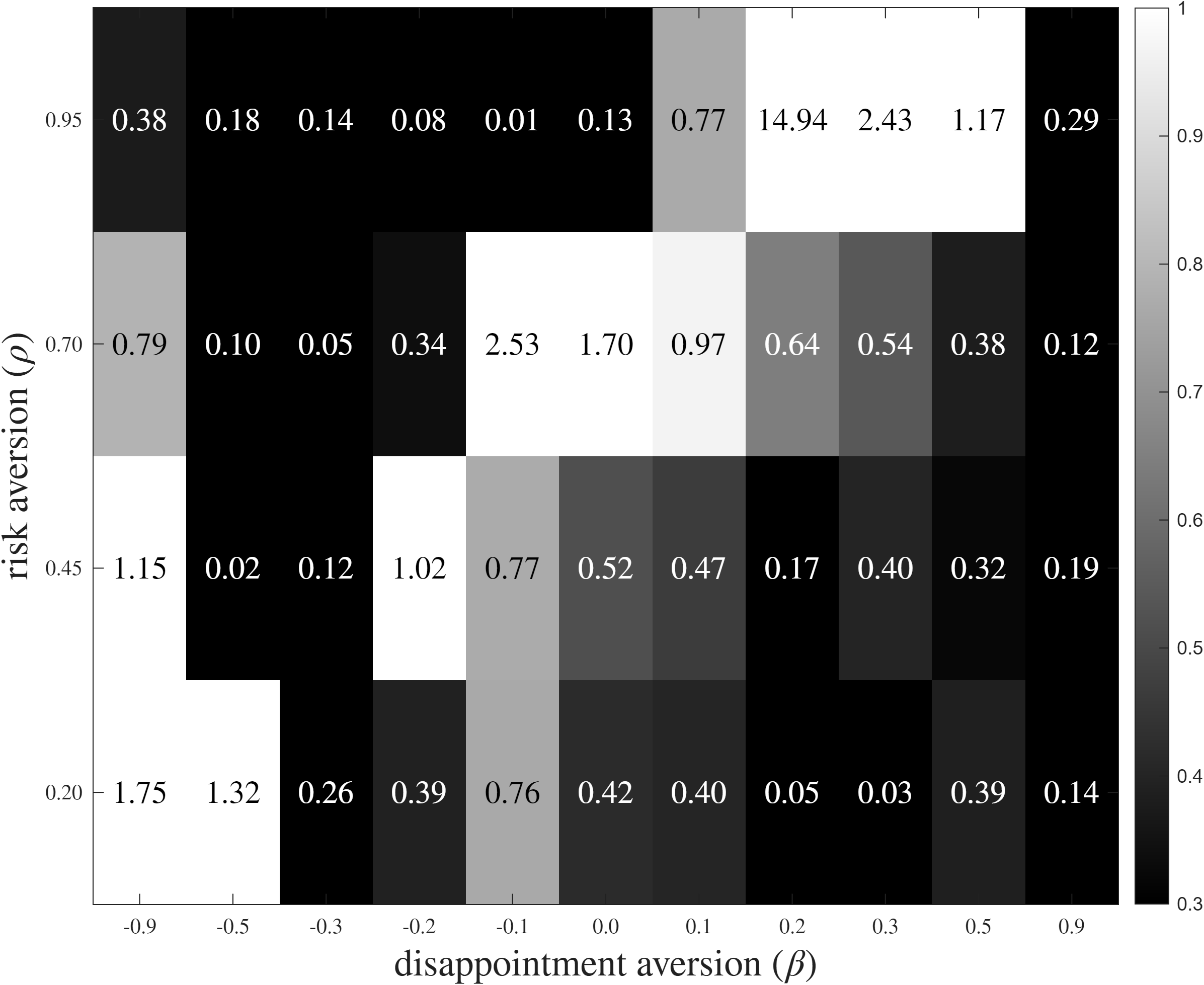}
}
\hfill
\subfigure[$\beta$ estimation error improvement ratio]{
\includegraphics[width=0.48\linewidth]{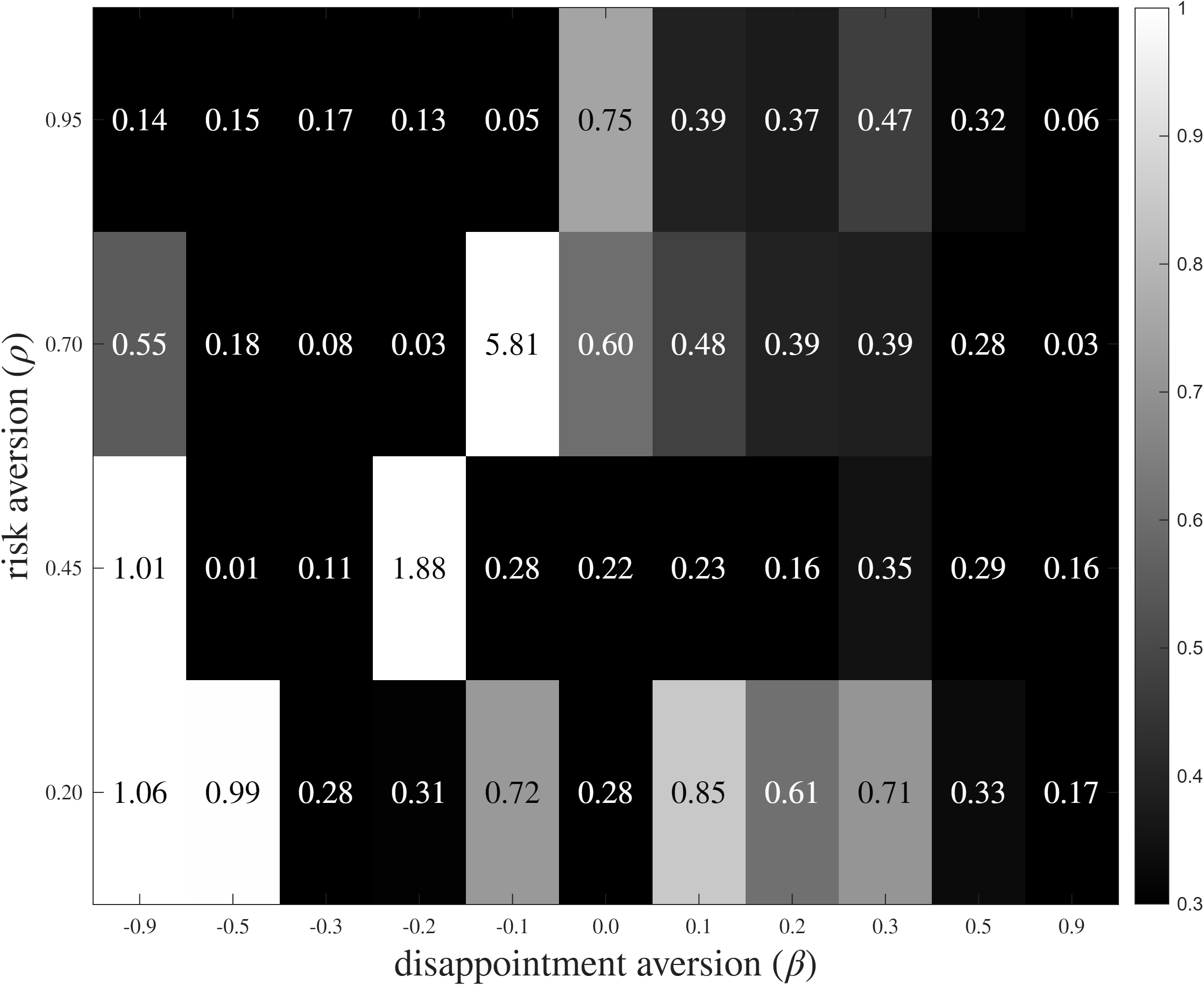}
}
\caption{The heatmaps display the estimation error improvement ratio (measure at $h = 125$ divided by that at $h = 5$) for risk aversion (Panel a) and disappointment aversion (Panel b). Darker cells indicate effective learning (reduced error), while lighter cells indicate little or no improvement. Values above 1 indicate that the estimation error increased as the history size grew from $h = 5$ to $h = 125$.}
\label{fig:kNN_additional_heatmaps2}
\end{figure}


\newpage

\noindent \textbf{Detailed variation.} For all the measures, the changes for history is gathered below.

\begin{figure}[!htbp]
\centering
\includegraphics[width=1.2\linewidth,angle=90]{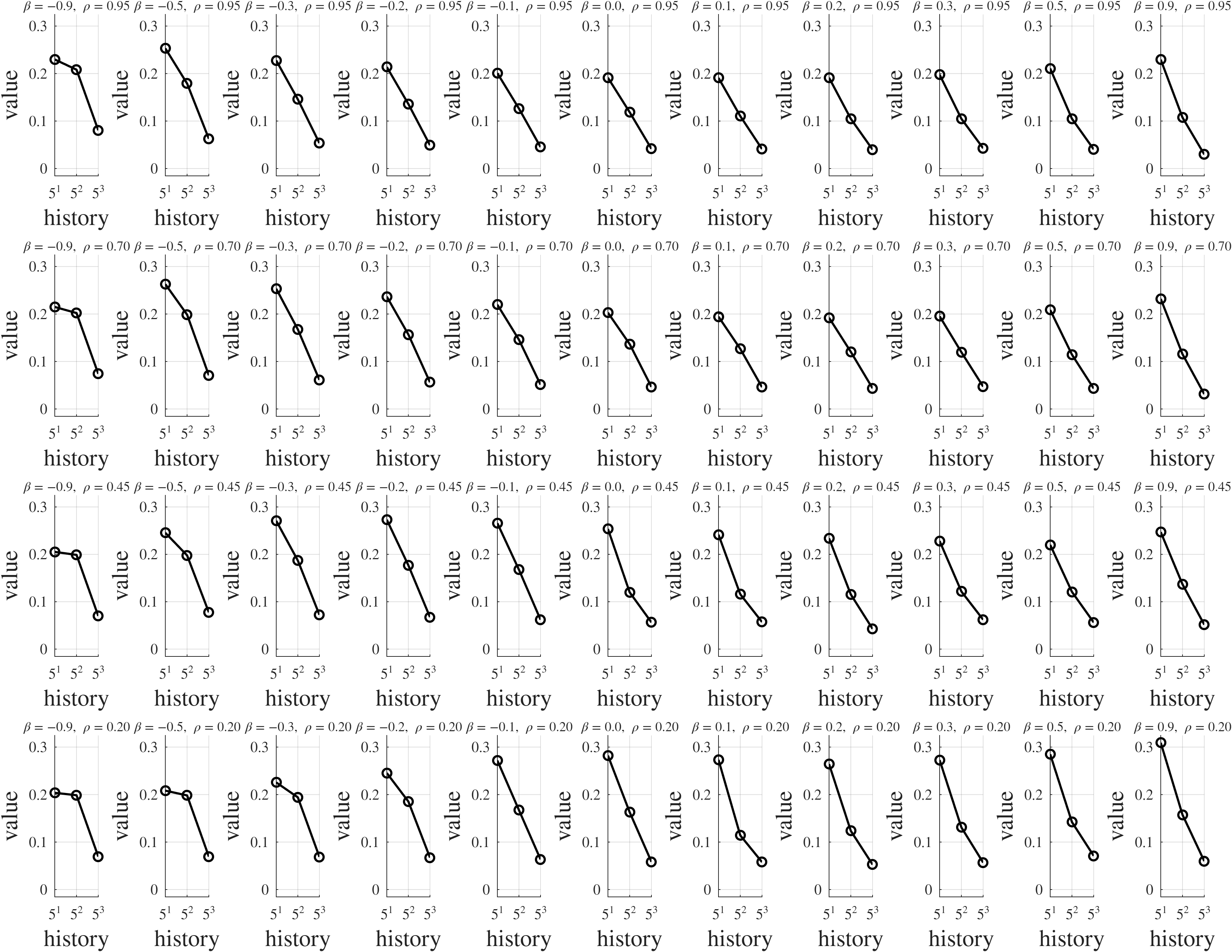}
\caption{Average normalized vector distance (AVD) by history size for each parameter combination ($\beta$,$\rho$) under kNN. Each panel plots the bootstrap mean of AVD across history sizes $h \in \{5^1, 5^2, 5^3 \}$, with bars representing 95\% bootstrap confidence intervals. Rows correspond to increasing values of $\rho$ (from bottom to top), and columns correspond to increasing values of $\beta$ (from left to right).}
\label{fig:kNN_additional_graphs1}
\end{figure}

\begin{figure}[!htbp]
\centering
\includegraphics[width=1.2\linewidth,angle=90]{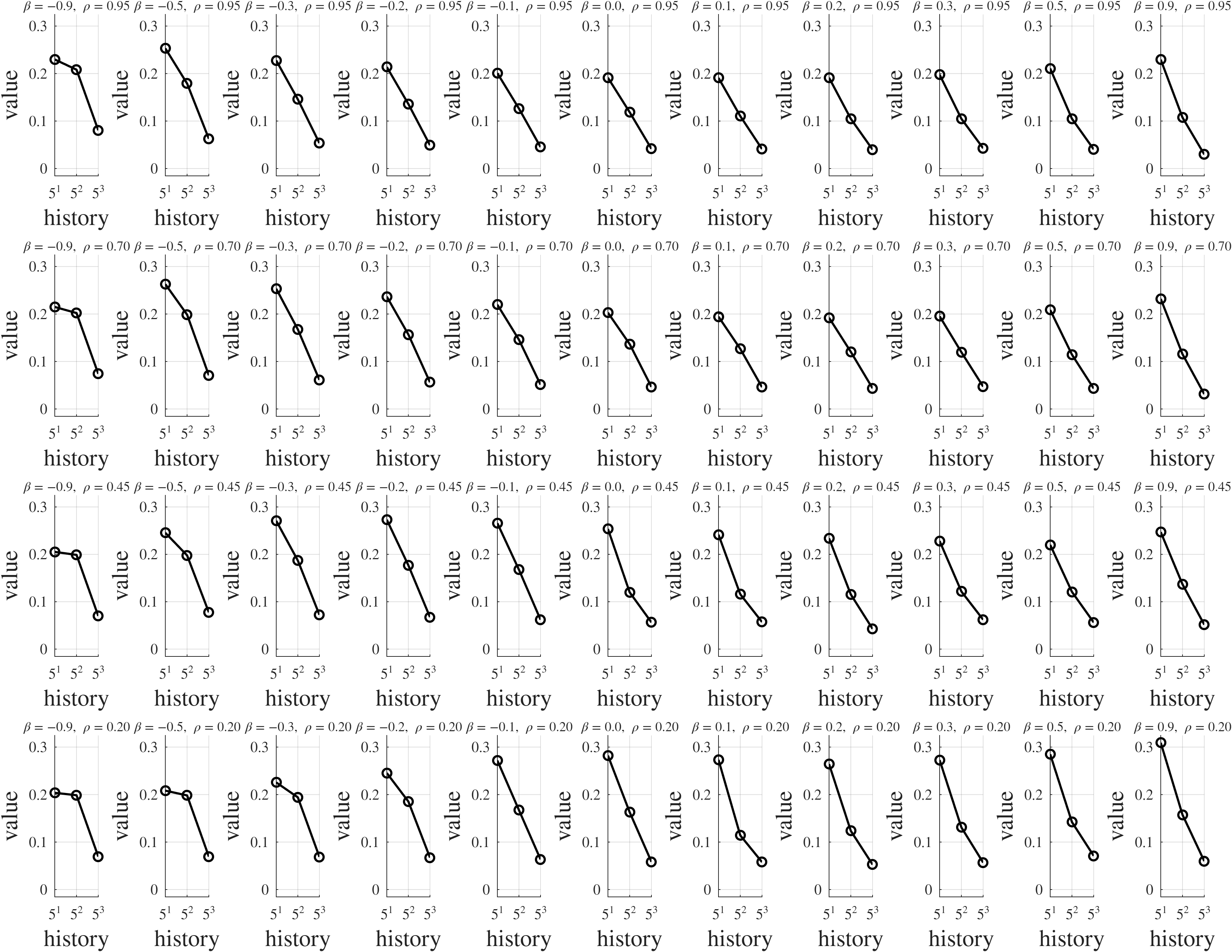}
\caption{Normalized learning error for risk aversion (NLE($\rho$)) by history size for each parameter combination ($\beta$,$\rho$) under kNN. Each panel plots the bootstrap mean of NLE($\rho$) across history sizes $h \in \{5^1, 5^2, 5^3 \}$, with bars representing 95\% bootstrap confidence intervals. Rows correspond to increasing values of $\rho$ (from bottom to top), and columns correspond to increasing values of $\beta$ (from left to right).}
\label{fig:kNN_additional_graphs2}
\end{figure}

\begin{figure}[!htbp]
\centering
\includegraphics[width=1.2\linewidth,angle=90]{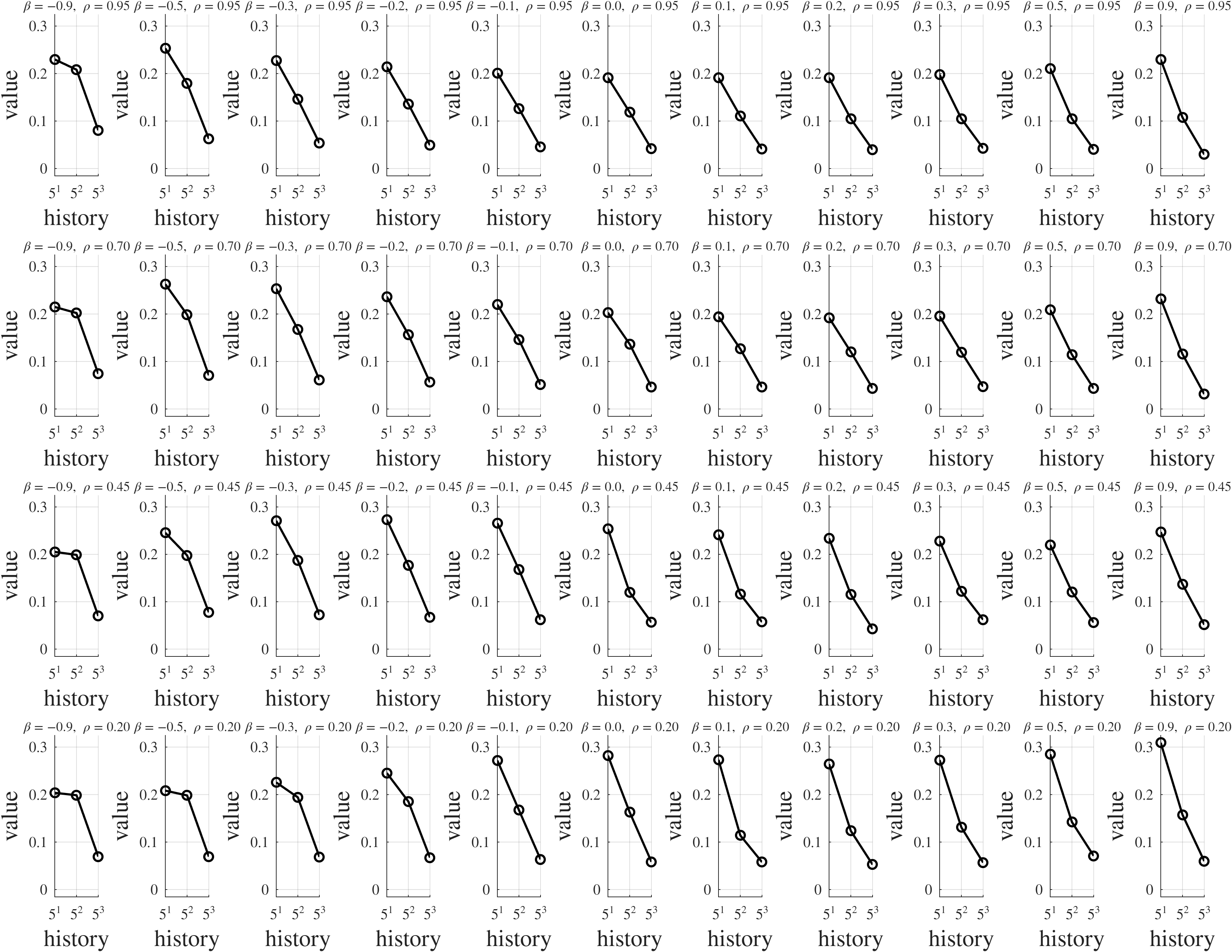}
\caption{Normalized learning error for disappointment aversion (NLE($\beta$)) by history size for each parameter combination ($\beta$,$\rho$) under kNN. Each panel plots the bootstrap mean of NLE($\beta$) across history sizes $h \in \{5^1, 5^2, 5^3 \}$, with bars representing 95\% bootstrap confidence intervals. Rows correspond to increasing values of $\rho$ (from bottom to top), and columns correspond to increasing values of $\beta$ (from left to right).}
\label{fig:kNN_additional_graphs3}
\end{figure}

\newpage

\subsection{Additional Figures for RF}
\label{label:subsection:RF}

\noindent \textbf{Additional heatmaps.} \autoref{fig:RF_additional_heatmaps1} presents the non-parametric improvement ratios for RF. The results are qualitatively similar to kNN, with uniformly dark heatmaps across the parameter space for all three measures. The improvement ratios are particularly small for the welfare loss measure (Panel c), where most cells fall below 0.1, indicating that near-complete convergence to the optimal allocation at $h=125$ is achieved.

\begin{figure}[ht]
\centering
\subfigure[vector distance improvement ratio]{
\includegraphics[width=0.48\linewidth]{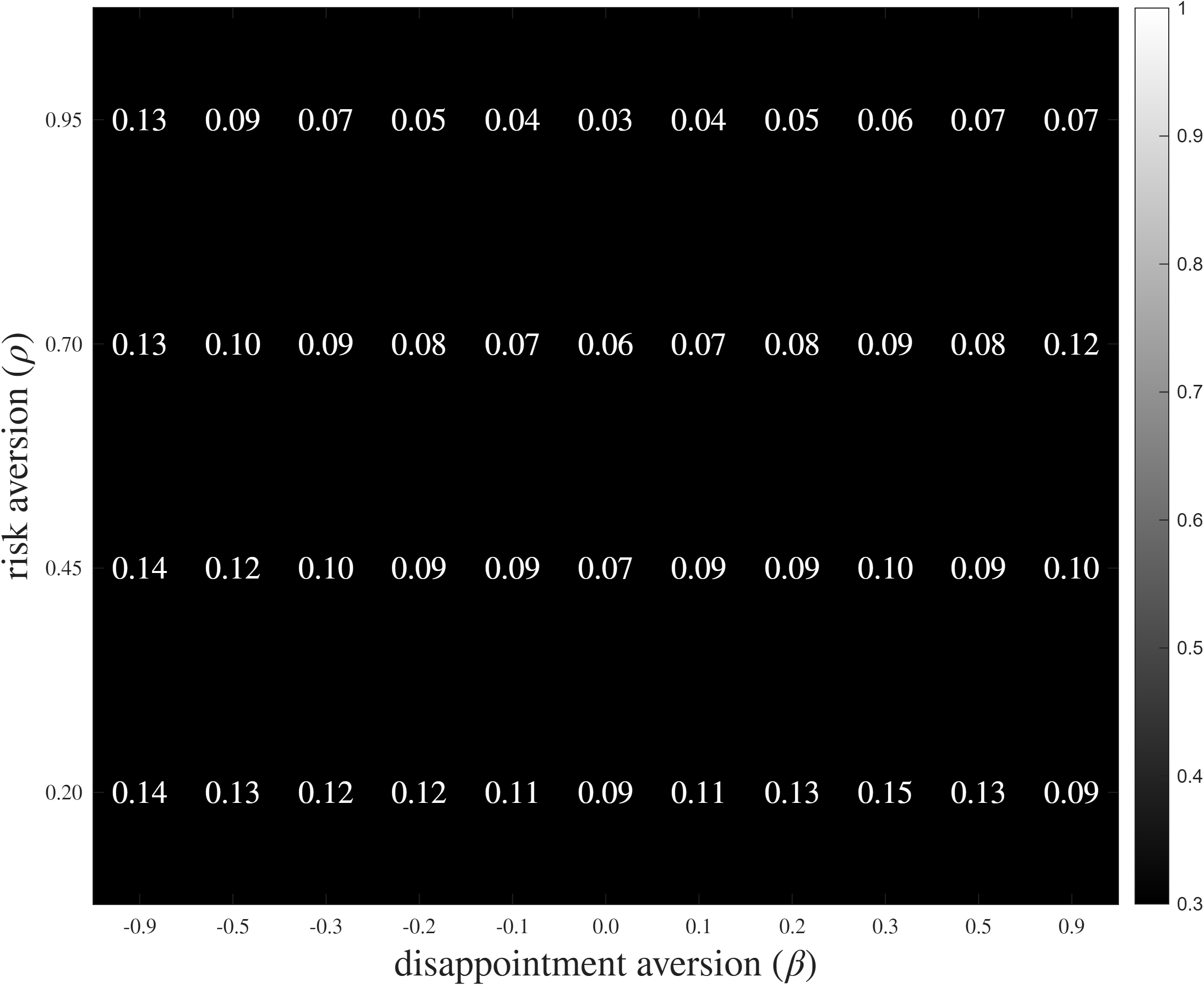}
}
\hfill
\subfigure[RN improvement ratio]{
\includegraphics[width=0.48\linewidth]{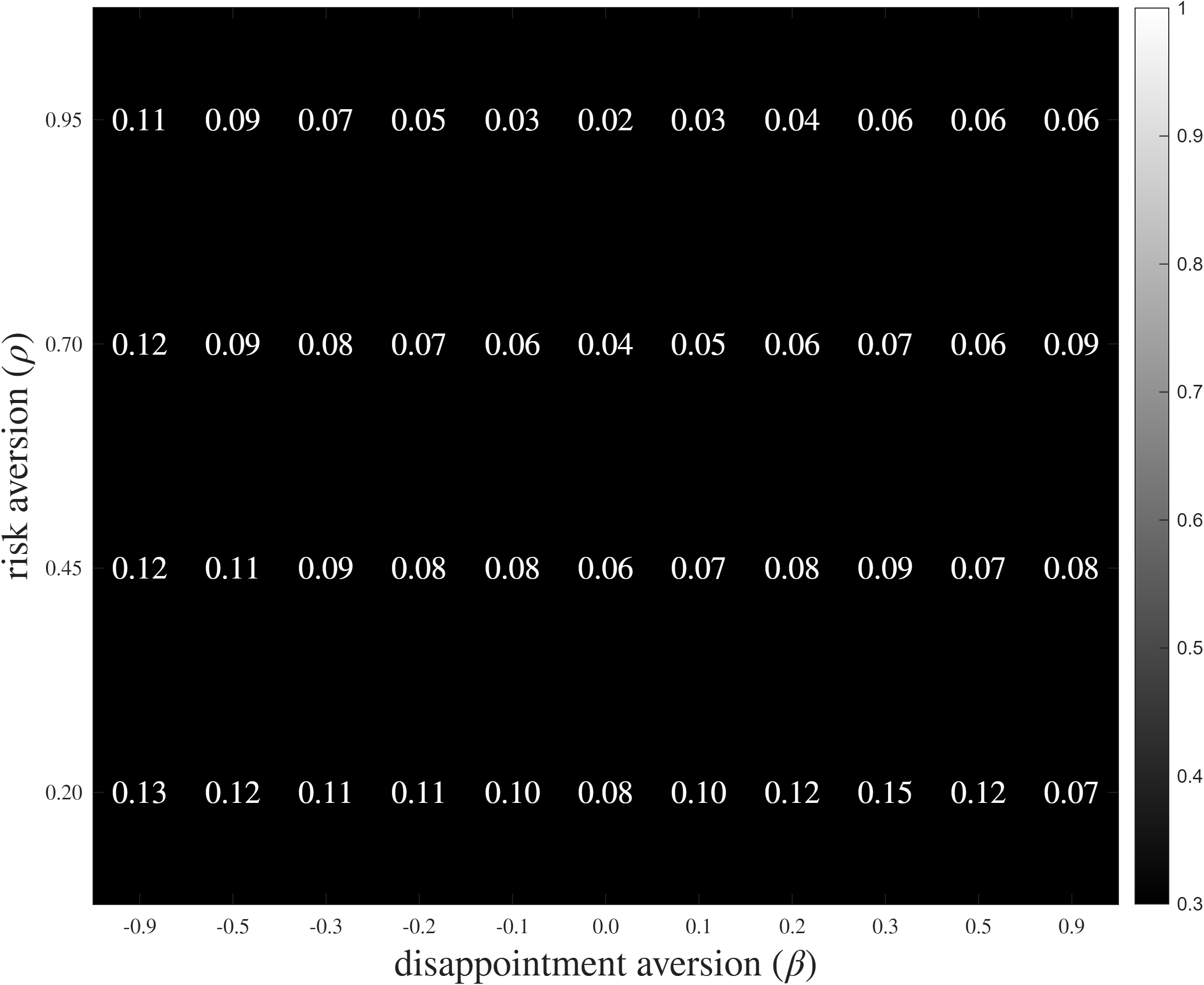}
}
\hfill
\subfigure[welfare improvement ratio]{
\includegraphics[width=0.48\linewidth]{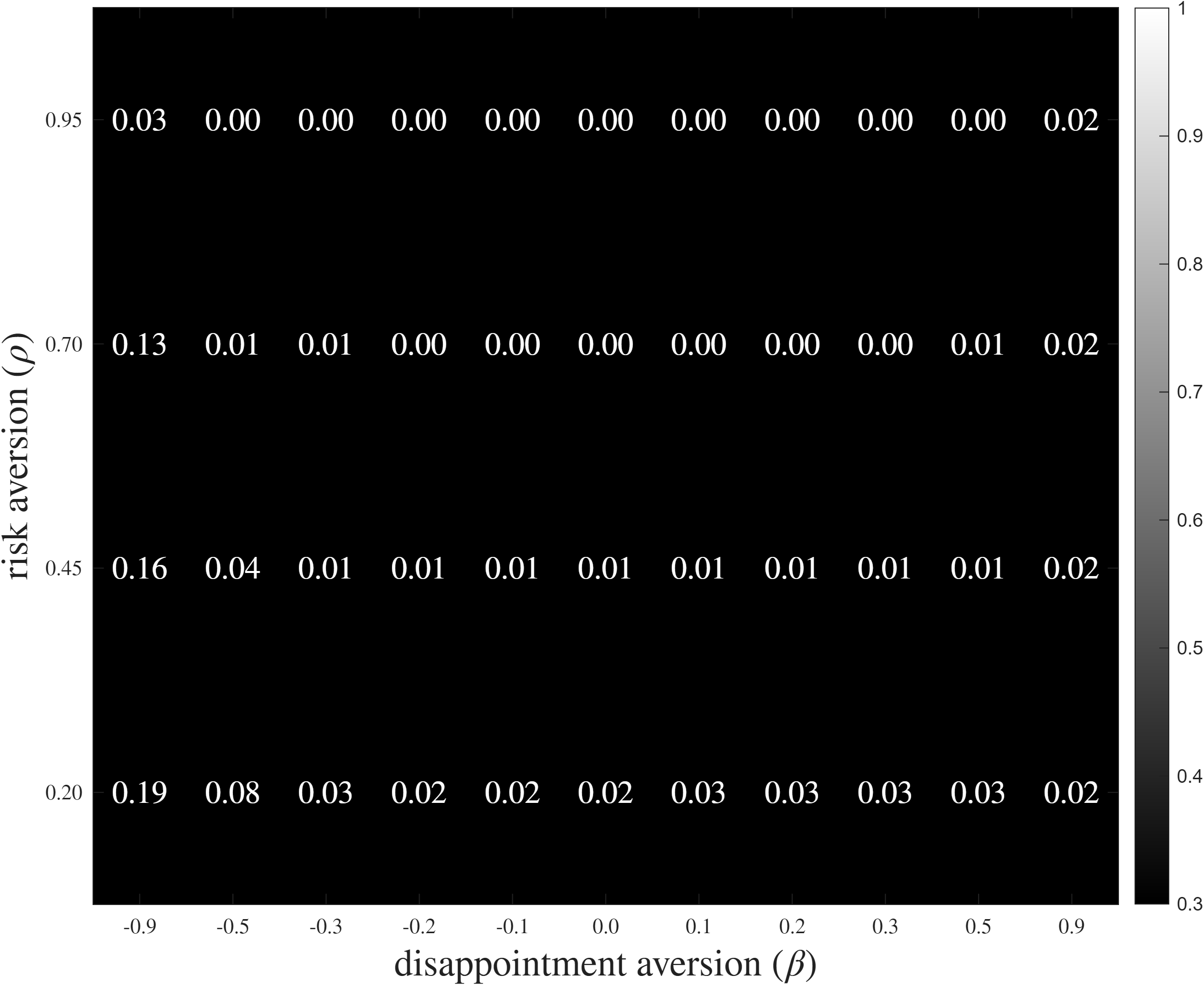}
}
\caption{The heatmaps display the improvement ratio (measure at $h=125$ divided by that at $h=5$) for (i) the vector distance measure (Panel a), (ii) the risk neutrality measure (Panel b), and (iii) the welfare loss measure (Panel c). Darker cells indicate effective learning (reduced error), while lighter cells indicate little or no improvement. Values above 1 indicate that recommendation quality worsened as the history size increased.}
\label{fig:RF_additional_heatmaps1}
\end{figure}

\autoref{fig:RF_additional_heatmaps2} presents the parametric estimation error improvement ratios for RF. Panel (a) shows that the estimation error for $\rho$ decreases across most of the parameter space, with the majority of cells well below 1. However, at high risk aversion combined with strong elation seeking (i.e., $\rho \in \{0.70, 0.95\}$ and $\beta = -0.9$), the ratios are substantially above 1, indicating that additional history does not improve the recovery of $\rho$. Panel (b) reveals that $\beta$ is well recovered across the elation-seeking region (i.e., $\beta < 0$), but a group of light cells appears at $\beta \in \{0, 0.1, 0.2, 0.3\}$. This suggests that RF struggles to recover $\beta$ when the true parameter is close to the expected utility benchmark.

\begin{figure}[ht]
\centering
\subfigure[$\rho$ estimation error improvement ratio]{
\includegraphics[width=0.48\linewidth]{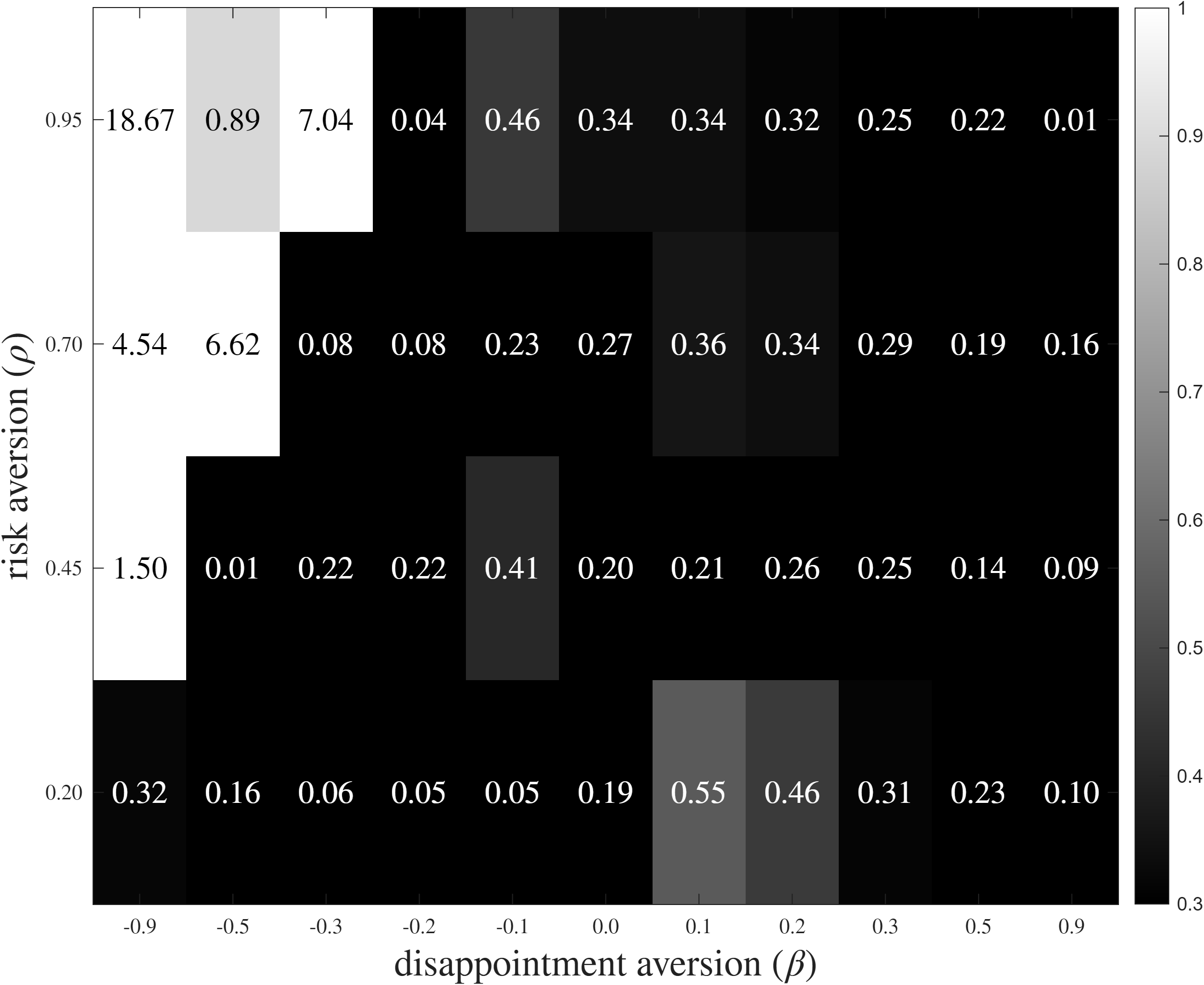}
}
\hfill
\subfigure[$\beta$ estimation error improvement ratio]{
\includegraphics[width=0.48\linewidth]{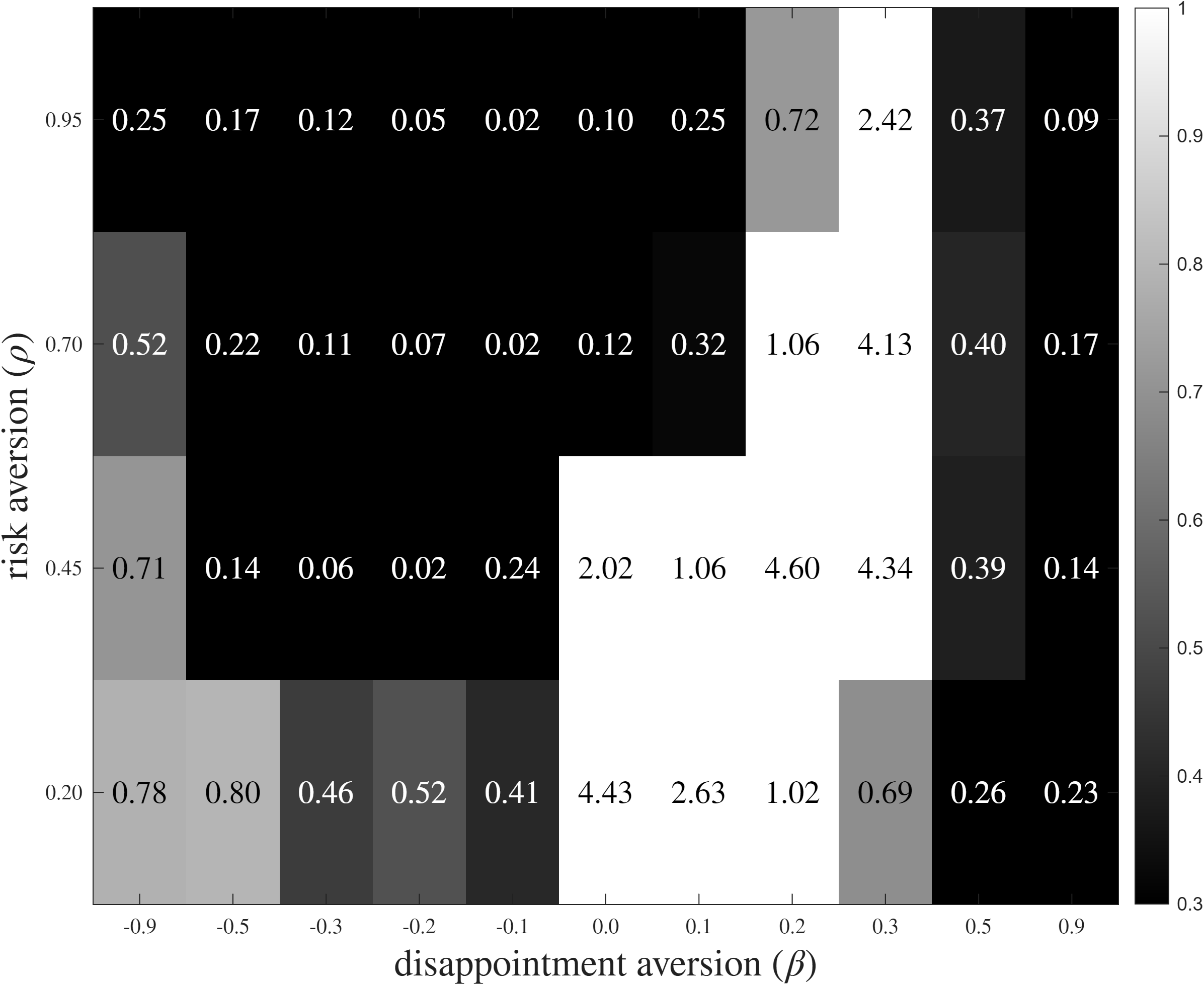}
}
\caption{The heatmaps display the estimation error improvement ratio (measure at $h = 125$ divided by that at $h = 5$) for risk aversion (Panel a) and disappointment aversion (Panel b). Darker cells indicate effective learning (reduced error), while lighter cells indicate little or no improvement. Values above 1 indicate that the estimation error increased as the history size grew from $h = 5$ to $h = 125$.}
\label{fig:RF_additional_heatmaps2}
\end{figure}

\newpage

\noindent \textbf{Detailed variation.} For all the measures, the changes for history is gathered below.

\begin{figure}[!htbp]
\centering
\includegraphics[width=1.2\linewidth,angle=90]{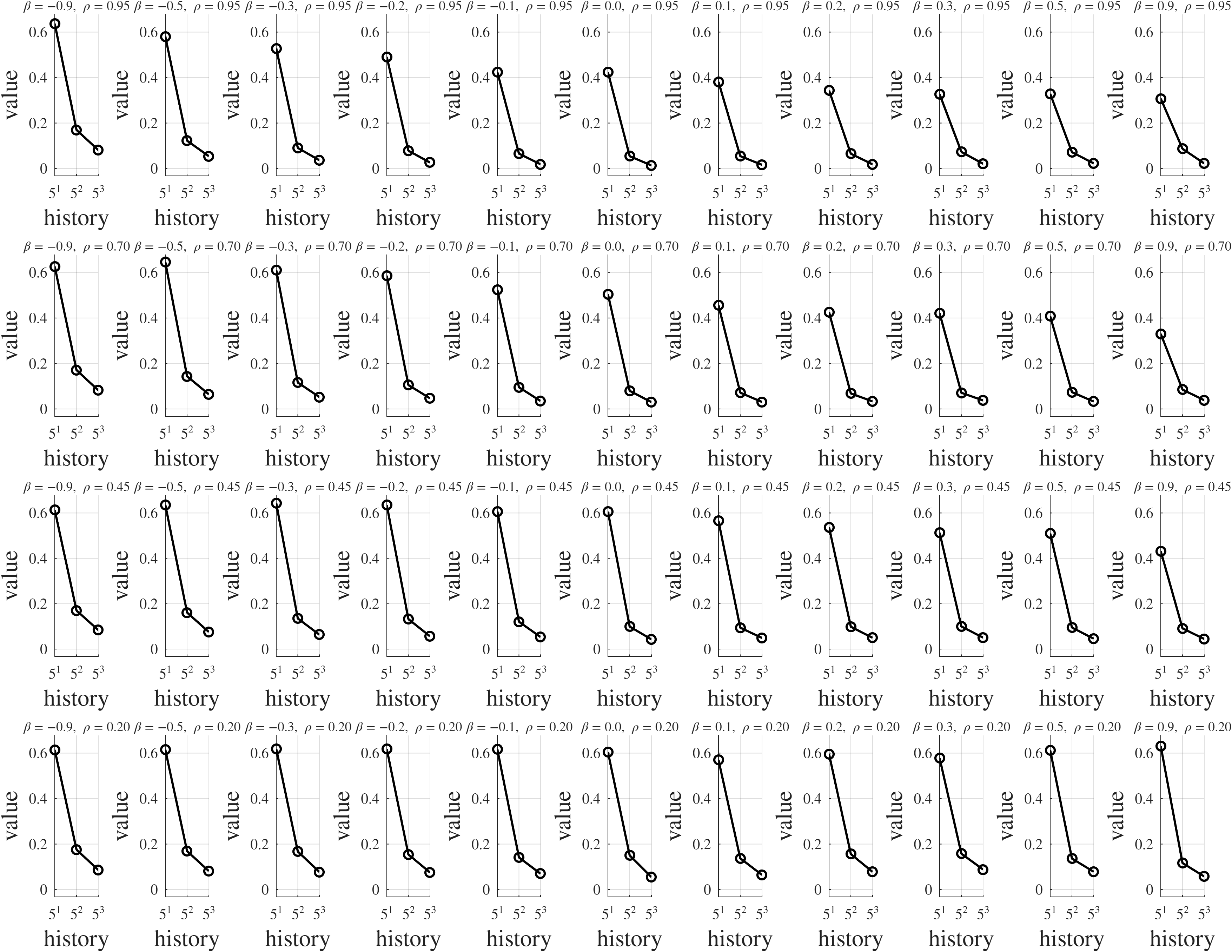}
\caption{Average normalized vector distance (AVD) by history size for each parameter combination ($\beta$,$\rho$) under RF. Each panel plots the bootstrap mean of AVD across history sizes $h \in \{5^1, 5^2, 5^3 \}$, with bars representing 95\% bootstrap confidence intervals. Rows correspond to increasing values of $\rho$ (from bottom to top), and columns correspond to increasing values of $\beta$ (from left to right).}
\label{fig:RF_additional_graphs1}
\end{figure}

\begin{figure}[!htbp]
\centering
\includegraphics[width=1.2\linewidth,angle=90]{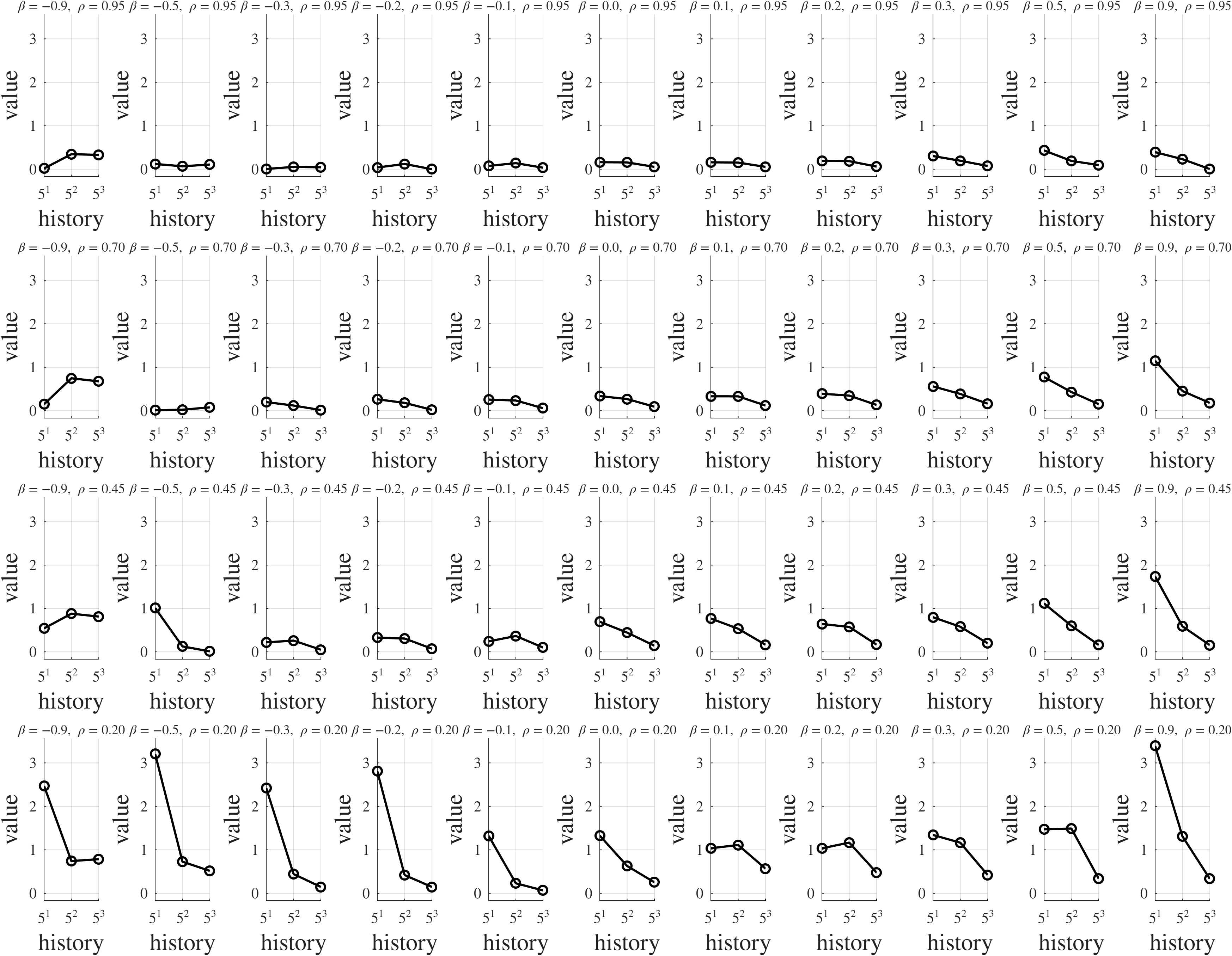}
\caption{Normalized learning error for risk aversion (NLE($\rho$)) by history size for each parameter combination ($\beta$,$\rho$) under RF. Each panel plots the bootstrap mean of NLE($\rho$) across history sizes $h \in \{5^1, 5^2, 5^3 \}$, with bars representing 95\% bootstrap confidence intervals. Rows correspond to increasing values of $\rho$ (from bottom to top), and columns correspond to increasing values of $\beta$ (from left to right).}
\label{fig:RF_additional_graphs2}
\end{figure}

\begin{figure}[!htbp]
\centering
\includegraphics[width=1.2\linewidth,angle=90]{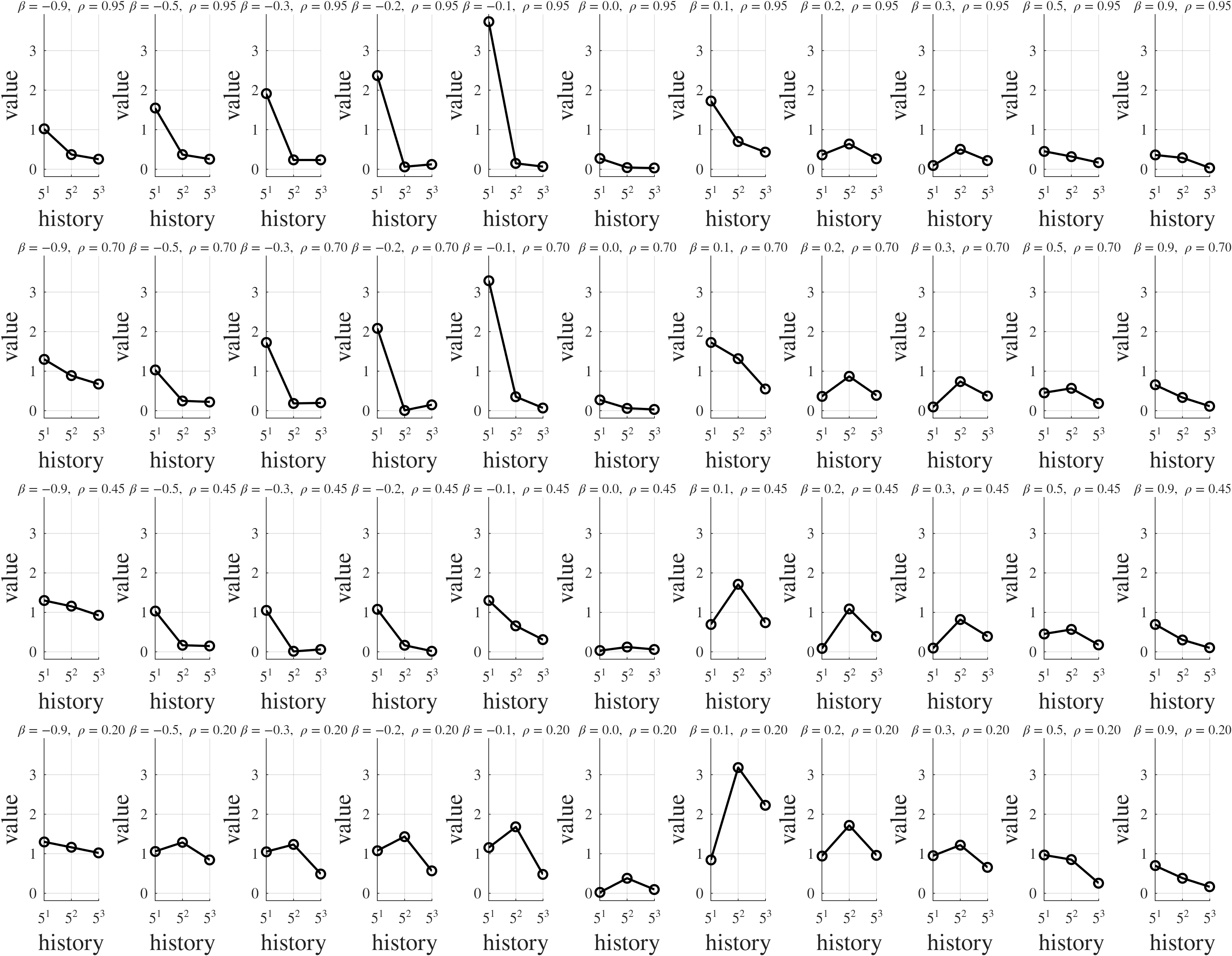}
\caption{Normalized learning error for disappointment aversion (NLE($\beta$)) by history size for each parameter combination ($\beta$,$\rho$) under RF. Each panel plots the bootstrap mean of NLE($\beta$) across history sizes $h \in \{5^1, 5^2, 5^3 \}$, with bars representing 95\% bootstrap confidence intervals. Rows correspond to increasing values of $\rho$ (from bottom to top), and columns correspond to increasing values of $\beta$ (from left to right).}
\label{fig:RF_additional_graphs3}
\end{figure}

\newpage
\subsection{Additional Figures for SVR}
\label{label:subsection:SVR}

\noindent \textbf{Additional heatmaps.} \autoref{fig:SVR_additional_heatmaps1} displays the corresponding results for SVR. Likewise, SVR exhibits uniform improvement across the full parameter grid for all measures. Notably, SVR achieves the smallest improvement ratios among the three ML models, with many cells falling below 0.05 for the welfare loss measure (Panel c), suggesting particularly strong convergence to the optimal allocations.

\begin{figure}[ht]
\centering
\subfigure[vector distance improvement ratio]{
\includegraphics[width=0.48\linewidth]{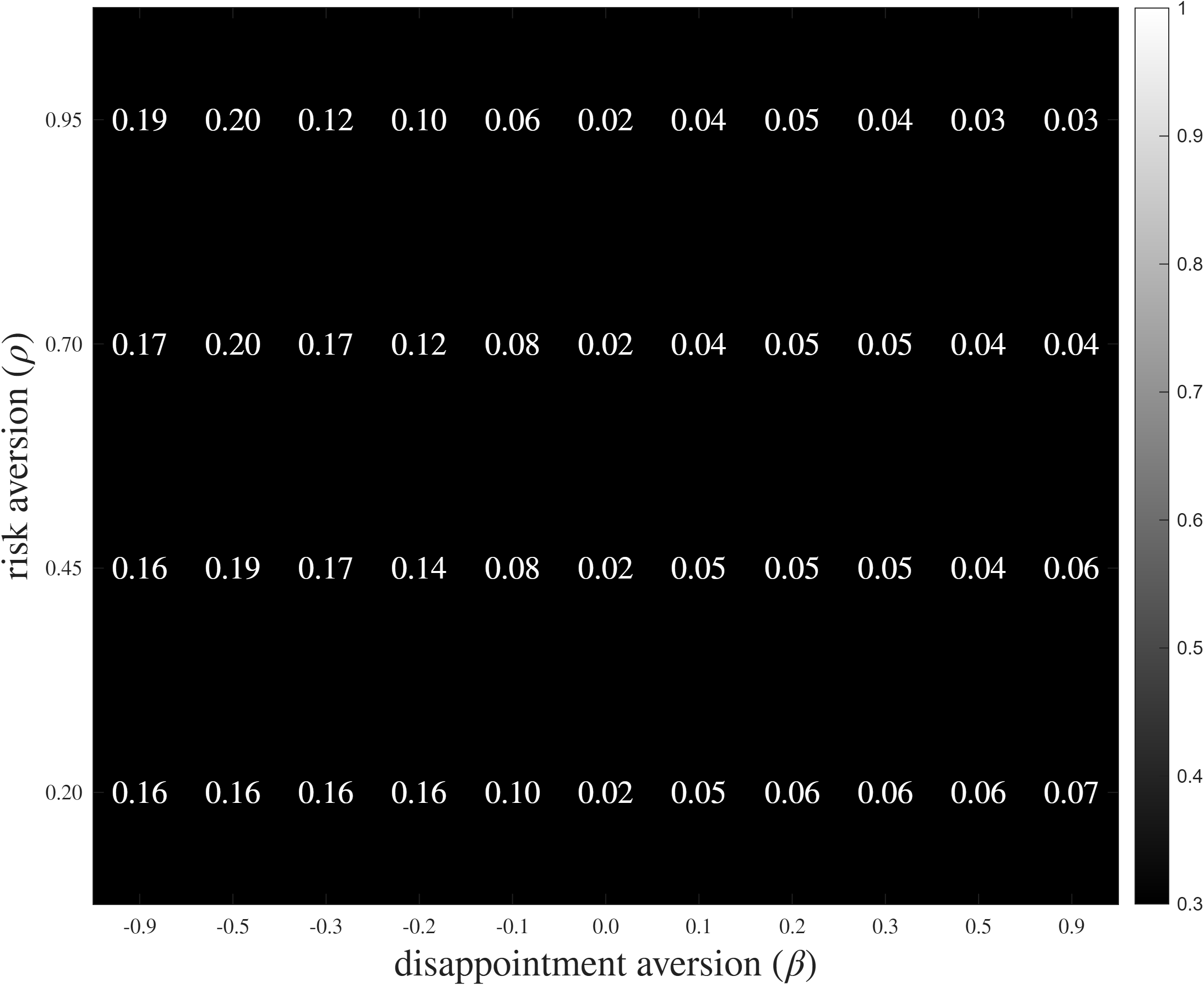}
}
\hfill
\subfigure[RN improvement ratio]{
\includegraphics[width=0.48\linewidth]{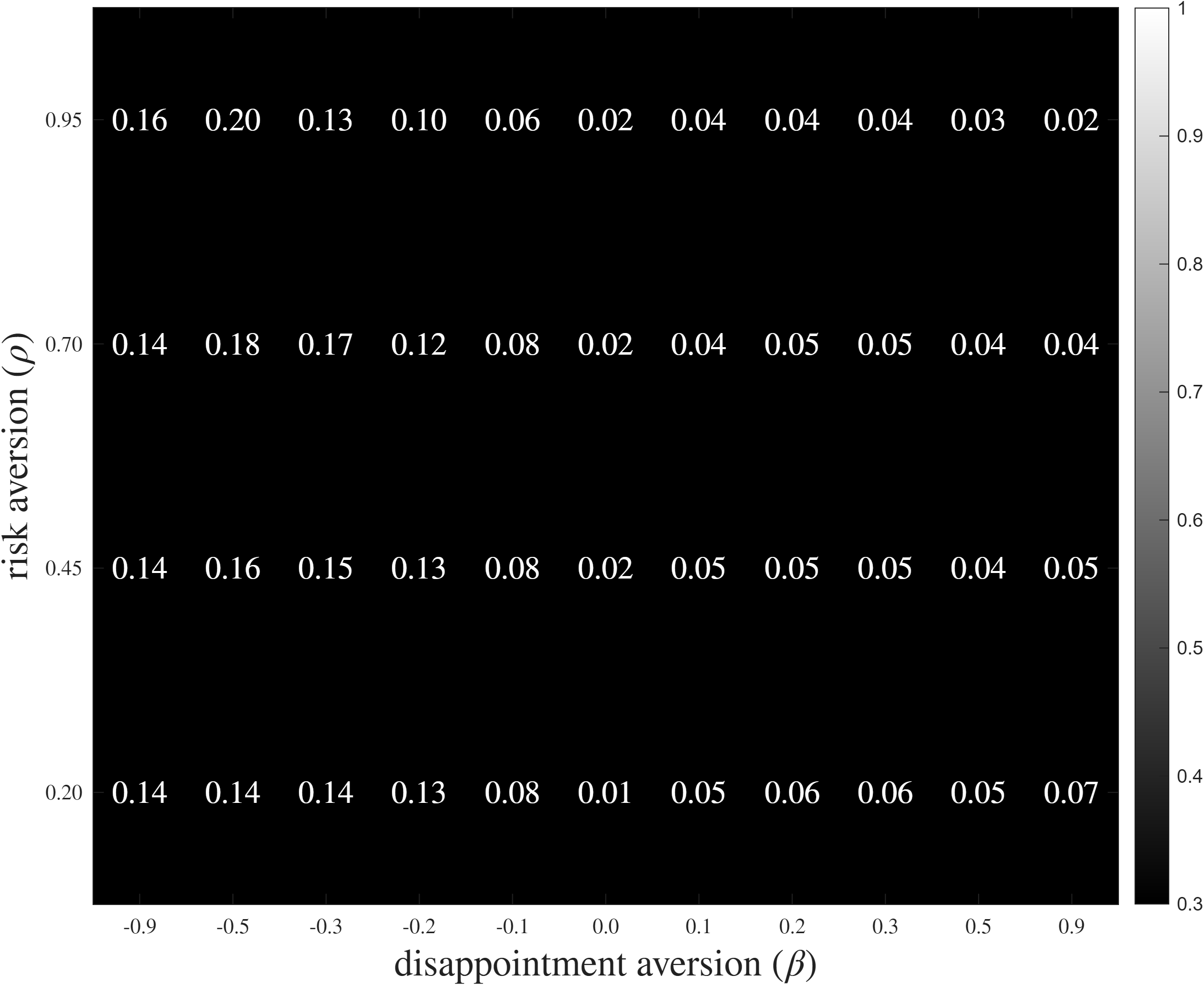}
}
\hfill
\subfigure[welfare improvement ratio]{
\includegraphics[width=0.48\linewidth]{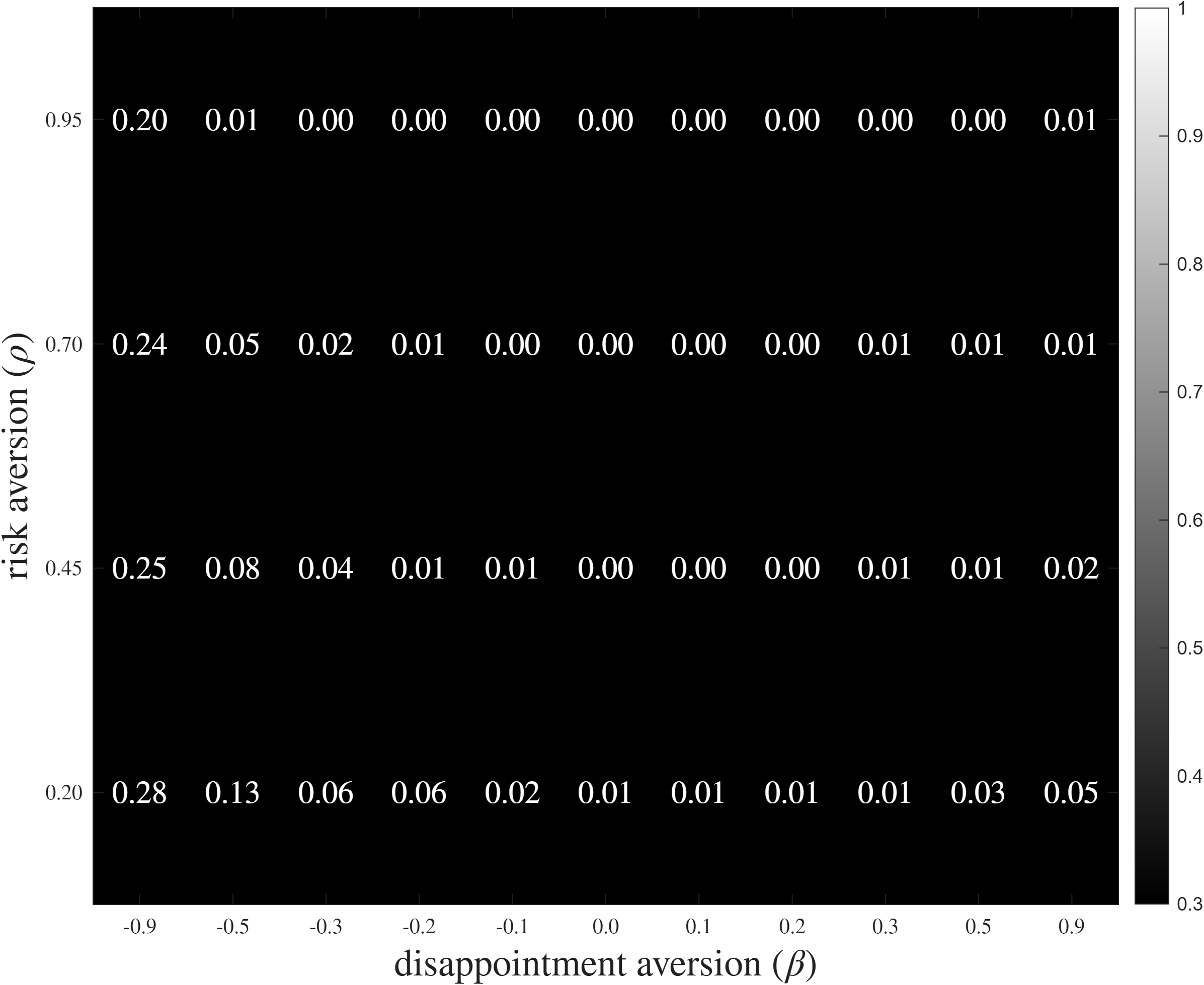}
}
\caption{The heatmaps display the improvement ratio (measure at $h=125$ divided by that at $h=5$) for (i) the vector distance measure (Panel a), (ii) the risk neutrality measure (Panel b), and (iii) the welfare loss measure (Panel c). Darker cells indicate effective learning (reduced error), while lighter cells indicate little or no improvement. Values above 1 indicate that recommendation quality worsened as the history size increased.}
\label{fig:SVR_additional_heatmaps1}
\end{figure}

\autoref{fig:SVR_additional_heatmaps2} presents the parametric estimation error improvement ratios for SVR. The heatmap for $\rho$ (Panel a) is nearly uniformly dark, indicating effective recovery of risk aversion across the entire parameter space. The heatmap for $\beta$ (Panel b) shows a similar pattern for most of the grid. However, in the elation-seeking region at low risk aversion ($\beta = - 0.9$ and $\rho = 0.20, 0.45$), the improvement ratios substantially large as $11.01$, indicating that the inherent difficulty of identifying $\beta$ from choice data in this region.


\begin{figure}[ht]
\centering
\subfigure[$\rho$ estimation error improvement ratio]{
\includegraphics[width=0.48\linewidth]{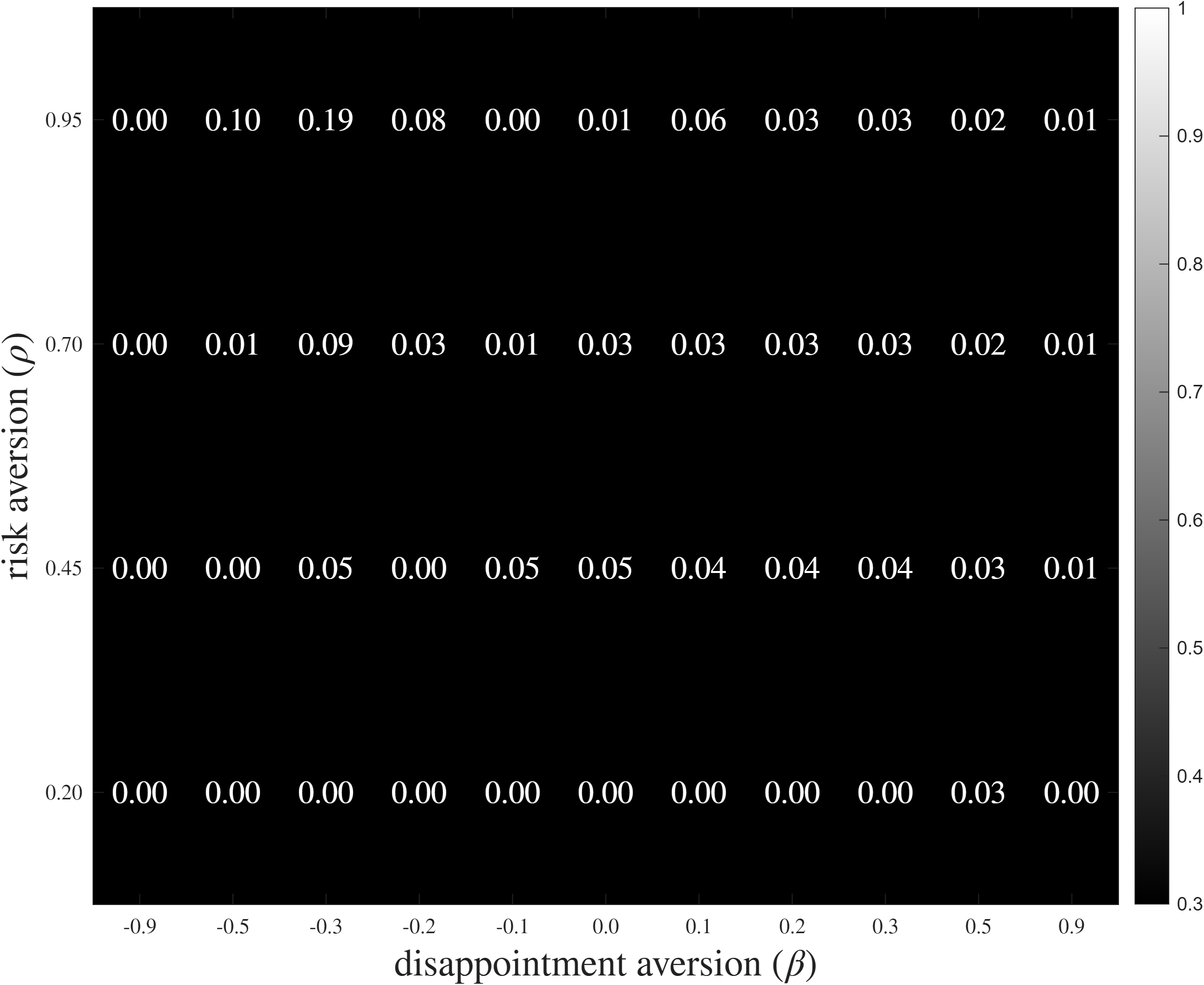}
}
\hfill
\subfigure[$\beta$ estimation error improvement ratio]{
\includegraphics[width=0.48\linewidth]{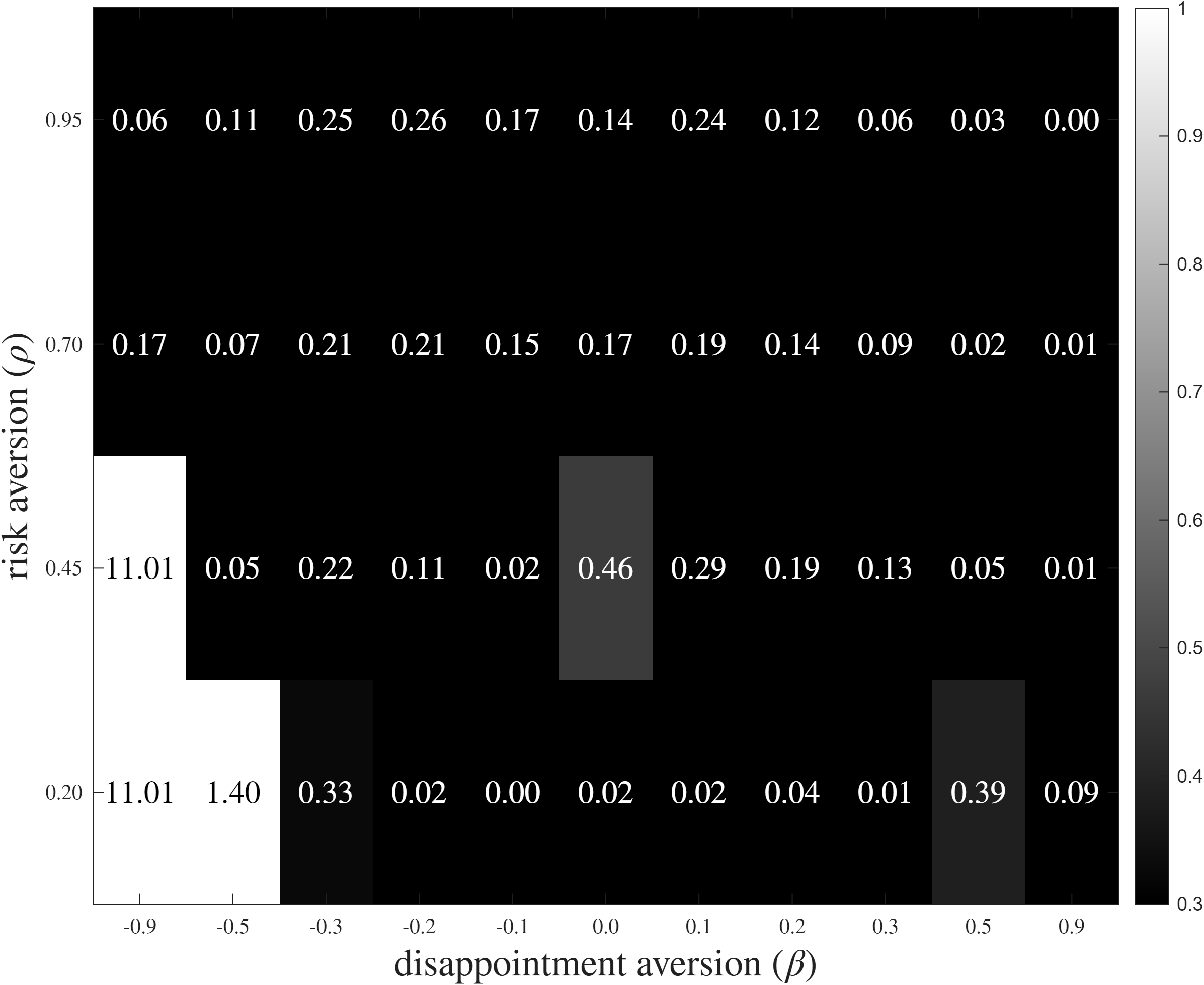}
}
\caption{The heatmaps display the estimation error improvement ratio (measure at $h = 125$ divided by that at $h = 5$) for risk aversion (Panel a) and disappointment aversion (Panel b). Darker cells indicate effective learning (reduced error), while lighter cells indicate little or no improvement. Values above 1 indicate that the estimation error increased as the history size grew from $h = 5$ to $h = 125$.}
\label{fig:SVR_additional_heatmaps2}
\end{figure}

\newpage

\noindent \textbf{Detailed variation.} For all the measures, the changes for history is gathered below.

\begin{figure}[!htbp]
\centering
\includegraphics[width=1.2\linewidth,angle=90]{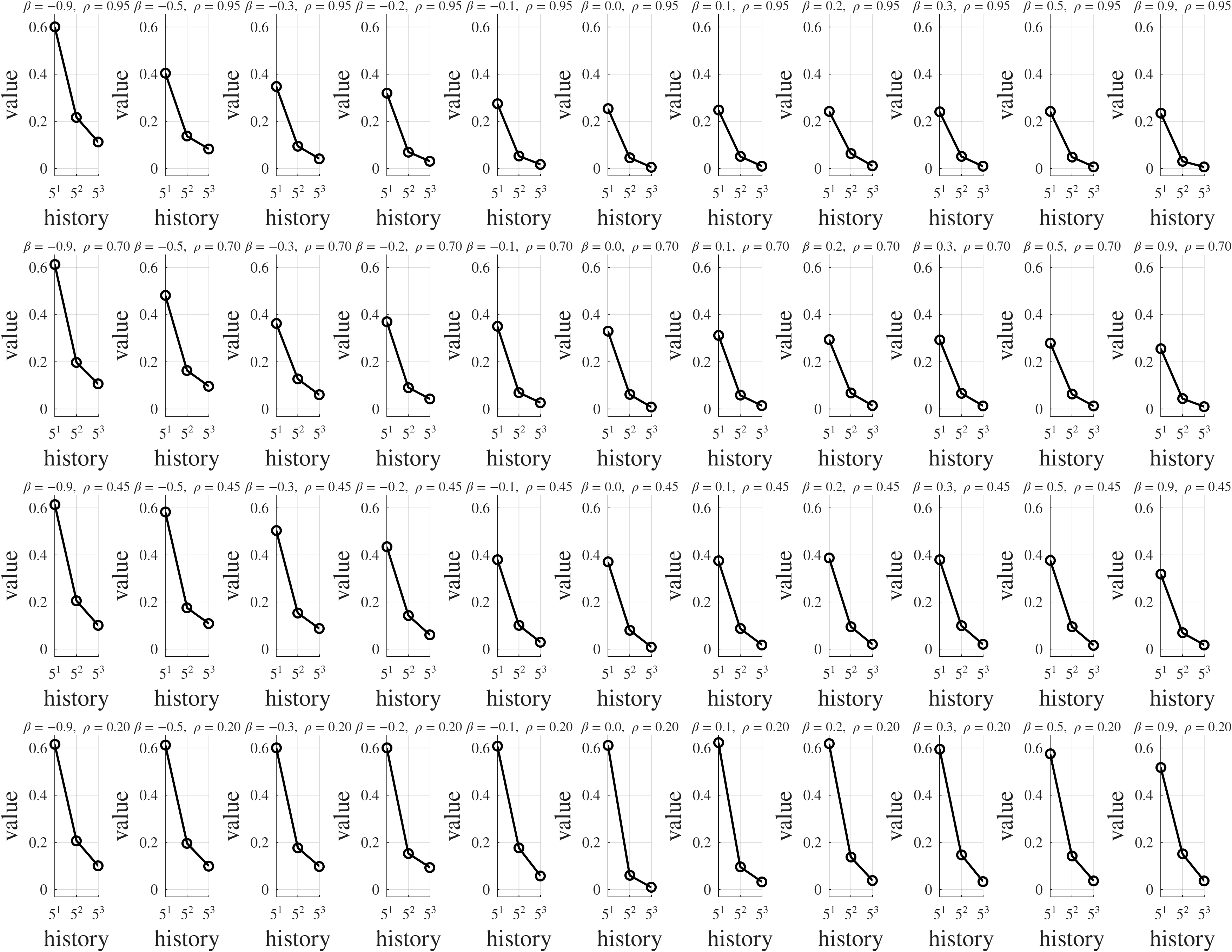}
\caption{Average normalized vector distance (AVD) by history size for each parameter combination ($\beta$,$\rho$) under SVR. Each panel plots the bootstrap mean of AVD across history sizes $h \in \{5^1, 5^2, 5^3 \}$, with bars representing 95\% bootstrap confidence intervals. Rows correspond to increasing values of $\rho$ (from bottom to top), and columns correspond to increasing values of $\beta$ (from left to right).}
\label{fig:SVR_additional_graphs1}
\end{figure}

\begin{figure}[!htbp]
\centering
\includegraphics[width=1.2\linewidth,angle=90]{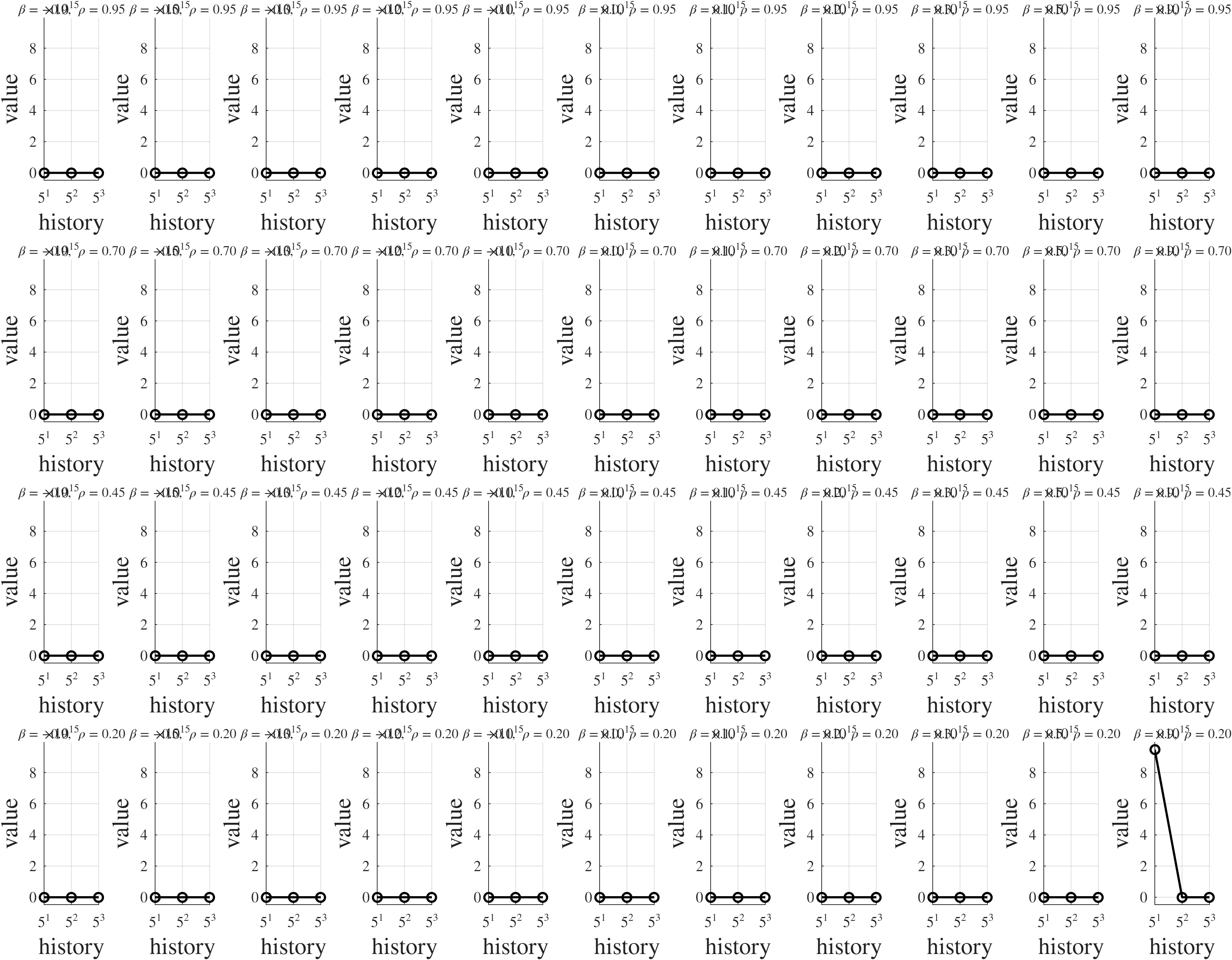}
\caption{Normalized learning error for risk aversion (NLE($\rho$)) by history size for each parameter combination ($\beta$,$\rho$) under SVR. Each panel plots the bootstrap mean of NLE($\rho$) across history sizes $h \in \{5^1, 5^2, 5^3 \}$, with bars representing 95\% bootstrap confidence intervals. Rows correspond to increasing values of $\rho$ (from bottom to top), and columns correspond to increasing values of $\beta$ (from left to right).}
\label{fig:SVR_additional_graphs2}
\end{figure}

\begin{figure}[!htbp]
\centering
\includegraphics[width=1.2\linewidth,angle=90]{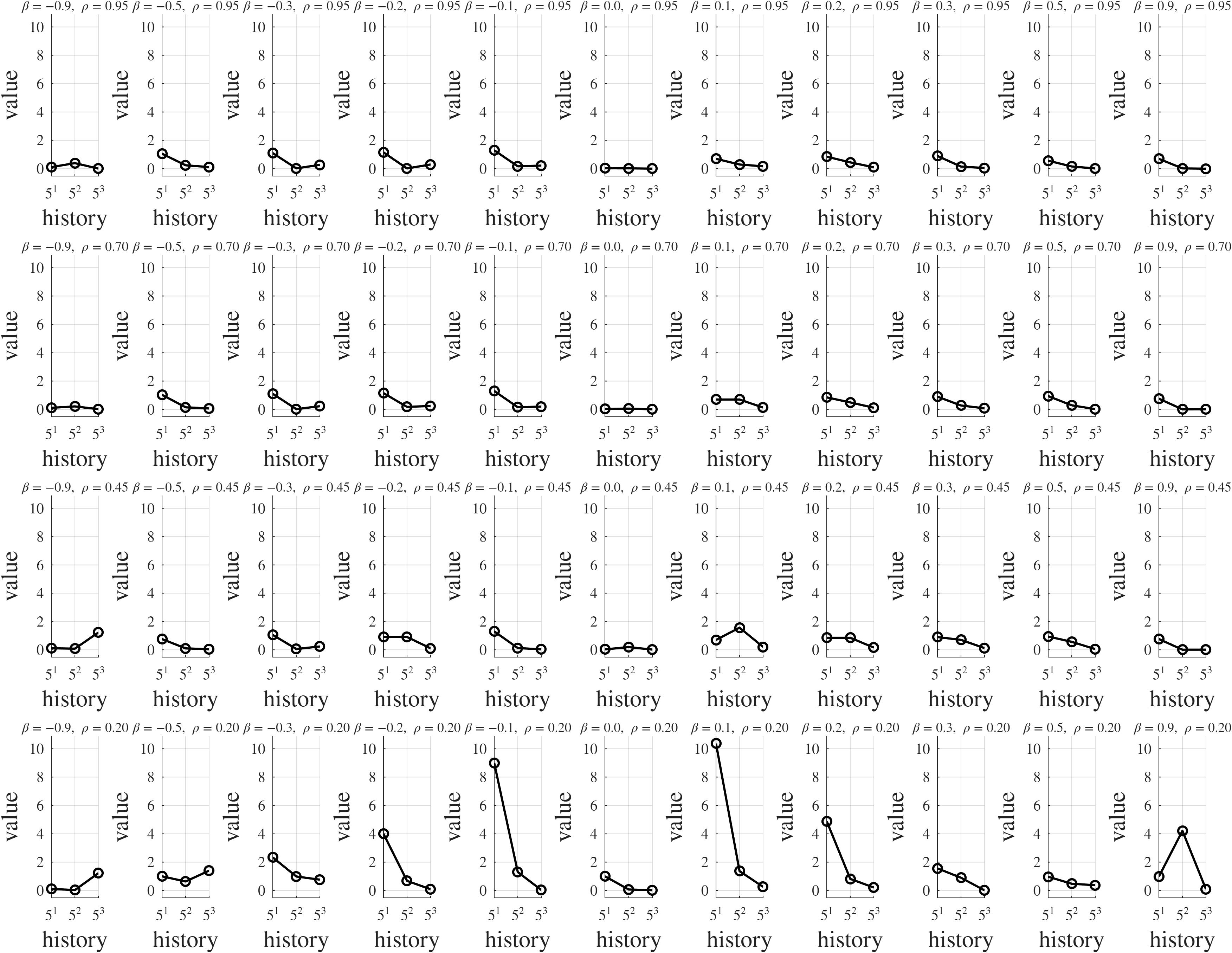}
\caption{Normalized learning error for disappointment aversion (NLE($\beta$)) by history size for each parameter combination ($\beta$,$\rho$) under SVR. Each panel plots the bootstrap mean of NLE($\beta$) across history sizes $h \in \{5^1, 5^2, 5^3 \}$, with bars representing 95\% bootstrap confidence intervals. Rows correspond to increasing values of $\rho$ (from bottom to top), and columns correspond to increasing values of $\beta$ (from left to right).}
\label{fig:SVR_additional_graphs3}
\end{figure}


\end{document}